\documentclass[useAMS,usenatbib]{mn2e}
\usepackage{graphicx}
\usepackage{color}

\newcommand{\be}{\begin{equation}}
\newcommand{\ee}{\end{equation}}

\def\no{\noindent}

\newcommand{\ifm}[1]{\relax\ifmmode#1\else$\mathsurround=0pt #1$\fi}
\newcommand{\kms}{\ifmmode\,{\rm km}\,{\rm s}^{-1}\else km$\,$s$^{-1}$\fi}

\newcommand{\ltsima}{$\; \buildrel < \over \sim \;$}
\newcommand{\lsim}{\lower.5ex\hbox{\ltsima}}
\newcommand{\gtsima}{$\; \buildrel > \over \sim \;$}
\newcommand{\gsim}{\lower.5ex\hbox{\gtsima}}

\definecolor{green}{rgb}{0,0.5,0}
\definecolor{grey}{rgb}{0.4,0.5,0.7}

\def\M11{M_{11}}
\def\V100{V_{100}}
\def\R1{R_{Mpc}}
\def\T6{T_6}

\begin{document}

\title[The new semianalytic code GalICS 2.0]{The new semianalytic code GalICS 2.0 
- Reproducing the galaxy stellar mass function and the Tully-Fisher relation simultaneously}

\pagerange{\pageref{firstpage}--\pageref{lastpage}} \pubyear{2016}

\author[Cattaneo et al.]{A.~Cattaneo$^{1,4}$, J.~Blaizot$^2$, J.~E.~G.~Devriendt$^3$, G.~A.~Mamon$^4$, E.~Tollet$^{1,5}$, 
\newauthor
A.~Dekel$^6$, B.~Guiderdoni$^2$, M.~Kucukbas$^{1,5}$, A.~C.~R.~Thob$^7$
\\
\\
$^1$Observatoire de Paris, GEPI, 77 av. Denfert-Rocherau, 75014 Paris, France \\
$^2$Centre de Recherche Astrophysique de Lyon, 9 avenue Charles Andr{\'e}, 69561 Saint-Genis-Laval, France\\
$^3$University of Oxford, Astrophysics, Keble Road, Oxford OX1 3RH, UK\\
$^4$Institut d'Astrophysique de Paris, 98bis Boulevard Arago, 75014 Paris, France\\
$^5$Universit{\'e} Paris VII, Sorbonne Paris Cit{\'e}, 5 rue Thomas-Mann, 75205 Paris Cedex 13, France \\
$^6$The Hebrew University, Racah Institute of Physics, Jerusalem 91904, Israel\\
$^7$Liverpool John Moores University, Astrophysics Research Institute, 146 Brownlow Hill, Liverpool L3 5RF, UK
}

\maketitle

\label{firstpage}

%%%%%%%%%%%%%%%%%%%%%%%%%%%%%%%%%%%%%%

\begin{abstract}

GalICS 2.0 is a new semianalytic code to model the formation and evolution of galaxies in a cosmological context.
N-body simulations based on a {\it Planck} cosmology are used to construct halo merger trees, track subhaloes, 
compute spins and measure concentrations.
The accretion of gas onto galaxies and the morphological evolution of galaxies are modelled with prescriptions derived from hydrodynamic
simulations.
Star formation and stellar feedback are described with phenomenological models (as in other semianalytic codes).
GalICS 2.0 computes rotation speeds
from the gravitational potential of the dark matter, the disc and the central bulge.
As the rotation speed depends not only on the virial velocity but also on the ratio of baryons to dark matter within a galaxy,
our calculation predicts a different Tully-Fisher relation from models in which $v_{\rm rot}\propto v_{\rm vir}$.
This is why GalICS 2.0 is able to reproduce the galaxy stellar mass function and the Tully-Fisher relation simultaneously.
Our results are also in agreement with halo masses from weak lensing and satellite kinematics,
gas fractions, the relation between star formation rate (SFR) and stellar mass, the evolution of the cosmic SFR density, 
bulge-to-disc ratios, disc sizes and the Faber-Jackson relation.

\end{abstract} 


%%%%%%%%%%%%%%%%%%%%%%%%%%%%%%%%%%%%%%%%%%%%%%
 
\section{Introduction}
 
Semianalytic models (SAMs) are a technique to model the formation and evolution of galaxies in a cosmological context.
Pioneered by \citet{white_frenk91} and \citet{lacey_cole93}, this technique is based on the notion that galaxy formation is a two-stage process \citep{white_rees78}.
The gravitational instability of primordial density fluctuations in the dark matter (DM) forms haloes. The dissipative infall of gas within haloes forms luminous galaxies.
SAMs follow these two stages separately. First, one constructs merger trees for the haloes in a representative cosmic volume.
Then, the evolution of baryons within haloes is broken down into a number of elementary processes, which are modelled analytically.

This article introduces the new SAM GalICS 2.0. 
An early version had already been presented in a comparison of all the main SAMs \citep{knebe_etal15}.
The models that participated to this comparison are those by \citet{bower_etal06}, \citet{font_etal08}, \citet{gonzalez_etal14}, \citet{croton_etal06}, \citet{delucia_blaizot07}, \citet{henriques_etal13}, \citet{benson12}, \citet{monaco_etal07} , \citet{gargiulo_etal15}, \citet{somerville_etal08} and \citet{lee_yi13}.

GalICS 2.0 builds on our previous experience with GalICS \citep{hatton_etal03,cattaneo_etal06,cattaneo_etal08,cattaneo_etal13} but is more than a new version. 
The entire code has been re-written from scratch.
One of the reasons is to enable a more extensive use of the cosmological N-body simulation used to contruct the merger trees.
In GalICS 2.0, we use the information on DM substructures (merger rates are more accurate) and the density profiles of DM haloes (we can compute realistic rotation curves).
Other advantages, besides an improved description of several physical processes, are a massive gain in computational speed 
and far greater modularity.

SAMs need complex baryon physics (radiative cooling, shock heating, active galactic nuclei, supernovae) to explain why the galaxy stellar mass function (SMF) 
has a knee at $M_{\rm stars}\sim 10^{11}\,M_\odot$ when the mass function of their host haloes is essentially a single power law up to $M_{\rm vir}\sim 10^{13}\,M_\odot$.
Yet the assumption that the growth histories of DM haloes determine the properties of galaxies underpins the entire semianalytic approach.

There are three main ways to measure halo masses and probe the galaxy - halo connection directly:
from rotation curves, from satellite kinematics and from weak lensing data.
The first method is the oldest. In fact, it is one of the ways DM was discovered.
Hence the importance of the relation between stellar mass and disc rotation speed (Tully-Fisher relation, TFR) as a key test for SAMs.
However, reproducing the TFR and the SMF simultaneously has been a main challenge for SAMs since their inception.
Either models are calibrated on the TFR and fail to fit the SMF/luminosity function \citep{kauffmann_etal93} 
or they are calibrated on the SMF/luminosity function and fail to reproduce the TFR \citep{heyl_etal95}. The discrepancy persists today, albeit to a lesser extent
(e.g., \citealp{guo_etal11}; also see \citealp{guo_etal10}, although the latter is based on abundance matching rather than semianalytic modelling).
The need to compare the predictions of SAMs to direct probes of halo masses has played a major role in the development of GalICS 2.0.
In this article, we show that modelling the rotation curves of disc galaxies accurately is necessary for a meaningful comparison with Tully-Fisher data.

The structure of the article is as follows. 
In Section~2, we describe GalICS 2.0 and we explain our strategy to set the values of the model parameters.
In Section~3, we compare its predictions with the observations (SMFs, baryonic mass function, halo masses from weak lensing and satellite kinematics, 
relation of SFR to gas and stellar mass, SFR function, evolution of the cosmic SFR density, gas fractions, bulge-to-disc ratios, disc sizes, stellar and baryonic TFR
at $z\sim 0$, stellar TFR at $z\sim 1$, the relation between disc rotation speed and virial velocity from rotation-curve studies, and the Faber-Jackson relation).
Section~4 summarises the conclusions of the article.
  
\section{The model}

Luminous matter is composed of galaxies and the intergalactic medium.  We could also say it is composed of gas and stars.
The objects we use to describe the Universe depend on the scale we are looking at.
GalICS 2.0 works the way we think. 
Different modules follow different objects, which capture the formation and evolution of galaxies on different scales.
Each is written to be as self-contained as possible.

On the largest scale, the {\sc tree} module follows the hierarchical formation and merging of DM haloes. 
For {\sc tree}, a halo is just a point in a network of relationships (progenitor, descendant, host, subhalo).
The flow of baryons in and out of haloes is followed by the {\sc halo} module.
{\sc halo} follows the exchanges of matter between the cold gas, the hot gas and the galaxy (e.g., the rate at which gas accretes onto the galaxy) but not the galaxy's internal structure.
The decision of what goes to the disc and what goes to the bulge is done in the {\sc galaxy} module, which computes all morphological, structural and kinematic properties.
The relation between {\sc halo} and {\sc galaxy} can be compared to that between a mill and a baker. There is an exchange of matter both ways (inflows and outflows, flour and money)
but the baker does not need to know if the flour has been ground with a water or a wind mill. Neither does the millman about the baker's recipies.
This philosophy explains some practical choices, such as that of the time substeps in Section~3.4.
A galaxy contains different components, such as a disc, a bulge or a bar (which we classify as a pseudobulge),
but star formation and feedback within a component are followed in the {\sc component} module.
The lowest scale corresponds to the {\sc star} (stellar evolution) and {\sc gas} (interstellar medium) modules.

GalICS 2.0 exists in both a {\sc Fortran 2003} and a {\sc C++} version.
Their quantitative agreement to several significant digits on a galaxy by galaxy basis
is one of the reasons why we are confident of the code’s quality and reliability.
Finally, GalICS 2.0 should not be confused with eGalICS \citep{cousin_etal15b}.
eGalICS started from an early version of GalICS 2.0 but the two codes have been developed independently in the last few years and should be considered as different SAMs.

\subsection{\sc tree}

\subsubsection{Cosmology and analysis of the N-body simulation}

GalICS 2.0 uses DM merger trees from cosmological N-body simulations. 
In this article, we use a simulation with $\Omega_M=0.308$, $\Omega_\Lambda=0.692$, $\Omega_b = 0.0481$,  $\sigma_8=0.807$ and $H_0=67.8{\rm\,km\,s}^{-1}{\rm Mpc}^{-1}$
(\citealp{planck14}, {\it Planck} + WP + BAO). 
The simulation has a volume of $(100{\rm\,Mpc})^3$ and contains $512^3$ particles (implying a particle mass of $2.9\times 10^8\,M_\odot$). 
As the Poisson equation is solved on a non-uniform mesh using a multi-grid method in RAMSES (\citealp{teyssier02} for details), the force
resolution is not spatially uniform. The simulation reaches a maximum force resolution of $1.5\,$kpc (physical units)
in the densest regions (the centres of dark matter haloes).
Outputs have been saved at $265$ timesteps equally spaced in the logarithm of the expansion factor between $z=13.2$ and $z=0$.

Haloes and subhaloes are identified with the halo finder HaloMaker, which is based on AdaptaHOP \citep{tweed_etal09}.
For each halo containing at least a hundred particles (corresponding to a minimum halo mass of $2.9\times 10^{10}\,M_\odot$),
we determine the inertia ellipsoid, centred on its centre of mass,  after iteratively removing gravitationally unbound particles,
and we keep rescaling it until the critical overdensity contrast inside the inertia ellipsoid
equals the one that we compute  with the fitting formulae of \citet{bryan_norman98} for a {\it Planck} cosmology\footnote{$3M_{\rm vir}/(4\pi\rho_cr_{\rm vir}^3)$, where $\rho_c$ is the critical density of the Universe, cannot be exactly equal to $\Delta_c$ because $M_{\rm vir}$
can only take values that are multiples of the N-body particle's mass. However, the only haloes for which the difference is significant are low-mass systems just above the detection threshold.}.
The virial mass $M_{\rm vir}$ is the mass of the gravitationally bound N-body particles contained within the virial ellipsoid (i.e., the rescaled inertia ellipsoid).
The virial radius $r_{\rm vir}$ is that of a sphere whose volume equals that of the virial ellipsoid.
At $z=0$, \citet{bryan_norman98}'s formulae give  $\Delta_c=102$.
Hence, a halo  mass of $M_{\rm vir}=10^{12}\,M_\odot$ corresponds to $v_{\rm vir}=\sqrt{{\rm G}M_{\rm vir}/r_{\rm vir}}=128{\rm\,km\,s}^{-1}$.

\begin{figure}
\begin{center}
\includegraphics[width=0.95\hsize]{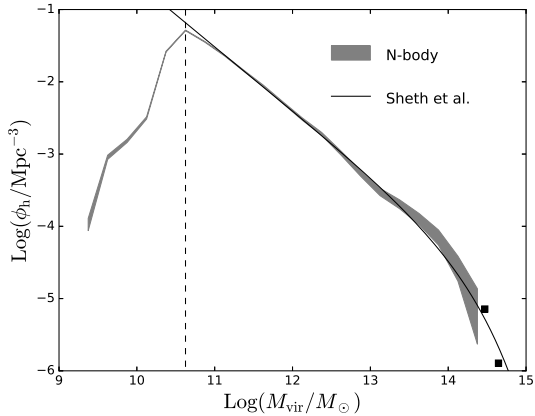} 
\end{center}
\caption{The halo mass function measured from our N-body simulation (shaded area) compared to a \citet{sheth_etal01} fit for the same cosmology 
(the solid curve corresponds to $a=0.6$ and $q=0.354$, where the parameters $a$ and $q$ are defined as \citealp{sheth_etal01}).
Here $\phi_{\rm h}$ is the number density of haloes per dex of virial mass.
If one defines the halo mass function $n$ so that $n(M_{\rm vir}){\rm\,d}M_{\rm vir}$ is the number density of haloes with mass between $M_{\rm vir}$ and $M_{\rm vir}+{\rm d}M_{\rm vir}$,
then the relation between $\phi_{\rm h}$ and $n$ is $\phi_{\rm h} = n(M_{\rm vir})M_{\rm vir}{\rm\,log}(10)$.
The thickness of the shaded curve is determined by the Poissonian error on the number of haloes in each mass bin.
The black squares are data points from \citet{wen_etal10} for the mass function of galaxy clusters.
The vertical dashed line at $M_{\rm vir} = 10^{10.625}\,M_\odot$ ($\sim 150$ particles) 
provides a measure of the halo-mass resolution.}
\label{HMF}
\end{figure}

Fig.~\ref{HMF} shows the halo mass function that we measure in our N-body simulation at $z=0$. The resolution mass 
is clearly visible as the $M_{\rm vir}$ below which the mass function drops (the peak is at $M_{\rm vir} = 10^{10.625}\,M_\odot$ rather than at  $M_{\rm vir} \simeq 10^{10.5}\,M_\odot$
simply because this is the midpoint of the logarithmic mass bin $10^{10.5}\,M_\odot<M_{\rm vir}<10^{10.75}\,M_\odot$).
The effects of mass variance/low-number statistics at high masses can be quantified by comparing our mass function (the gray shaded area) with the analytic fit of 
\citet{sheth_etal01}. This fit contains two free parameters, which the original article adjusted on N-body simulations by \citet{kauffmann_etal99}, and
which we have recalibrated on ours. The fit is consistent with the mass function of galaxy clusters from Wen et al. (2010; black squares in Fig.~\ref{HMF}).
The comparison of our halo mass function (gray shaded area) with this fit (black solid curve) shows good agreement up to $M_{\rm vir}\sim 10^{13.3}\,M_\odot$.
However, at $M_{\rm vir}\sim 10^{13.8}-10^{14}\,M_\odot$, the centre of the gray shaded area is above the black solid curve by about $\sim 40\%$.
Number densities for the central galaxies of haloes in this mass range may overestimated by a similar factor.

For each halo, we measure the virial angular momentum ${\bf J}_{\rm vir}$, which we use to compute the spin parameter $\lambda$ (Section~2.3.2).
\citet{bett_etal07} argued that they needed at least three hundred particles to measure halo spins accurately but, going from three hundred to one hundred particles, as we do here, 
the median spin parameter changes from $\lambda=0.0425$ to $\lambda=0.044$ (Fig.~7 of \citealp{bett_etal07}).
The difference ($<4\%$) is well within the uncertainties of the results presented in this work and is comparable to the uncertainties that derive from different unbinding procedures (e.g., \citealp{onions_etal13}).

We also fit the mass distribution of each halo with an NFW profile \citep{navarro_etal97} to measure its concentration $c$.
\citet{neto_etal07} found that they needed at least $10^4$ particles to measure concentrations accurately.
Indeed, our concentrations drop below the fitting formulae by \citet{munoz_etal11} and \citet{dutton_maccio14} for $M_{\rm vir}<10^{12}M_\odot$
(corresponding to $\sim 3000$ particles), suggesting that our measurements are affected by N-body resolution in a manner that may be significant for haloes with $M_{\rm vir}<10^{11.5}\,M_\odot$.
Most of the figures shown in this article are not sensitive to the value of $c$ but the TFR is.
We could use the concentrations that we measure in our N-body simulation for haloes with more than $3000$ particles and the fitting formulae of
\citet{dutton_maccio14} for haloes with less than $3000$ particles but this approach neglects the large scatter in measured concentration values and poses the problem that 
our $c$ - $M_{\rm vir}$ relation exhibits systematic differences with respect to that of \citet{dutton_maccio14} even for haloes with more than $10^5$ particles.
There is also the problem of applying fitting formulae for haloes to subhaloes.
In this article, we do all calculations with concentrations from our N-body simulation for self-consistency.
However, in Section~3.7, we explore how the TFR varies when our concentrations measurements are replaced with values from
the fitting formulae of \citet{dutton_maccio14}.

In the case of a subhalo, we use the same procedure applied to haloes to obtain a first estimate of $r_{\rm vir}$. Then, we shrink the subhalo by peeling off its outer layers
until the density at the recomputed virial radius is at least as large as the host density at the position where the subhalo is located.
The concentration parameter $c$ is recomputed accordingly.
The particles peeled off the outer layers of a subhalo are reassigned to the host halo if they are gravitationally bound to it.
The host halo masses used in GalICS 2.0 are exclusive, i.e., they do not include those of subhaloes. 
By construction, a host halo is always more massive than its most massive subhalo.
 
The TreeMaker algorithm \citep{tweed_etal09} is used  to link haloes/subhaloes identified at different redshifts and to generate merger trees.
A halo is identified as the descendent of another when it inherits more than half of its progenitor's particles.
Because of this definition, a halo can have many progenitors but at most one descendent. The main progenitor is always the one with the largest virial mass.
A halo/subhalo is found to have no descendent if it looses more than half its mass but no single halo accretes enough mass from it to qualify as its descendent.
Haloes that disappear in this manner exist in our merger trees but they are so rare that they are not statistically significant (tidally stripped particles are normally accreted by the halo that causes the tide).

\subsubsection{Halo representation}

The {\sc tree} module reads the halo catalogues and the tree structure. The properties that are read and used for each halo are: 
1) virial mass, radius and angular momentum, $M_{\rm vir}$, $r_{\rm vir}$ and ${\bf J}_{\rm vir}$,
2) concentration $c$, 
3) position and velocity of the centre of mass,
4) position in the hierarchy of substructures (host halo/subhaloes) and
5) position in the tree (progenitors/descendant).		

The virial mass $M_{\rm vir}$ measured in the N-body simulation is a total mass of DM and baryons, treated as if they were both collisionless.
Assuming that the DM distribution is described by the NFW profile, the mass of DM enclosed within a sphere of radius $r$ is
\begin{equation}
M_{\rm dm}(x)={\log(1+cx)-{cx/(1+cx)}\over \log(1+c)-{c/(1+c)}}\cdot{\Omega_M-\Omega_b\over\Omega_M}M_{\rm vir},
\label{M_nfw}
\end{equation}
where $x \equiv r/r_{\rm vir}$.
Here and throughout this paper, $\log$ is the natural logarithm. The decimal logarithm is ${\rm Log}$.

\subsubsection{Scheme to evolve baryons along merger trees}
							
The code loops over all timesteps. At each timestep, it loops over all haloes and calls the routines that compute the evolution of the baryons within them.
The transition from one timestep to the next is handled as follows.
Let $t_s$ be the age of the Universe at timestep $s$.
If a halo detected at timestep $s$ has one or more progenitors at timestep $s-1$, we compute a random merging time 
\begin{equation}
t_{s-1}<t_{\rm m}<t_{s}.
\label{t_m}
\end{equation}
If the halo has no progenitors, we assume that $t_{\rm m}=t_{s-1}$ is its formation time.
The baryons are evolved using the DM properties at $t_{s-1}$ between $t_{s-1}$ and $t_m$. If there are more than one progenitor, 
they are merged at $t_{\rm m}$. The merger remnant, or the halo if there is only one progenitor, is then
evolved between $t_m$ and $t_s$ using the DM properties measured at
$t_s$.

In the code’s current version, there is a one-to-one correspondence
between satellite galaxies and subhaloes. When the latter merge, so do the former.
However, we have developed a beta version 
that includes the possibility of delayed mergers. 
The beta version computes the delay using \citet{jiang_etal08}'s formula for the dynamical friction timescale (see \citealp{cattaneo_etal11} for a description of the method).
Preliminary investigations with the beta version show little difference with respect to the conclusions of this article.

\subsection{\sc halo}

The {\sc halo} module follows the accretion of gas onto a halo and the exchanges of matter between its baryonic components (the cold gas in the halo, the hot-gas halo and the central galaxy).

\subsubsection{Accretion}

The mass $M_b$ of the baryons within a halo is updated from its value at $t_{\rm m}$ (with $t_{s-1}<t_{\rm m}<t_{s}$) to its values at $t_s$ with the equation:
\begin{equation}
\Delta M_b =\max[f_bM_{\rm vir}(t_s) -M_b(t_m),0],
\label{Delta_Mb}
\end{equation}
where $M_b(t_m)$ is the sum of the baryon masses of all progenitors at the time of merging. 

There is no accretion onto subhaloes or their descendents.
This prescription models the physical phenomenon of strangulation through ram-pressure and tidal stripping \citep{gunn_gott72,abadi_etal99,balogh_etal00,balogh_morris00,peng_etal15}.
It is also imposed because a halo may become a subhalo, evade detection by the halo finder as it passes through the centre of its host, reappear as 
a subhalo one or two timesteps later, and finally be identified as a halo again when it comes out on the other side (``backsplash haloes'').
When the subhalo disappears, the associated satellite galaxy merges with the central one 
unless the subhalo is not bound to the host, in which case the satellite galaxy and its baryons
disappear from the model universe. However, this case is rare and has no impact on our statistical predictions. 
Backsplash haloes are much more frequent. Allowing them to accrete gas would duplicate baryons by reinstating galaxies that have just merged.
Hence, in GalICS 2.0, backsplash haloes are haloes with no galaxies (we are working on a new version that improves the description of these systems).

In Eq.~(\ref{Delta_Mb}),  $f_b = f_b(M_{\rm vir},z)$ is a function that models reionisation feedback (gas will not accrete onto haloes with virial temperature lower than the temperature of the intergalactic medium).  
We assume
that the intergalactic medium has a Maxwellian velocity distribution and that the escape speed is $v_{\rm esc}=v_{\rm vir}\sqrt 2$, where $v_{\rm vir}(M_{\rm vir},z)$ is the halo virial velocity. This assumption gives:
\begin{equation}
f_b ={\Omega_b\over\Omega_M} \int_0^{v_{\rm vir}\sqrt 2} 4\pi\left({\mu\over 2\pi kT_{\rm reio}}\right)^{3\over 2}v^2e^{-{\mu v^2\over 2kT_{\rm reio}}}{\rm\,d}v,
\label{maxwell}
\end{equation}
where $T_{\rm reio}$ is the temperature at which the intergalactic medium is reionised,  $\mu$ is the mean particle mass and $k$ is the Boltzmann constant.
The cosmic baryon fraction $\Omega_b/\Omega_M$ appears in front of the right-hand side of Eq.~(\ref{maxwell}) because the integral gives
the fraction of the {\it baryonic} mass with $v<v_{\rm esc}$, while $f_b$ is expressed in terms of the total mass $M_{\rm vir}$.
Let $\sigma_{\rm reio}$ be the one-dimensional thermal velocity dispersion of the intergalactic medium, so that $\mu \sigma_{\rm reio}^2 = kT_{\rm reio}$.
Then, Eq.~(\ref{maxwell}) becomes:
\begin{equation}
f_b={\Omega_b\over\Omega_M}\left[
{\rm erf}\left({v_{\rm vir}\over \sigma_{\rm reio}}\right) -{2\over\sqrt{\pi}} \left({v_{\rm vir}\over \sigma_{\rm reio}}\right)e^{-\left({v_{\rm vir}\over \sigma_{\rm reio}}\right)^2}
\right].
\label{f_b} 
\end{equation}
Eq.~(\ref{f_b}) implies that $f_b = 0.5\Omega_b/\Omega_M$ for $v_{\rm vir}/\sigma_{\rm reio}=1.088$.

Our assumption for $v_{\rm esc}$ is accurate only for a singular isothermal sphere. In the NFW model, $v_{\rm esc}\sim 2.5v_{\rm vir}$ (the exact value depends on concentration).
However, the difference in $v_{\rm esc}$ will simply be reabsorbed by $\sigma_{\rm reio}$, which is a free parameter of the model.
For this reason, $\sigma_{\rm reio}$ may differ from the physical thermal velocity dispersion of the intergalactic medium by a factor of order unity.
Finally, reionisation feedback has been added with a view to running GalICS 2.0 on higher-resolution N-body simulations
because we expect that $\sigma_{\rm reio}\sim 20-30{\rm\,km\,s}^{-1}$
and the current simulation cannot resolve haloes with $v_{\rm vir}< 40{\rm\,km\,s}^{-1}$.

A fraction $f_{\rm hot}$ of the accreted baryon mass $\Delta M_b$ is
shock heated to the virial temperature while falling in and is added to the hot halo.
The rest accretes onto the central galaxy through cold filamentary flows. 
The fraction $f_{\rm hot}(M_{\rm vir})$ is assumed to be $f_{\rm hot}=0$ for $M_{\rm vir} \le M_{\rm shock}$,
\begin{equation} 
f_{\rm hot}={{\rm Log\,}M_{\rm vir}-{\rm Log\,}M_{\rm shock} \over  {\rm Log\,}M_{\rm shutdown}-{\rm Log\,}M_{\rm shock}}
\label{f_hot}
\end{equation}
for $M_{\rm shock}<M_{\rm vir}<M_{\rm shutdown}$, and $f_{\rm hot}=1$ for $M_{\rm vir}\ge M_{\rm shutdown}$.
Here $M_{\rm shock}$ is the halo mass at which the accreted gas begins to be shock-heated, 
while $M_{\rm shutdown}$ is the halo mass above which shock heating is complete and any residual cold gas in the filaments is evaporated by the hot phase. 
While $M_{\rm shock}$ and $M_{\rm shutdown}$ are in principle free parameters of the model, to be determined by fitting the galaxy SMF,
the functional form in Eq.~(\ref{f_hot}) is based on work by \citet{ocvirk_etal08}, 
who measured the flow rates of cold ($T<10^5\,$K) and hot  ($T>10^5\,$K) gas through a spherical surface of radius $0.2r_{\rm vir}$
in the Horizon-Mare Nostrum cosmological hydrodynamical simulation.
The results of \citet{ocvirk_etal08} for
\begin{equation}
f_{\rm hot} ={\dot{M}_{\rm hot}\over \dot{M}_{\rm cold}+\dot{M}_{\rm hot}}
\label{ocvirk}
\end{equation}
at $0.2r_{\rm vir}$ are very similar to a ramp between
${\rm Log}(M_{\rm shock}/M_\odot)= 10.7$ and ${\rm Log}(M_{\rm shutdown}/M_\odot)= 12.7$, 
at least in the redshift range $2<z<4$
(the Horizon-Mare Nostrum simulation stops at $z=2$).
In GalICS, \citet{cattaneo_etal06} found a good fit to SDSS data for a sharper cut-off of the form $M_{\rm shock}=M_{\rm shutdown}=10^{12.3}\,M_\odot$.

The equations for the variations of the masses  of the hot gas and
the cold filaments are:
\begin{equation}
\Delta M_{\rm hot} = f_{\rm hot}\Delta M_b,
\label{Delta_Mhot}
\end{equation}
\begin{equation}
\Delta M_{\rm  fil} = (1-f_{\rm hot})\Delta M_b-{M_{\rm fil}\over t_{\rm ff}}\Delta t,
\label{Delta_Mfil}
\end{equation}
where $\Delta M_b$ is the accreted mass calculated in Eq.~(\ref{Delta_Mb}).
The second term on the right hand side of Eq.~(\ref{Delta_Mfil}) is the accretion rate from the filaments onto the galaxy,
which we assume to take place on a freefall timescale $t_{\rm ff}=r_{\rm vir}/v_{\rm vir}$.

Following \citet{dekel_birnboim06} and \citet{dekel_etal09}, we assume that the accretion of cold gas is the main mode of galaxy formation and that hot gas never cools. 
Hence, there is no cooling term  in Eq.~(\ref{Delta_Mhot}).
The assumption of a total shutdown of gas accretion in massive haloes is extreme
(see, e.g., \citealp{bildfell_etal08}),
but we know from previous work \citep{cattaneo_etal06} that its predictions are in good agreement
with the galaxy colour-magnitude distribution, while letting the hot gas cool leads to results in clear disagreement with the observations. 
Introducing cooling makes sense only if one has a physical model of how AGN feedback mitigates it (see \citealp{cattaneo_etal09} for a review).
Attempts in this direction have been made, starting with \citet{croton_etal06}, \citet{bower_etal06} and \citet{somerville_etal08}.
Following \citet{benson_babul09}, \citet{Fanidakis_etal11} have gone as far as to compute the mechanical luminosity of the jets from the accretion rate and the spin of the black holes that power them
(an approach pioneered in SAMs by \citealp{cattaneo02}). However, the physics of these models are uncertain.
Hence, we have considered that it would be premature to include AGN feedback in our SAM, especially in an article focused on spiral galaxies, which tend to live in haloes
with $M_{\rm vir}\lsim 10^{12}\,M_\odot$.

In these lower-mass systems, however, the assumption that hot gas never cools may be even more extreme 
because it prevents the reaccretion of ejected gas (see the discussion in Section~2.2.2) and thus the possibility of substantially delaying star formation
with respect to gas accretion onto the halo.
It is important to realize that, in this article,  we are not arguing for the absence of cooling on physical ground.
This is the simplest possible assumption within the cold-flow paradigm and we want to explore how far it can take us.
In Section~3, we shall show that the results are reasonably good, although we have not compared them to all possible observations.
(\citealp{bower_etal06} have suggested that the gradual reaccretion of ejected gas is necessary in
order to predict a large enough fraction of galaxies on the blue sequence at low stellar masses).

The effects of introducing cooling and therefore reaccretion will be explored in a future publication.
However, we note that, to an extent,
Eq.~(\ref{f_hot}) already includes some of the effects of cooling implicitly because the cold gas fraction $1-f_{\rm hot}$ that \citet{ocvirk_etal08} measure 
in their simulation at $0.2r_{\rm vir}$
includes any gas that may have been shock-heated and cooled before reaching $0.2r_{\rm vir}$ (the outer boundary of the {\sc Hi} disc; C. Pichon, private communication),
although it does not account for the possibility that a cooling-flow may develop inside the galaxy itself and for the reaccretion of ejected gas.
  
\subsubsection{Outflows}

The gas that is blown out of the galaxy is either mixed to the hot gas in the halo or expelled  from the halo altogether. 
In the second case, we store it in an outflow component, so that we always know how much gas and how many metals have been expelled from each halo.

The baryon mass $M_b$ that enters Eq.~(\ref{Delta_Mb}) is
\begin{equation}
M_b = M_{\rm hot} + M_{\rm fil} + M_{\rm gal} + M_{\rm out},
\label{M_b}
\end{equation}
where $M_{\rm gal}$ is the total baryonic mass of the galaxy. 
This definition of $M_b$ includes the ejected mass $M_{\rm out}$ to prevent its re-accretion.
Hence, the actual halo baryon fraction $(M_{\rm hot} + M_{\rm fil} + M_{\rm gal})/M_{\rm vir}$ may be 
lower than
$M_b/M_{\rm vir}$. 
Had we defined $M_b$ as $M_b = M_{\rm hot} + M_{\rm fil} + M_{\rm gal}$,
$M_{\rm out}$ would have been available for immediate reaccretion onto the halo.

In this paper, it makes no difference whether galactic winds mix with the hot gas or escape from the halo
because hot gas is not allowed to cool. Physically, however, cooling can only be important if the mass of hot gas is at least comparable
to the mass of cold gas in the filaments, in which case
the material blown out of the galaxy will almost certainly mix with it.

\subsection{\sc galaxy}

The {\sc galaxy} module deals with the structural properties of galaxies (morphologies, scale lengths, kinematics) and 
is organized around two main routines. {\sc galaxy evolve} follows the evolution of an individual galaxy over an interval of time.
It computes disc radii, speeds, and the formation of pseudobulges through disc instabilities.
{\sc galaxy merge} models morphological transformations induced by mergers.

\subsubsection{Galaxy structure}

A galaxy is modelled as the sum of three components: a disc, built through gas accretion and minor mergers, 
a pseudobulge, built through disc instabilities, and a classical bulge, built through major mergers. 

The disc has an exponential surface-density profile:
\begin{equation}
\Sigma(r)=\Sigma_0{\rm\,exp}\left(-{r\over r_d}\right).
\label{exp_pro}
\end{equation}
The pseudobulge originates from the buckling and bar instability of the disc within radius $r_{\rm pseudo}$.
We therefore assume that, while instabilities transfer matter from the disc to the pseudobulge, 
the mass redistribution is mainly vertical and azimuthal, and that
the radial exponential profile of the disc plus pseudobulge system is largely unaffected by this process.

We assume that the bulge is spherical and that its density distribution is described by a \citet{hernquist90} profile.
For this model, the bulge mass within radius $r$ is 
\begin{equation}
M_{\rm bulge}(r) = \left({  {r/r_{\rm bulge}}\over 1+ {r/r_{\rm bulge}}  }\right)^2M_{\rm bulge},
\label{hernquist}
\end{equation}
where $r_{\rm bulge}$ is the scale radius of the bulge.

Having described the components with which we model a galaxy, we are ready to enter the details of the physical processes that determine their formation and characteristic quantities.

\subsubsection{Disc radii and rotation speeds}

Any gas that accretes onto a galaxy is automatically added to the disc component.
Its scale radius $r_d$ is determined by solving
the disc angular momentum equation
\begin{equation}
J_d = 2\pi\Sigma_0\int_0^\infty e^{-{r/r_d}}r^2v_c{\rm\,d}r =M_dr_d\int_0^\infty e^{-x} x^2v_c{\rm\,d}x,
\label{J_d}
\end{equation}
where $v_c(r)$ is the the disc rotation curve (see Eq.~\ref{v_c} below) and $M_d$ is the total baryonic mass (stars and gas) of the disc and the pseudobulge combined
(this is what $M_d$ stands for throughout this article).

Following \citet{fall_efstathiou80} and \citet{mo_etal98},
SAMs compute disc sizes by assuming that baryons and DM have the same initial angular momentum distribution and that specific angular momentum is conserved, i.e., that
\begin{equation}
{J_d\over M_d} = {J_{\rm vir}\over M_{\rm vir}}.
\label{ang_mom_cons}
\end{equation}
Cosmological hydrodynamic simulations have shown that the angular momentum of the gas is not conserved during infall \citep{kimm_etal11,danovich_etal15}. However,
even though Eq.~(\ref{ang_mom_cons}) can give incorrect results when applied to individual galaxies,
it retains a {\it statistical} validity because the distribution of specific angular momentum of discs is similar to that of DM haloes \citep{danovich_etal15}
This statistical validity is backed by observations both in the local Universe \citep{tonini_etal06} and at high redshift \citep{burkert_etal16}.

Eq.~(\ref{J_d})  can be re-written as
\begin{equation}
r_d={ \lambda r_{\rm vir}\over \int_0^\infty e^{-x} x^2{v_c\over v_{\rm vir}}{\rm\,d}x},
\label{r_d}
\end{equation}
where
\begin{equation} 
\lambda \equiv {J_h\over M_h r_{\rm vir} v_{\rm vir}}
\end{equation} 
is the halo spin parameter as defined by \citet{bullock_etal01}. This definition differs from the usual one by \citet{fall_efstathiou80}
by a factor equal to $\sqrt 2$ for a truncated singular isothermal sphere.
If $v_c$ were equal to $v_{\rm vir}$ at all radii, then the integral in Eq.~(\ref{r_d}) would make $2$ and
Eq.~(\ref{r_d}) would reduce to $r_d=\lambda r_{\rm vir}/2$. However, $v_c(r)$ is not flat.

The disc rotation curve 
$v_c(r)$ is determined by the sum in quadrature of three terms:
\begin{equation}
v_c^2(r)={GM_{\rm dm}(r)\over r} +{GM_{\rm bulge}(r)\over r}+v_d^2(r),
\label{v_c}
\end{equation}
where $M_{\rm dm}(r)$ is the DM mass with radius $r$ (Eq.~\ref{M_nfw}), $M_{\rm bulge}(r)$ is the bulge mass within radius $r$ (Eq.~\ref{hernquist}) and $v_d(r)$ is the contribution from the disc, which
has a more complicated form because the disc has a cylindrical rather than spherical symmetry. 
We compute $v_d(r)$ exactly by using Bessel functions as in \citet{freeman70}.
The dependence of $v_d(r)$ on $r_d$ is the reason why solving Eq.~(\ref{J_d}) is not trivial.
The pseudobulge does not appear as a fourth contribution to $v_c(r)$ in Eq.~(\ref{v_c}) because it is simply the inner disc that buckles up
(Section~2.3.3). There is no radial migration in our model.

\subsubsection{Adiabatic contraction}

Adiabatic contraction is the contraction of the DM halo in response to the infall of the baryons.
\citet{blumenthal_etal86} estimated it from the adiabatic invariance of the specific angular momentum $rv_c(r)\propto\sqrt{rM(r)}$, where $M(r)$ is the total mass enclosed in a sphere of radius $r$.
If $r_i$ is the initial radius of a DM shell that contracts to radius $r$ and $M_{\rm NFW}(r_i)$ is the initial DM profile (assumed to be described by the NFW model), then this assumption 
gives the equation:
\begin{equation}
r[M_{\rm NFW}(r_i)+M_d(r)+M_{\rm bulge}(r)] = r_i{M_{\rm NFW}(r_i)\over 1-\Omega_b/\Omega_M},
\label{adiabatic_contraction}
\end{equation}
from which $r_i(r)$ and thus $M_{\rm dm}(r) = M_{\rm NFW}(r_i)$ can be computed.

Eq.~(\ref{adiabatic_contraction}) is the standard description of adiabatic contraction in models of galaxy formation
(e.g., \citealp{mo_etal98}; \citealp{cole_etal00}; \citealp{somerville_etal08}).
However, cosmological hydrodynamic simulations have shown that it overestimates its importance \citep{gnedin_etal04,abadi_etal10}. 
It is also likely that adiabatic contraction may be compensated by adiabatic expansion during massive outflows \citep{pontzen_governato12,teyssier_etal13,tollet_etal16}
because the haloes of dwarf galaxies have shallow cores rather than the central cusps predicted by cosmological simulations of dissipationless hierarchical clustering 
(\citealp{moore94} and \citealp{flores_primack96};  but also see \citealp{swaters_etal03}).
Hence, in the standard version of GalICS 2.0, there is no adiabatic contraction. 

A version with adiabatic contraction (computed with Eq.~\ref{adiabatic_contraction}) has however been explored. 
Its results will be briefly discussed in Section~3.7, when we talk about the possible effects of adiabatic contraction on the TFR.

\subsubsection{Disc instabilities}

Following \citet{efstathiou_etal82}, \citet{christodoulou_etal95} and \citet{vandenbosch98}, we assume that discs are unstable when
their self-gravity contributes more than a critical fraction of the circular velocity  (see, however, \citealp{athanassoula08} for a criticism of this model).
In formulae, our instability condition is:
\begin{equation}
v_d>\epsilon_{\rm inst}v_c,
\label{instability_condition}
\end{equation}
where $\epsilon_{\rm inst}\le 1$ is a free parameter of the model that sets the instability threshold
($\epsilon_{\rm inst}= 1$ corresponds to the unphysical assumption that all discs are always stable, that is, to turning off disc instabilities).

The pseudobulge radius $r_{\rm pseudo}$ is the largest radius at which the instability condition (Eq.~\ref{instability_condition}) is satisfied.
Its value is used to increment the pseudobulge mass with the algorithm
\begin{equation}
\Delta{M}_{\rm pseudo} = {\rm max}[M_d(r_{\rm pseudo})-M_{\rm pseudo},0],
\label{pseudo}
\end{equation}
where $M_{\rm pseudo}$ is the pseudobulge mass before incrementation.
Eq.~(\ref{pseudo}) guarantees that the pseudobulge mass never decreases (except in major mergers, where all components form one large classical bulge).
Gas and stars transferred from the disc to the pseudobulge contribute to $\Delta{M}_{\rm pseudo}$ in a ratio equal to the disc gas-to-stellar mass ratio.
The pseudobulge characteristic speed is $v_{\rm pseudo}=v_c(r_{\rm pseudo})$. 

We note that, in our model, the pseudobulge is any structure formed by disc instabilities, be it a peanut-shaped pseudobulge, a bar or an oval.

\subsubsection{Mergers}

A merger of two galaxies is major if ${\cal M}_1<\epsilon_m{\cal M}_2$ with ${\cal M}_1>{\cal M}_2$, where $\epsilon_m$ is a parameter of the model. 
Here
${\cal M}_1$ and ${\cal M}_2$ are total masses of the two galaxies (baryons and DM) within their
baryonic half-mass radii $r_1$ and $r_2$, which we compute numerically from the profiles of their discs and bulges.
 
In a major
merger, the two galaxies are destroyed and all their baryons are put into one large bulge, the size of which is determined by energy conservation \citep{cole_etal00,hatton_etal03,shen_etal03}:
\begin{equation}
{1\over 2}U_{\rm bulge} = {1\over 2}U_1+{1\over 2}U_2+E_{12},
\label{en_cons}
\end{equation}
where $U_{\rm bulge}$, $U_1$ and $U_2$ are the gravitational potential energies of the resulting bulge, galaxy~$1$ and galaxy~$2$, respectively, while $E_{\rm 12}$ is the interaction energy of the two-body system.
The $1/2$ coefficient in front of the potential-energy terms comes from the virial theorem, since, for each galaxy, the total energy equals half the gravitational potential energy.
Eq.~(\ref{en_cons}) can be re-written as:
\begin{equation}
{1\over 2}C_{\rm remn}{{\rm G}({\cal M}_1+{\cal M}_2)^2\over r_{1/2}}=
\label{en_cons2}
\end{equation}
$$={1\over 2}C_1{{\rm G}{\cal M}_1^2\over r_1}+{1\over 2}C_2{{\rm G}{\cal M}_2^2\over r_2}+{1\over 2}C_{\rm 12}{{\rm G}{\cal M}_1{\cal M}_2\over r_1+r_2}.$$
Here $r_{1/2}$ is the merger remnant's half-mass radius.
$C_{\rm remn}$, $C_1$ and $C_2$ are form factors that relate the gravitational potential energy of each system to its mass and half-mass radius.
$C_{12}\equiv 2E_{12}/U_{12}$ where $U_{12}=-{\rm G}{\cal M}_1{\cal M}_2/(r_1+r_2)$ is the two-body system gravitational binding energy.
The factor of two in the definition of $C_{12}$ comes, once agains, from the virial theorem.
$C_{12}=1$ corresponds to merging from circular orbits.
$C_{12}=0$ corresponds to a motion that starts with zero speed at infinity.

The form factor that relates the gravitational potential energy of a system to its mass and half-mass radius is about $0.4$ for a \citet{hernquist90} profile and $0.49$ for 
the purely academic case of a self-gravitating exponential disc. Real form factors are complicated by the presence of several components but can be simplified by assuming they are all equal.
\citet{covington_etal08} have shown that Eq.~(\ref{en_cons2}) with $C_{\rm remn}=C_1=C_2=0.5C_{12}$
is in good agreement with the sizes of the remnants of dissipationless mergers in hydrodynamic simulations,
though the best fit for $C_{\rm remn}/C_{12}$ depends on the simulations' assumptions for the initial orbits.
Here, we use $C_{12}=0$ because \citet{shankar_etal13,shankar_etal14} showed that this assumption is in better agreement with the size evolution of spheroids.
\citet{covington_etal08} have also shown that energy dissipation through radiation in gas-rich mergers
can result in final values of $r_{1/2}$ smaller than those obtained from energy conservation
(Eq.~\ref{en_cons2}) by up to a factor of two.

The scale radius of the bulge $r_{\rm bulge}$ is computed from the relation \citep{hernquist90}:
\begin{equation}
r_{1/2}= (1+\sqrt{2})r_{\rm bulge}.
\label{r_bulge}
\end{equation}
This equation assumes that the half-mass radius for the stars is equal to the half-mass radius for the stars and the DM within the galaxy 
(see the definition of  ${\cal M}_1$ and  ${\cal M}_2$ at the beginning of this subsection).
This assumption is reasonable, inasmuch as we know that DM makes a negligible contribution to the stellar dynamics of elliptical galaxies.

The radius $r_{1/2}$ computed with Eq.~(\ref{r_bulge}) is the half-mass radius in three dimensions.
The radius that contains half of the mass in a two-dimensional projection is $R_e= r_{1/2}/1.33=1.8153r_{\rm bulge}$ assuming a Hernquist profile
\citep{hernquist90}.

The one-dimensional stellar velocity dispersion averaged over a cylindrical aperture on the sky of projected radius
$R_e$ is given by:
\begin{equation}
\sigma_e^2= {\rm G}{{\cal M}_1+{\cal M}_2\over CR_e},
\label{sigma_e}
\end{equation}
where $C$ is a structure coefficient. For a Hernquist profile with isotropic velocity dispersion,
$C=3.31$ (\citealp{courteau_etal14}, chapter~5, section~B). However, the Hernquist profile is just a phenomenological model.
It has the advantage that its total mass, gravitational potential and velocity dispersion can be computed analytically, but there is no physical reason why galaxies should follow it.
In fact, systematic departures from $C=3.31$ are observed \citep{cappellari_etal13,courteau_etal14}.
We therefore treat $C$ as a parameter of the model to be constrained by observations.

A fraction $\epsilon_\bullet$ of the total gas mass of the merging galaxies falls to the centre and feeds the growth of a supermassive black hole. The post-merger black hole mass is computed with the formula
\begin{equation}
M_\bullet = M_{\bullet 1} + M_{\bullet 2} +\epsilon_\bullet (M_{\rm gas 1}+M_{\rm gas 2}),
\label{m_bh}
\end{equation}
where $ M_{\bullet 1}$, $ M_{\bullet 2}$,  $M_{\rm gas 1}$, $M_{\rm gas 2}$ are the black hole and gas masses of the merging galaxies prior to the merging event.

In  a minor merger, the gas and the stars in the disc of the smaller
galaxy are added to the disc of the larger galaxy, while the gas and the stars in the bulge of the smaller galaxy are added to the bulge of the larger galaxy.

\subsection{\sc component}

A galactic component (a disc, a pseudobulge or bulge) is composed of a stellar population and its interstellar medium.
The {\sc component} module follows the exchanges of matter between gas and stars, 
as well as the ejection of gas from a component, i.e., the processes of star formation and feedback.

The equation that governs the evolution of the gas mass within a component is:
\begin{equation}
\dot{M}_{\rm gas} = \dot{M}_{\rm accr}+\dot{M}_{\rm sml}-{\rm SFR}-\dot{M}_{\rm out}.
\label{mdot_gas1}
\end{equation}
Here $\dot{M}_{\rm accr}$ is the accretion rate onto the galaxy, which is entirely due cold flows since threre is no cooling in our model (Section~2.2.1).
$\dot{M}_{\rm sml}$ is the gas deposited into interstellar medium by the later stages of stellar evolution (stellar mass loss),
SFR is the star formation rate and
$\dot{M}_{\rm out}$ is the rate at which gas is blown out by stellar feedback (outflow rate).

The time separation between two output timesteps $t_{s-1}$ and $t_s$ in the N-body simulation used to construct the merger trees ranges from $6\,$Myr at $z\simeq 13$ to $144\,$Myr at $z\simeq 0$,
but we follow star formation and feedback on smaller substeps of $\Delta t=1\,$Myr.
We have chosen this value because it is short compared to the timescale on which a stellar population evolves, i.e., the timescale on which $\dot{M}_{\rm sml}$ varies
(even massive OB stars spend $>6\,$Myr on the main sequence).
$\dot{M}_{\rm accr}$ varies on a timescale (set by $t_{\rm dyn}$) that is even longer.
Hence, without loss of generality, we can write
${\rm SFR}= M_{\rm gas}/t_{\rm sf}$ and $\dot{M}_{\rm out}=\eta{\rm SFR}$, where our only assumption
about the star formation timescale $t_{\rm sf}$ and the mass-loading factor $\eta$ is that they are constant during a $\Delta t=1\,$Myr substep,
even though they vary on longer timescales in response to stellar evolution or changes in the structural properties of the component (computed in {\sc galaxy}).

In conclusion, the equations for $M_{\rm gas}$ and $M_{\rm stars}$ take the form:
\begin{equation}
\dot{M}_{\rm gas} = \dot{M}_{\rm accr}+\dot{M}_{\rm sml}-(1+\eta){M_{\rm gas}\over t_{\rm sf}},
\label{mdot_gas2}
\end{equation}
\begin{equation}
\dot{M}_{\rm stars}={\dot{M}_{\rm gas}\over t_{\rm sf}}-\dot{M}_{\rm sml}.
\label{mdot_stars}
\end{equation}
These equations have an analytic solution \citep{cole_etal00,lilly_etal13,dekel_mandelker14,peng_maiolino14}\footnote{The analytic solution has the interesting feature that $M_{\rm gas}\rightarrow \dot{M}_{\rm accr}t_{\rm sf}/(1+\eta)$ for 
$t\rightarrow\infty$ when $\dot{M}_{\rm sml}=0$.}, which we use to evolve $M_{\rm gas}$ and $M_{\rm stars}$ on substeps of $\Delta t=1\,$Myr 
(a direct numerical integration would require $\Delta t\ll 1\,$Myr to provide the required accuracy).

In this article, we assume instantaneous recycling (a fraction $R$ of the gas that forms stars is immediately returned to the interstellar medium).
This assumption is coded in the {\sc star} module (Section~2.5).
Since $\dot{M}_{\rm sml}=RM_{\rm gas}/t_{\rm sf}$, the presence of an analytic solution means that there is no need for any substepping whatsoever.
However, we retain the substepping with $\Delta t=1\,$Myr because, in the future, 
we may want to replace the instantaneous recycling approximation with a stellar evolution model in which $\dot{M}_{\rm sml}$ does depend on the age of the stellar population 
({\sc component} assumes no knowledge of what is assumed in {\sc stars}).

This description hides the complexity star formation and feedback in the values of $t_{\rm sf}$ and $\eta$, which may depend on several galaxy properties. 
Their calculation is done by the star formation model (Section~2.4.1) and the feedback model (Section~2.4.2), respectively.

\begin{figure*}
\begin{center}$
\begin{array}{cc}
\includegraphics[width=0.5\hsize]{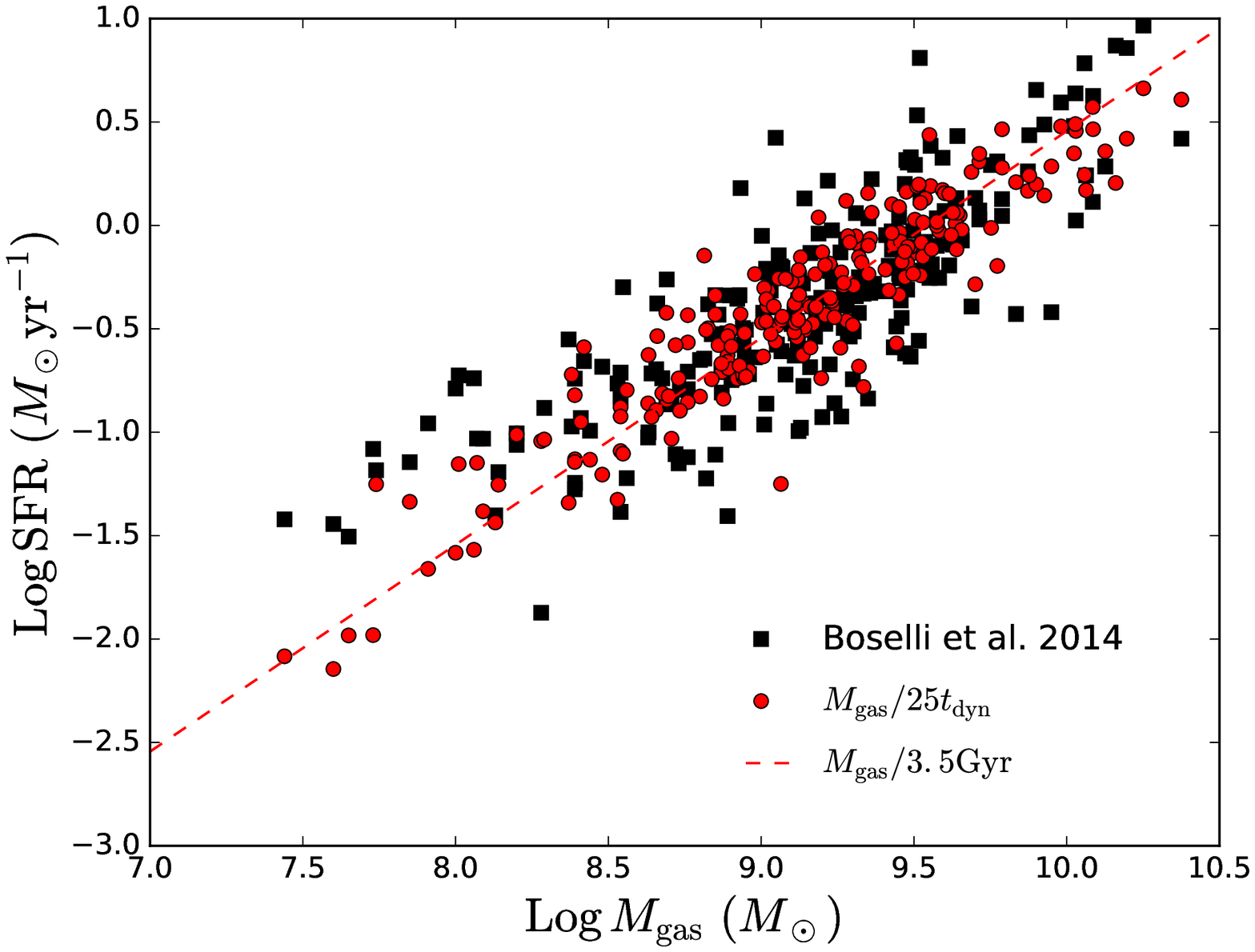} &
\includegraphics[width=0.5\hsize]{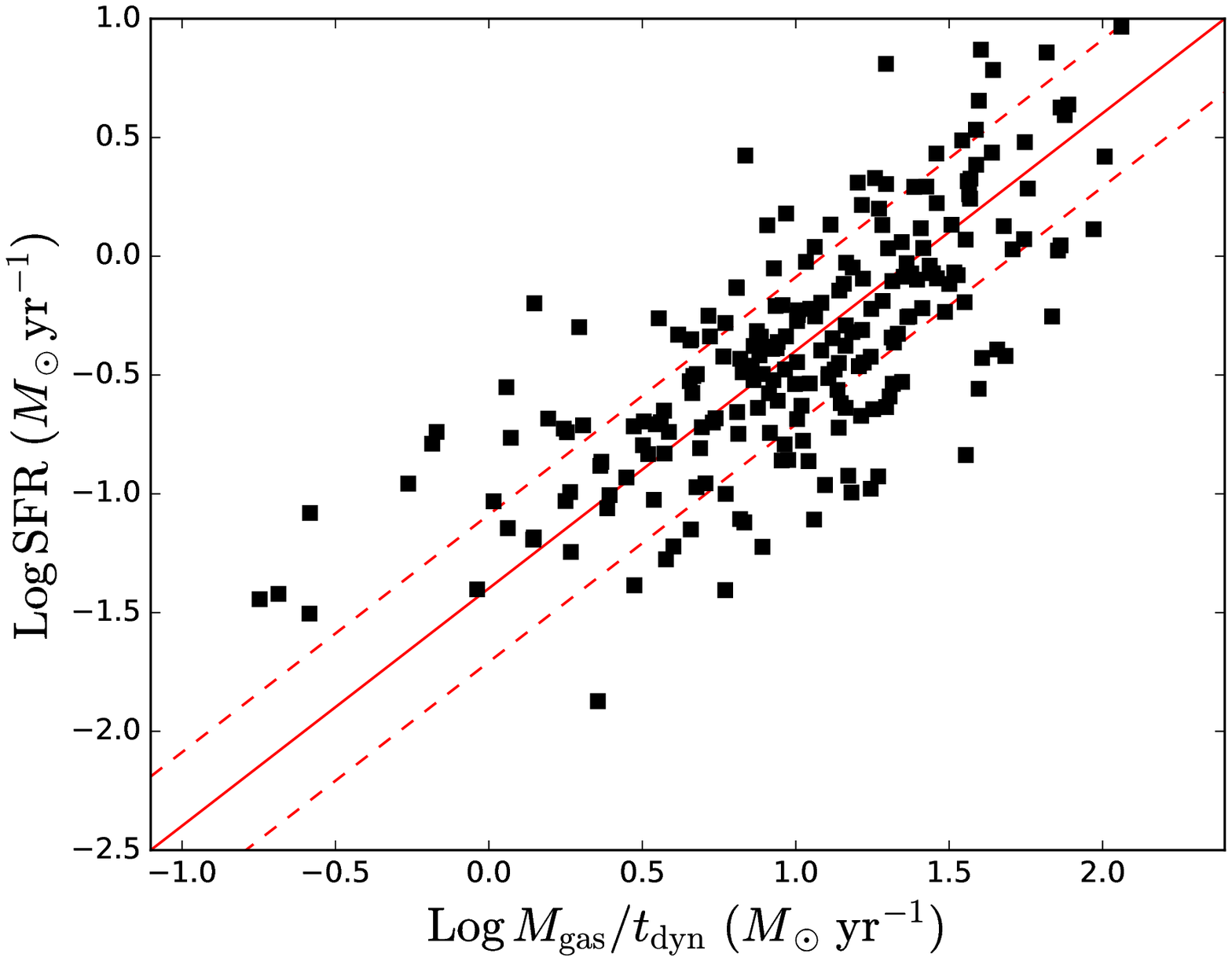} 
\end{array}$
\end{center}
\caption{Relation between total gas mass ({\sc Hi} plus {\sc H}$_2$; left) and goodness of $M_{\rm gas}/t_{\rm dyn}$ as a SFR estimator (right) 
in a local sample from Boselli et al. (2014, black squares).
Here,  $t_{\rm dyn}=2\pi r_d/v_{\rm rot}$, where $r_d$ and $v_{\rm rot}$ are observational measurement from  \citet{boselli_etal14}'s sample, from which
we have retained only galaxies classified as S or Irr for which there is a  measurement of the exponential scale-length $r_d$.
In the left panel, the observed SFR - $M_{\rm gas}$ relation is compared what one would expect for SFR $=M_{\rm gas}/25t_{\rm dyn}$  (red circles).
The red dashed line corresponds to a constant star-formation timescale of 3.5$\,$Gyr.
The right panel shows that \citet{boselli_etal14}'s data (black squares) follow the SFR $=M_{\rm gas}/25t_{\rm dyn}$ relation (red solid line)
within a scatter of a factor of two (red dashed lines). }
\label{tSFR}
\end{figure*}

\subsubsection{The star formation model}
  
The star formation model computes the star formation timescale $t_{\rm sf}$.
For discs, we assume that $t_{\rm sf}$ is proportional to $t_{\rm dyn}$, where $t_{\rm dyn}$ is the orbital time at the disc scale radius $r_d$
($t_{\rm sf}=t_{\rm dyn}/\epsilon_{\rm sf}$; see, e.g., \citealp{kauffmann_etal93} and \citealp{kennicutt98}). Hence:
\begin{equation}
{\rm SFR} = \epsilon_{\rm sf} {M_{\rm gas}\over t_{\rm dyn}}.
\label{epsilon_sf}
\end{equation}

The star formation efficiency $\epsilon_{\rm sf}=1/25$ adopted in this article is calibrated on a local galaxy sample by \citet{boselli_etal14}.
With this efficiency, our model reproduces the statistical relation between $M_{\rm gas}$ and SFR in local galaxies (Fig.~\ref{tSFR}, left) and the SFRs 
of individual galaxies within a factor of two  (Fig.~\ref{tSFR}, right).
The gas masses plotted in Fig.~\ref{tSFR} are total masses of cold neutral gas. Had we used {\sc H}$_2$ masses instead, 
the observed $M_{\rm gas}$ - SFR relation would have had a higher normalization and
woud have been tighter because the molecular gas forms stars.
However, deciding what fraction of the cold gas is molecular would add another layer of complexity and uncertainty to our model, possibly larger than the difference in scatter between
the SFR - $M_{\rm gas}$ relation and the SFR - $M_{{\rm H}_2}$ relation.

Explaining why $t_{\rm sf}$ is so long compared to the local freefall time of the gas is a hot topic in star formation theory (e.g., \citealp{renaud_etal12,hopkins_etal14,kraljic_etal14,forbes_etal16,gatto_etal16}).
By calibrating Eq.~(\ref{epsilon_sf}) on observational data, we are effectively short-circuiting these complex physics, whose outcome is summarised in the value of $\epsilon_{\rm sf}$.

It is also important to remark that our star formation model is based on observations of disc galaxies.
We assume that we can generalize it to pseudobulges and bulges by simply redefining the dynamical time.
In GalICS 2.0, $t_{\rm dyn}$ is the orbital time at $r_{\rm pseudo}$ for pseudobulges and the half-crossing time of the starbursting region
$t_{\rm dyn}= r_{\rm starburst}/\sigma$ for bulges.

Mergers are the only mechanism through which bulges can accrete gas. When they are gas-rich, they
induce intense starbursts. The SFR has a first peak at the first pericentic passage and another later when the two galaxies coalesce (e.g., \citealp{dimatteo_etal08}).
At the first pericentric passage, the system's morphology is completely irregular and star formation is concentrated in a series of knots along the galaxies' spiral arms.
By the time the galaxies coalesce, most of the gas has sunk to the centre of the merger
remnant. The scale-length of its distribution is of order $r_{\rm starburst}\sim 0.1r_{\rm bulge}$ \citep{cattaneo_etal05a}. 
In SAMs, mergers are instantaneous and galaxies jump from their initial morphologies directly to this final state.
However crude, this assumption is in line with
observational evidence that the star formation timescale $t_{\rm sf}=M_{\rm gas}/\dot{M}_{\rm sf}$
 for starburst galaxies is about ten times shorter
than it is for normal galaxies \citep{bigiel_etal08}.

Observationally, galaxies begin to depart from the mean Schmidt-Kennicutt law \citep{kennicutt98} between SFR surface density $\Sigma_{\rm SFR}$ and gas surface density $\Sigma_{\rm gas}$
for $\Sigma_{\rm gas}<\Sigma_{\rm th}\sim 9\,M_\odot{\rm\,pc}^{-2}$ \citep{bigiel_etal08}, where $\Sigma_{\rm gas}$ is the mean gas surface density 
({\sc Hi} plus {\sc H}$_2$) within the optical radius $r_{\rm opt}=3.2r_d$.
However, there are galaxies on the relation (including some of the
black squares in Fig.~\ref{tSFR}) 
with values of $\Sigma_{\rm gas}$ as low as $\sim 2\,M_\odot{\rm\,pc}^{-2}$. Therefore, the threshold is not sharp.
In GalICS 2.0, we set SFR $=0$ for $\Sigma_{\rm gas}<\Sigma_{\rm th}$, where $\Sigma_{\rm th}$ is a parameter of the model.
We set it to the relatively low value $\Sigma_{\rm th}=2\,M_\odot{\rm\,pc}^{-2}$ because higher values suppress star formation too much in low-mass haloes,
leading to galaxies that are all gas and no stars, though this may be a resolution artifact.
The surface area on which we spread the gas to compute $\Sigma_{\rm gas}$ is
$2\pi(r_{\rm opt}^2-r_{\rm pseudo}^2)$ for discs, $2\pi r_{\rm pseudo}^2$ for pseudobulges and $2\pi r_{\rm starburst}^2$ for bulges.

\subsubsection{The feedback model}

Feedback is a generic term for the effects that star formation and black hole accretion exert on the surrounding gas.
These effects influence the processes that cause them and can regulate their rates.
This section is on stellar feedback but even that is multifaceted because it results from the synergy of different processes (SNe, radiation pressure, photoionization and photoelectric heating)
that act on different scales.

\citet{mathews_baker71} and \citet{larson74} were the first to suggest that gas may be strongly heated by supernova (SN) blastwaves and driven out of galaxies in hot winds.
While SNe have certainly the energy to this, and have become for this reason a standard ingredient of galaxy formation theory, 
their efficiency and the mass scale at which they become important are affected by the fraction of SN energy that is radiated \citep{dekel_silk86}.
If SNe explode inside dense molecular clouds, most of their energy will be quickly lost to X-rays.
Radiation pressure and stellar winds from massive OB stars must disperse giant molecular clouds rapidly, after they have turned just a few percent of their mass into stars,
for this not to occur (\citealp{hopkins_etal13} and references therein).
Photoelectrons extracted from dust grains by ultraviolet radiation are the primary source of heating for the neutral interstellar medium 
and suppress star formation by preventing its overcooling and overcondensation into dense molecular clouds \citep{forbes_etal16}.

These complex physics are beyond the scope of our feedback model, 
whose purpose is to computes the mass-loading factor $\eta=\dot{M}_{\rm  out}/{\rm SFR}$, i.e., the rate at which cold gas is removed from galaxies.
Any feedback mechanism that regulates star formation without removing gas from galaxies is already incorporated phenomenologically in our star formation efficiency $\epsilon_{\rm sf}$ (Section~2.4.1).
Similarly, the fraction $\epsilon_{\rm SN}$ of the power output from SN explosions that is converted into wind kinetic energy and/or thermalized in the hot atmosphere is chosen to reproduce the observation
and therefore includes the effects of all the other processes (e.g., radiation pressure, stellar winds, photoionization, photoelectric heating) that may affect the outflow rate.

If $\Psi_{\rm SN} \simeq 1/(140\,M_\odot)$ is the number of SNe per unit stellar mass formed 
(assuming a \citealp{chabrier03} initial mass function in the stellar mass range $0.1-100\,M_\odot$ and a minimum mass for core-collapse SNe of $8\,M_\odot$) and 
$E_{\rm SN}\sim 10^{51}\,$erg is the energy released by one SN, then the power output from SNe will be $E_{\rm SN}\Psi_{\rm SN}{\rm SFR}.$
If a fraction $\epsilon_{\rm SN}$ of this power is used to drive a wind with speed $v_{\rm w}$, then
the outflow rate $\dot{M}_{\rm out}$ from the component will satisfy:
\begin{equation}
{1\over 2}\dot{M}_{\rm out}v_{\rm w}^2 \sim \epsilon_{\rm SN}E_{\rm SN}\Psi_{\rm SN}{\rm SFR}.
\label{sn1}
\end{equation}
\citep{silk03}. 
Expulsion from the gravitational potential well of the DM requires $v_{\rm w}\ge v_{\rm esc}\sim 2.5v_{\rm vir}$ 
(the numerical coefficient in front of $v_{\rm vir}$ depends on halo concentration),
but here we make no assumption as to whether the wind
escapes from the halo or settles into a hot circumgalactic medium.
We therefore reabsorbe the uncertainty on $v_{\rm w}/v_{\rm vir}$ into the free parameter $\epsilon_{\rm SN}$
and define mass-loading factor $\eta$ so that:
\begin{equation}
\eta\equiv {\dot{M}_{\rm out}\over{\rm SFR}} = {2\epsilon_{\rm SN}E_{\rm SN}\Psi_{\rm SN}\over v_{\rm vir}^2}.
\label{sn2}
\end{equation}
The only inconvenient of this definition is that $\epsilon_{\rm SN}$ underestimates the real SN efficiency required to produce
the mass-loading factors assumed by our model.
The difference is small (a factor of $\sim 1.5$) is the gas blown out of the galaxy if heated to the virial temperature and mixed with the hot atmosphere
(for a singular isothermal sphere, ${3\over 2}kT_{\rm vir} = {3\over 2}\cdot{1\over 2}\mu v_{\rm vir}^2$; \citealp{white_frenk91}).
Much larger energetic efficiencies ($\gsim 2.5^2\epsilon_{\rm SN}$) are required if the gas expelled from galaxies is also blown out of the halo.

\begin{table*}
\begin{center}
\caption{Model parameters: symbols, units (for dimensional quantities), default values and alternative models (where they differ).}
\begin{tabular}{ l l l l l l l}
\hline
\hline 
Parameter                                  & Symbol                & Units          & Default                           & Model1               & Model 2             & Model 3 \\
\hline
{\bf  Cosmology}                           &                       &                &                                   &                      &                     &          \\
Matter density                             & $\Omega_M$            &                &$0.308$                             &                       &                    &          \\
Baryon density                             & $\Omega_b$            &                &$0.0481$                            &                       &                    &           \\ 
Cosmological constant                      & $\Omega_\Lambda$       &                &$0.692$                             &                       &                    &           \\
Hubble constant                            & $H_0$                 &$100{\rm\,km\,s}^{-1}{\rm Mpc}^{-1}$ &$0.678$                             &                       &                    &           \\
\hline
{\bf N-body simulation}                    &                       &                &                                     &                       &                    &           \\
Box size                                   & $L_{\rm box}$          & Mpc            &$100$                                &                       &                     &           \\
Resolution                                 & $N_{\rm part}$         &                &$512^3$                            &                       &                    &            \\
\hline
{\bf Dimensional parameters}               &                      &                 &                                  &                      &                     &            \\
Thermal velocity dispersion of IGM         & $\sigma_{\rm reio}$    & km$\,$s$^{-1}$  &$25$                              &                      &                    &             \\
Star formation threshold                   & $\Sigma_{\rm th}$  & $M_\odot{\rm\,pc}^{-2}$ &$2$                            &                       &                    &             \\
Minimum shock heating mass                 & $M_{\rm shock}$    & $M_\odot$           &$10^{10.7}$                       & $10^{11.3}$             &                    & $10^{11.0}$ \\
Shutdown mass                              & $M_{\rm shutdown}$  & $M_\odot$          &$10^{12.7}$                        & $10^{12.4}$            &                    & $10^{12.3}$  \\
SN feedback saturation scale               & $v_{\rm SN}$       & ${\rm\,km\,s}^{-1}$&$24$                              &                       &                    &  $43$ \\
SN energy                                  & $E_{\rm SN}$       & erg               &$10^{51}$                          &                       &                    &               \\
SN rate                                    & $\Psi_{\rm SN}$    & $M_\odot^{-1}$     &$1/140$                            &                       &                     &              \\
\hline
{\bf Adimensional efficiency factors}      &                   &                   &               &                        &                   &                \\
Star formation                             & $\epsilon_{\rm sf}$&                   & $0.04$                             &                        &                    &                \\
Disc instabilities                       &$\epsilon_{\rm inst}$ &                   &$0.9$                               &                       &                    &                 \\
Mass ratio for major mergers                &$\epsilon_{\rm m}$ &                   &$4$                                &                       &                     &     $3$            \\
Structure coefficient of bulges             &$C$               &                   &$2.5$                              &                       &                     &                 \\
Black hole accretion                       &$\epsilon_\bullet $ &                   &$0.01$                             &                       &                     &                 \\
Maximum SN feedback efficiency           & $\epsilon_{\rm max}$ &                   &$1$                                &                       & $0.12$                & $0.12$                 \\           
Returned fraction                          & $R$               &                   &$0.45$                            &                       &                      &                  \\
Metal yield                                & $y$               &                   &$0.06$                             &                      &                      &                   \\
\hline
{\bf SN feedback scaling exponents}        &                   &                    &              &                      &                       &                   \\
%\hline
$v_{\rm vir}$ scaling                       & $\alpha_v$        &                   & $-4$                              &                       &                      & $-6.2$                   \\
$z$ scaling                                & $\alpha_z$        &                   & $3$                               &                      &                       &                    \\
\hline
\hline
\end{tabular}
\end{center}
\label{model_parameters}
\end{table*}

The problem of this simple scaling with $v_{\rm vir}^{-2}$ is that it cannot reproduce the shallow slope of the low-mass end of the galaxy SMF 
(unless we use merger trees from a low-resolution N-body simulation that misses low-mass haloes, as in \citealp{cattaneo_etal06}, 
but there we focussed on massive galaxies). 
A phenomenological solution is to introduce a SN efficiency $\epsilon_{\rm SN}$ that depends on both $v_{\rm vir}$ and redshift $z$,
and to impose a plausible maximum $\epsilon_{\rm max}$ to the values that $\epsilon_{\rm SN}$ can take:
\begin{equation}
\epsilon_{\rm SN}={\rm min}\left[\left({v_{\rm vir}\over v_{\rm SN}}\right)^{\alpha_{\rm v}}(1+z)^{\alpha_z},\epsilon_{\rm max}\right],
\label{epsilon_sn} 
\end{equation}
where $v_{\rm SN}$, $\alpha_v$ and $\alpha_z$ are free parameters of the model to be determined by fitting the galaxy SMF.
The speed $v_{\rm SN}$ corresponds to the virial velocity for which $\epsilon_{\rm SN}=1$ at $z=0$ if no maximum efficiency is imposed.

As the laws of physics do not vary with time, one could argue that a physical model should not contain any explicit dependence on $z$.
A simple answer is that this objection does not apply to a phenomenological model 
(see \citealp{peirani_etal12} for evidence from cosmological hydrodynamic simulations supporting more efficient feedback at high $z$).
We also remark that the values with which we fit the data ($\alpha_v=-4$, $\alpha_z=3$, $v_{\rm SN}=24{\rm\,km\,s}^{-1}$; Table~1 and Section~3) give a simple relation between mass-loading factor and halo mass:
\begin{equation}
\eta\simeq 3.8\left({M_{\rm vir}\over 10^{11}\,M_\odot}\right)^{-2},
\label{eta_sn}
\end{equation}
since $v_{\rm vir}\propto M_{\rm vir}^{1/3}(1+z)^{1/2}$. $M_{\rm vir}$ is a physical quantity, though it is not clear why 
the outflow rate should scale with $M_{\rm vir}$ rather than $v_{\rm vir}$.
The mass resolution of the N-body simulation used to construct the merger trees is $M_{\rm vir}\sim 3\times 10^{10}\,M_\odot$. 
In this article, we formulate our model in terms of $v_{\rm vir}$ and $z$ rather than $M_{\rm vir}$ to ease comparison with previous work, for the sake of greater generality and because, with approach, it is easier to check that our feedback model is energetically plausible.

Eq.~(\ref{eta_sn}) corresponds to a very strong dependence of the mass-loading factor on the virial velocity ($\eta\propto v_{\rm vir}^{-6}$). 
For comparison, the exponents used by other SAMs are $-5.5$ \citep{cole_etal94}, $-3.5$ \citep{guo_etal11}, $-2.5$ \citep{somerville_etal12}, 
$-0.92$ \citep{henriques_etal13} and $-3.2$ with an allowable range between $0$ and $-5.5$ \citep{lacey_etal16}, though the details of how stellar feedback is implemented vary from one model to another
(see \citealp{hirschmann_etal16} for a discussion of the mass-loading in different SAMs and simulations).

Our normalization of $\eta$ at $M_{\rm vir}=10^{11}\,M_\odot$, $\eta=3.8$, is comparable to those of Guo et al. ($\eta\sim 1$) and Henriques et al. ($\eta\sim 2.5$), but 
much lower than that of \citet{lacey_etal16}.
As the mass $M_{\rm vir}=10^{11}\,M_\odot$ is only a factor of three larger that our resolution limit, our normalization combined to our much steeper dependence on $v_{\rm vir}$
implies that our mass-loading factors are lower than those assumed by \citep{guo_etal11}, 
\citet{henriques_etal13}, and \citet{lacey_etal16} at all but the smallest halo masses probed in this article.
It is possible that we fit the observations with lower mass-loading factors for a given halo mass because our current model neglects the reaccretion of ejected gas.

Physically, $\eta$ is limited by the maximum energetic efficiency of supernovae $\epsilon_{\rm max}$.
Without such maximum, Eq.~(\ref{epsilon_sn}) implies $\epsilon_{\rm SN}\rightarrow\infty$ for $v_{\rm vir}\rightarrow 0$, which is absurd
(the wind cannot contain more energy than it is available). In the most generous case, $\epsilon_{\rm max}=1$.
The real efficiency will probably be much lower.
In practice, $\epsilon_{\rm SN}$ is limited by the mass resolution of the N-body simulation,  $M_{\rm vir}\sim 3\times 10^{10}\,M_\odot$.
Inserted into Eq.~(\ref{eta_sn}), this mass gives a maximum mass-loading factor of $\eta=30-40$.
As $M_{\rm vir}\sim 3\times 10^{10}\,M_\odot$ corresponds to $v_{\rm vir}\sim 40{\rm\,km\,s}^{-1}$ at $z=0$,
our default parameter values (Table~1) imply $\epsilon_{\rm SN}\lsim 0.1$ at $z\sim 0$ for all haloes that we can resolve.
At high $z$, however, $\epsilon_{\rm SN}$ can take much larger values if no maximum efficiency is prescribed.

\subsection{\sc star}

The {\sc star} module follows the evolution of a component's stellar
population. In the code's current version, this evolution is computed
based on the instantaneous recycling approximation. 
Stellar evolution is, therefore, entirely described by two parameters:
the returned fraction $R$ and the metal yield $y$. 
The explicit equations for the stellar mass loss rate $\dot{M}_{\rm sml}$ and the mass loss rate in metals  $\dot{M}_{\rm sml,\,Z}$ are
\begin{equation}
\dot{M}_{\rm sml} = R\cdot{\rm SFR},
\end{equation}
and
\begin{equation}
\dot{M}_{\rm sml,\,Z} = y(1-R)\cdot{\rm SFR}.
\label{MsmlZ}
\end{equation}
In Eq.~(\ref{MsmlZ}), $1-R$ is the fraction of the star-forming gas that remains in stars and contributes to the final stellar masses of galaxies, while
$y$ is the metal mass ejected into the interstellar medium per unit mass locked into stars.

Metal enrichment has been included in GalICS 2.0 to pave the way future developments but has no effect whatsoever on any of the results presented in this article
because we are not computing cooling or any properties that depend on the spectral energy distribution of galaxies, such as magnitudes and colours.

\subsection{\sc gas}

The {\sc gas} module defines what composes a gas. Currently, an object of type gas has only two attributes: total mass and metal mass.
The metal yield $y$ in {\sc star} determines the metallicity of gas returned to the interstellar medium.
This is the only place where metals enter GalICS 2.0 explicitly outside the {\sc gas} module.
The reason is that, whenever an object of type gas is transferred from one gas component to another, its metals are transferred with it 
automatically in a manner completely transparent to the other modules.

\subsection{Summary of parameters and models explored}

Table~1 summarizes the parameters of the models considered in this article.
The first two sets of parameters (cosmology and N-body simulation)
are set by the N-body simulation used to build the halo catalogues and merger trees. They are not free parameters of the SAM.

There are sixteen free parameters in GalICS 2.0.
We have separated them into three groups: dimensional parameters, efficiency factors and scaling exponents. 
The dimensional parameters set the characteristic surface-density,
velocity and mass scales for star formation, stellar feedback and and shock heating.
Efficiency factors and scaling exponents are dimensionless.
The former are multiplicative factors that set the efficiency of a
process (star formation, bar formation, bulge formation,
black hole accretion). The latter determine the exponent of the power-law
with which a quantity depends on another (in our case, how the energetic efficiency of SN feedback scales with
virial velocity and redshifts).

Some parameters, such as those related to SN feedback, are
highly uncertain. Others are reasonably well constrained by
observations, previous models and simulations.

$E_{\rm SN}$, $\Psi_{\rm SN}$, $R$ and $y$ are determined by stellar evolution.
We have used the values for a \citet{chabrier03} initial mass function with \citet{romano_etal10}'s stellar yields (see \citealp{vincenzo_etal16}) and we have not allowed them to vary.

\citet{okamoto_etal08} find that cosmic reionization suppresses gas accretion onto haloes up to $M_{\rm reio} = 6.5\times 10^9h_0^{-1}M_\odot$
at $z=0$. This mass corresponds to $v_{\rm vir}\sim 25{\rm\,km\,s}^{-1}$. We therefore assume $\sigma_{\rm reio}=25{\rm\,km\,s}^{-1}$ but notice that this assumption will have little consequence on our results
as this scale is below the resolution of our N-body simulation.

Cosmological hydrodynamic simulations by \citet{ocvirk_etal08} find $M_{\rm shock}\sim 10^{10.7}\,M_\odot$ and $M_{\rm shutdown}\sim 10^{12.7}\,M_\odot$.  

The two parameters $\epsilon_{\rm m}$ and $\epsilon_{\rm inst}$ affect
morphology only. 
For the formation of bulges,
simulations of galaxy mergers have consistently found that the critical mass ratio that separates major and minor mergers is of $\epsilon_{\rm m}=3-5$.
We choose $\epsilon_{\rm m}=4$ as our default value.
For disc instabilities, \citet{efstathiou_etal82} used N-body simulations and showed that $\epsilon_{\rm inst}\simeq 0.91$.
 \citet{christodoulou_etal95} argued for a lower value ($\epsilon_{\rm inst}\simeq 0.83$) in stellar discs but remarked that the presence of gas could raise $\epsilon_{\rm inst}$.
The values reported above are the inverse of those contained in the original articles because of our different definition of $\epsilon_{\rm inst}$.
Furthermore, the original articles used
a global instability criterion at the radius $r = 2.15r_d$ at which the rotation curve of a self-gravitating exponential disc peaks.
Hence, there is no reason why their values should apply to our SAM, in which Eq.~(\ref{instability_condition}) is applied at each radius to find the one, if any, at which an instability develops. 
However, in Section~3.5, we shall show that assuming $\epsilon_{\rm inst}= 0.9$ leads to morphologies in reasonable agreement with observations.

The structure coefficient of bulges (defined so that $\sigma_e$ is the average one-dimensional velocity dispersion within an aperture $R_e$; Section~2.3.5)
is a parameter of the model but not a free one. 
Observations of early-type galaxies find $C=2.5$ on average \citep{cappellari_etal06,cappellari_etal13}.
\citet{cappellari_etal06} remarked that, in a self-consistent model, this value corresponds to a profile with \citet{sersic63} index $n=5.5$, 
but warned that this conclusion is based on assuming a spherical isotropic system
with uniform mass-to-light ratio.

We have set the gas fraction that accretes onto the central black hole in major mergers to
$\epsilon_\bullet=0.01$ because we know from experience
with different codes
\citep{cattaneo01,cattaneo_etal05b} that this value is in
good agreement with the black hole - bulge mass relation \citep{magorrian_etal98,marconi_hunt03,haering_rix04}.

The greatest uncertainty is in the parameters that control the efficiency of SN feedback ($v_{\rm SN}$, $\alpha_v$, $\alpha_z$, $\epsilon_{\rm max}$).
They are the only true free parameters of our model and they have been calibrated on the galaxy SMF in the local Universe.
Our default assumption is $\epsilon_{\rm max}=1$.
When all the other parameters are kept fix to their default values, the best fit to the local SMF is found for
 $v_{\rm SN}=24{\rm\,km\,s}^{-1}$, $\alpha_v=-4$ and $\alpha_z=3$  (Section~3.1).

In this article, we have explored three models in addition to our default parameter combination.
The corresponding parameters are listed in Table~1 
when they differ from the default values.
Model~1 corresponds to a more abrupt shutdown of cold accretion  
($M_{\rm shock}$ has been increased to $M_{\rm shock}\sim 10^{11.3}\,M_\odot$ and $M_{\rm shutdown}$ has been lowered to $M_{\rm shutdown}\sim 10^{12.4}\,M_\odot$). 
Model~2 limits the efficiency of SN feedback to $\epsilon_{\rm max}=0.12$ (models with $\epsilon_{\rm max}>0.2$ are indistinguishable from the
default model because there are few galaxies with an energetic efficiency of SNe $>20\%$). 
Model~3 produces a SMF that is flat around $M_{\rm stars}\sim 10^{10}\,M_\odot$ and rises steeply at lower masses 
in agreement with observations by \citet{baldry_etal08,baldry_etal12}. The default model predicts a shape of the SMF in better agreement with that of Bernardi et al. (2013; Section~3.1).
 
In addition to the four main models in Table~1,  we have run four other models to test our sensitivity to specific assumptions.
Their results are shown only in connection with the relevant figures.
They are: a model without disc instabilities (Section~3.5, morphologies),
a model in which all haloes have the same spin parameter $\lambda=0.05$ (Section~3.6, disc sizes),
a model in which the halo concentration parameter $c$ is computed with the fitting formulae of
\citet{dutton_maccio14} rather than by using the values measured in our N-body simulation,
and the same model when we also include adiabatic contraction
(Section~3.7, TFR).

\begin{figure*}
\begin{center}$
\begin{array}{cc}
\includegraphics[width=0.5\hsize]{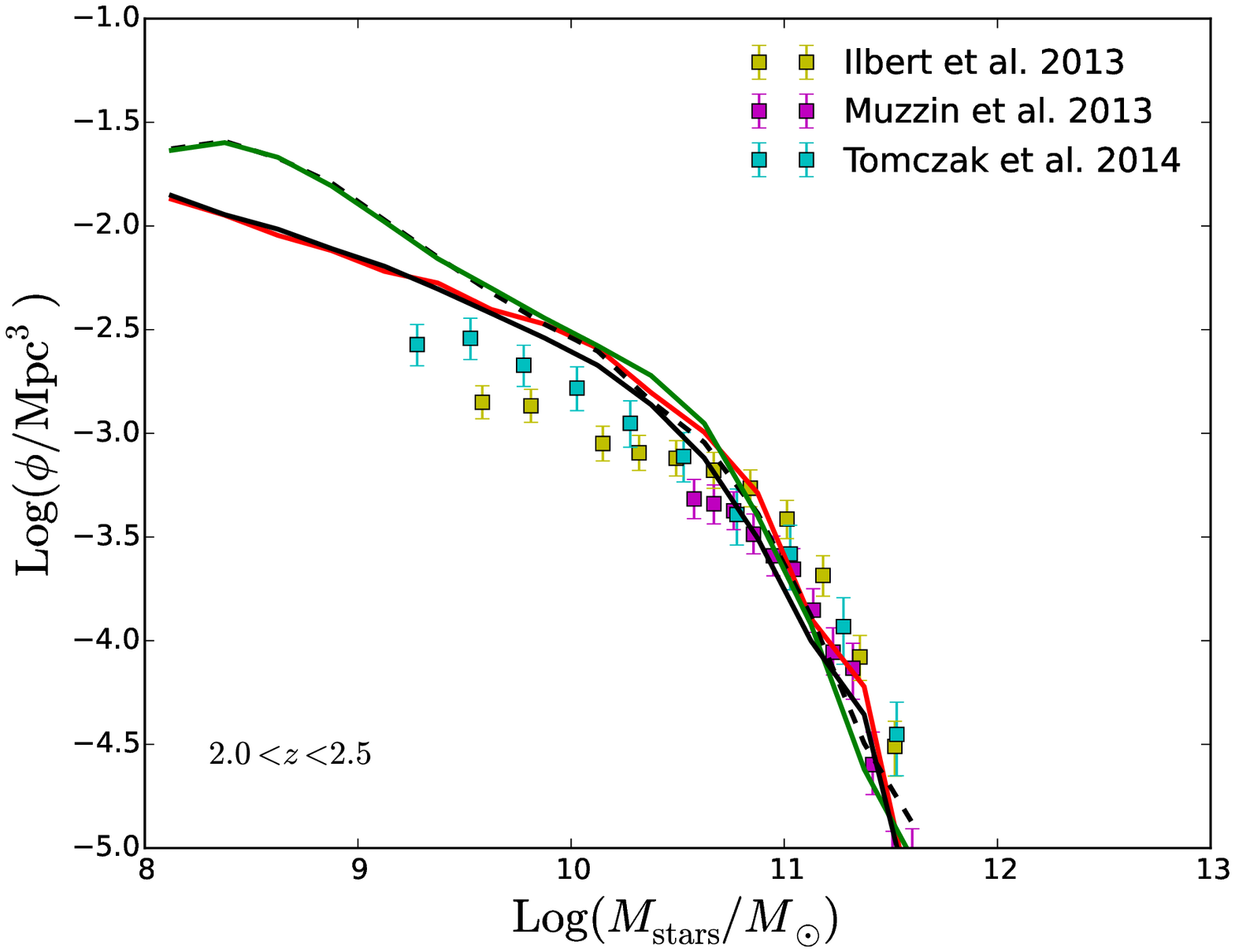} &
\includegraphics[width=0.5\hsize]{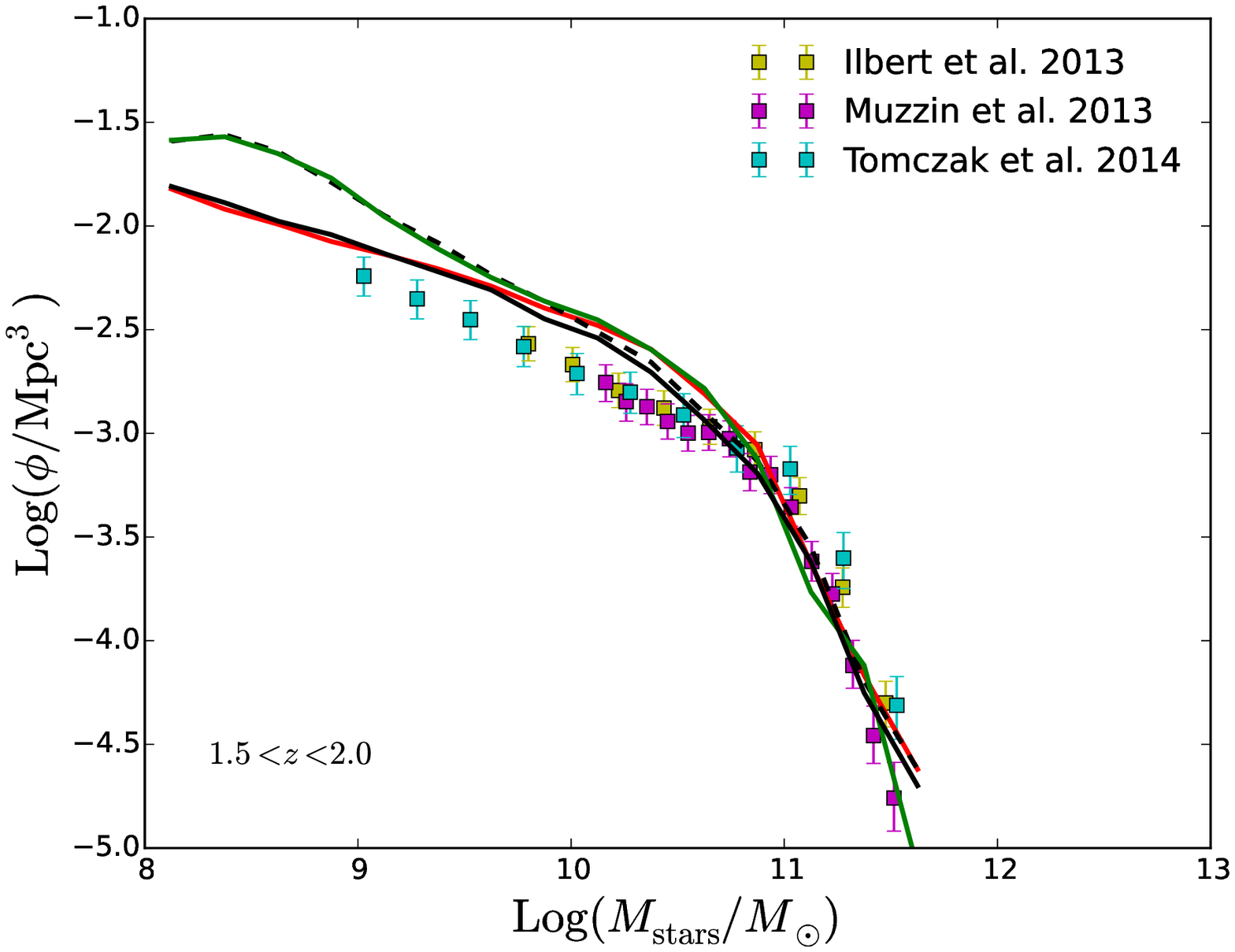} \\
\includegraphics[width=0.5\hsize]{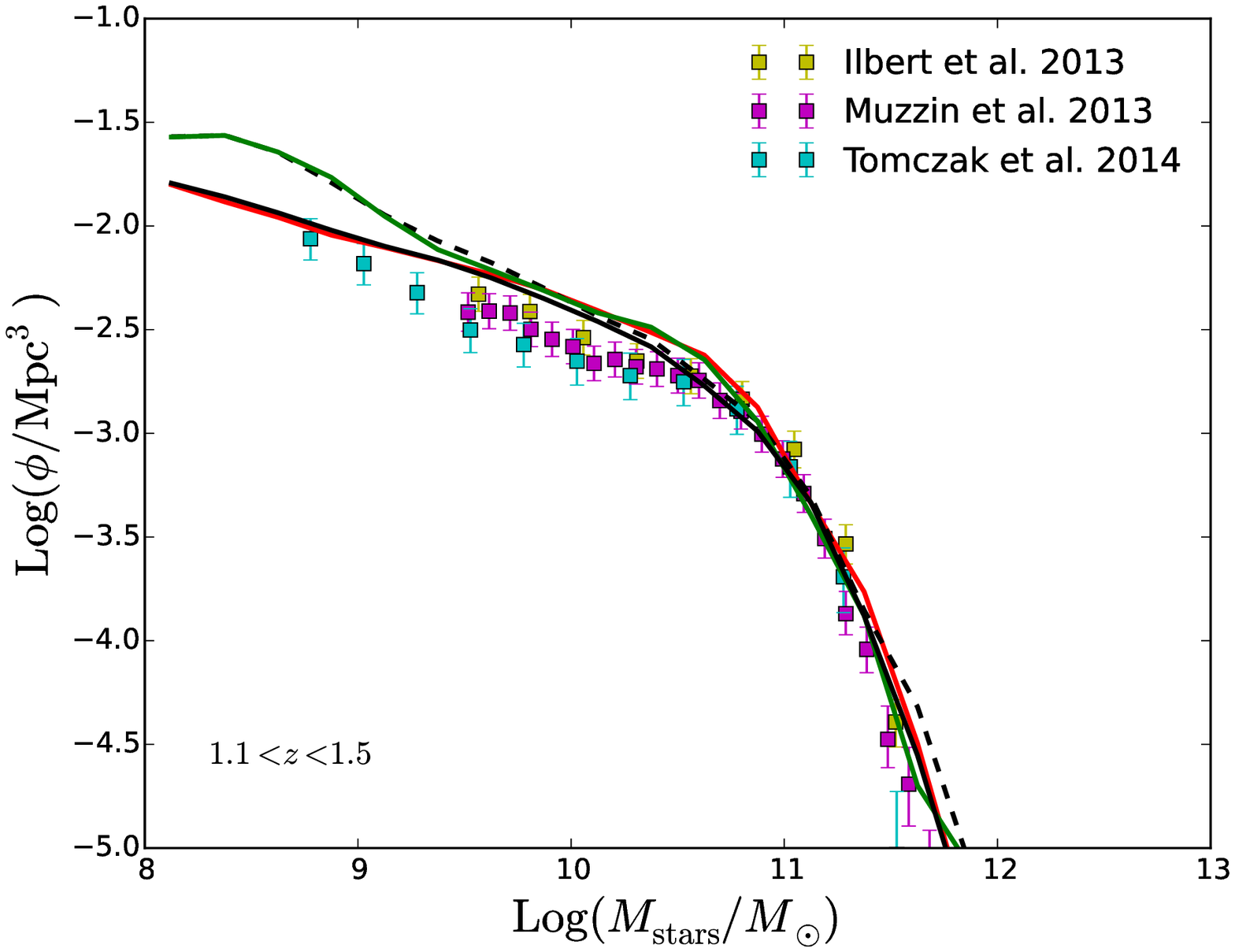}&
\includegraphics[width=0.5\hsize]{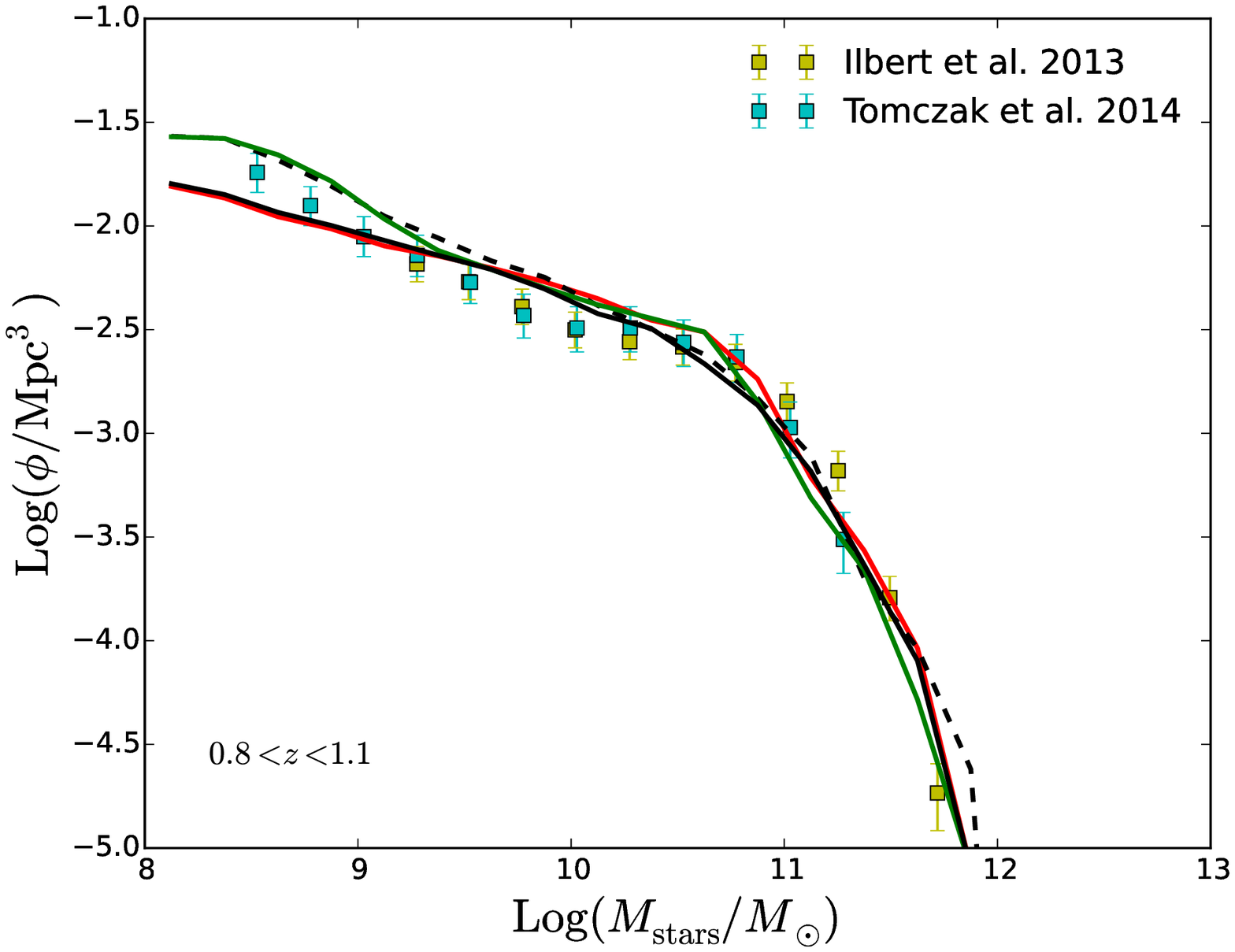} \\
\includegraphics[width=0.5\hsize]{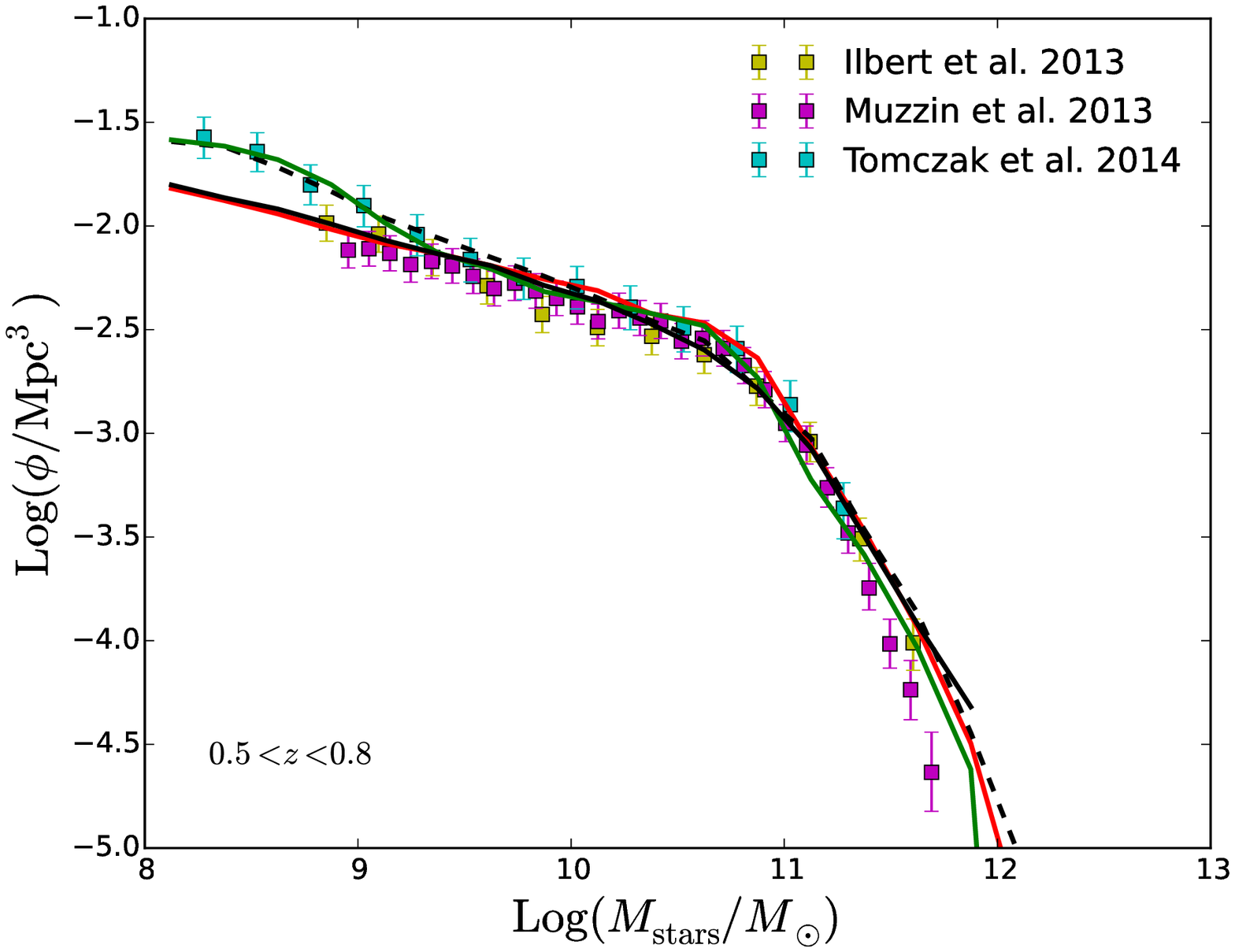} &
\includegraphics[width=0.5\hsize]{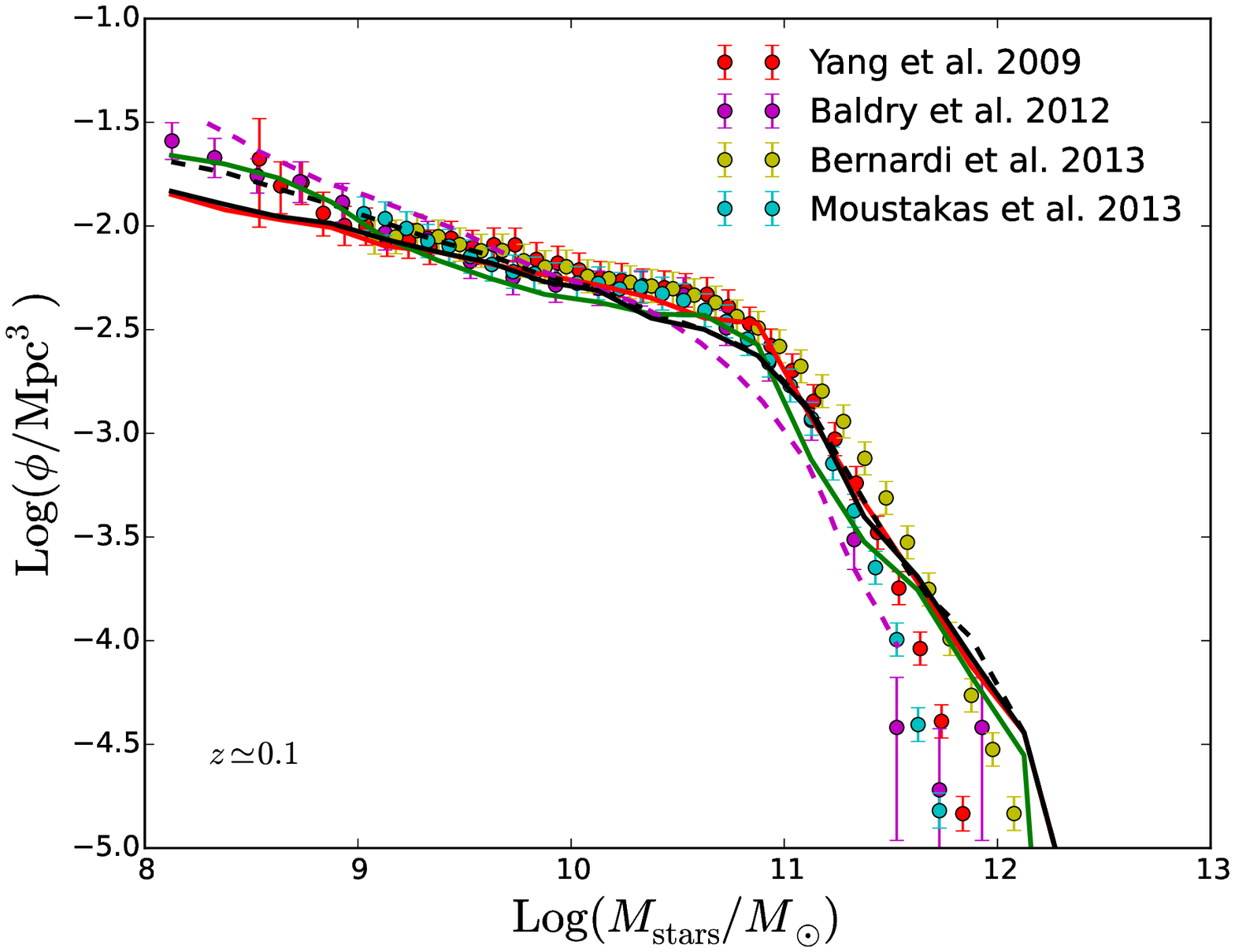} 
\end{array}$
\end{center}
\caption{Model SMFs at $0<z<2.5$ (curves) compared with observations (data points with error bars) from several groups
\citep{baldry_etal12,bernardi_etal13,moustakas_etal13,yang_etal09,ilbert_etal13,muzzin_etal13,tomczak_etal14}.
Here, $\phi$ is the number density of galaxies per dex of stellar mass.
Black solid curves, red curves, black dashed curves and greeen curves correspond to the default model, model~1, model~2 and model~3, respectively. 
The parameter values for each model are in Table~1.
The magenta curve at $z\simeq 0.1$ is the SMF in the EAGLE simulation (\citealp{schaye_etal15}, Ref-L100N1504).
All models have been convolved with a Gaussian random error of $0.04(1+z)\,$dex on stellar masses.}
\label{SMFs}
\end{figure*}

\begin{figure}
\begin{center}
\includegraphics[width=0.95\hsize]{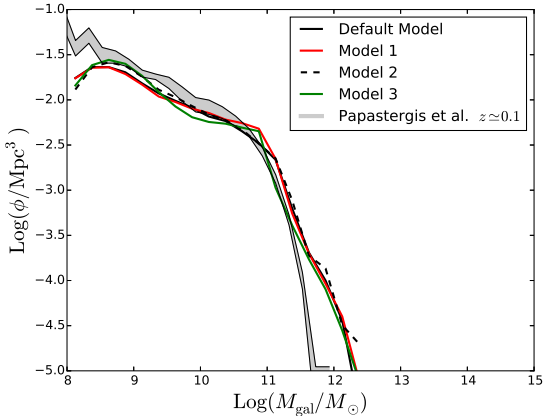} 
\end{center}
\caption{The baryonic mass function in GalICS 2.0 (curves) and in the observations by Papastergis et al. (2012, gray shaded area). 
Here, $\phi$ is the number density of galaxies per dex of baryonic mass.
The black solid curve corresponds to our default model. The red curve, the black dashed curve and the green curve correspond to models~1, 2 and~3 of Table~1,
respectively.
In the observations, $M_{\rm gal}=M_{\rm stars}+1.4M_{\rm HI}$. The lower boundary of shaded area is based on measured {\sc Hi} masses (ALFALFA). 
The upper boundary is computed based on the maximum {\sc Hi} masses that the galaxies could contain
given the survey's detection limit.}
\label{BMF}
\end{figure}

\begin{figure*}
\begin{center}$
\begin{array}{cc}
\includegraphics[width=0.5\hsize]{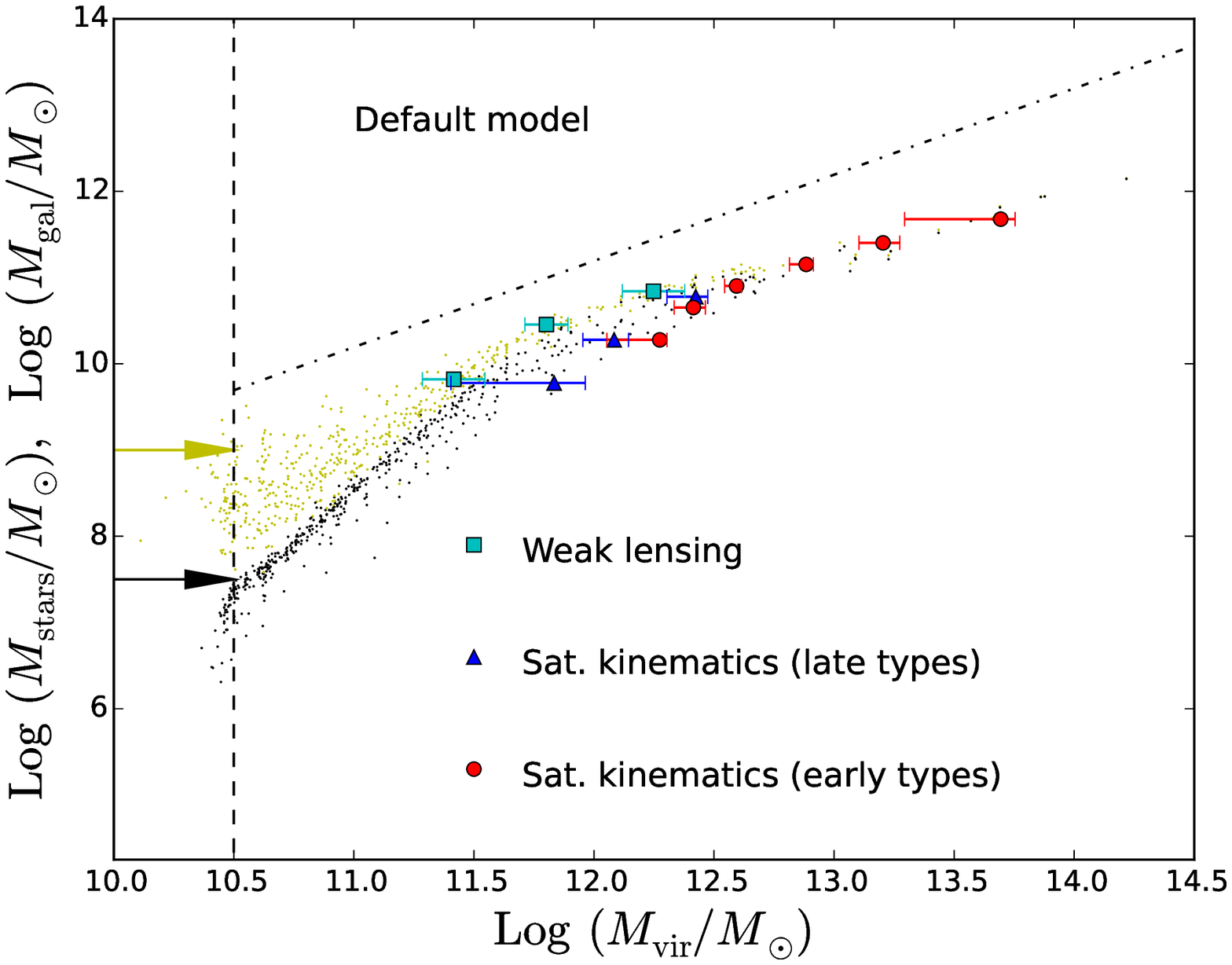}&
\includegraphics[width=0.5\hsize]{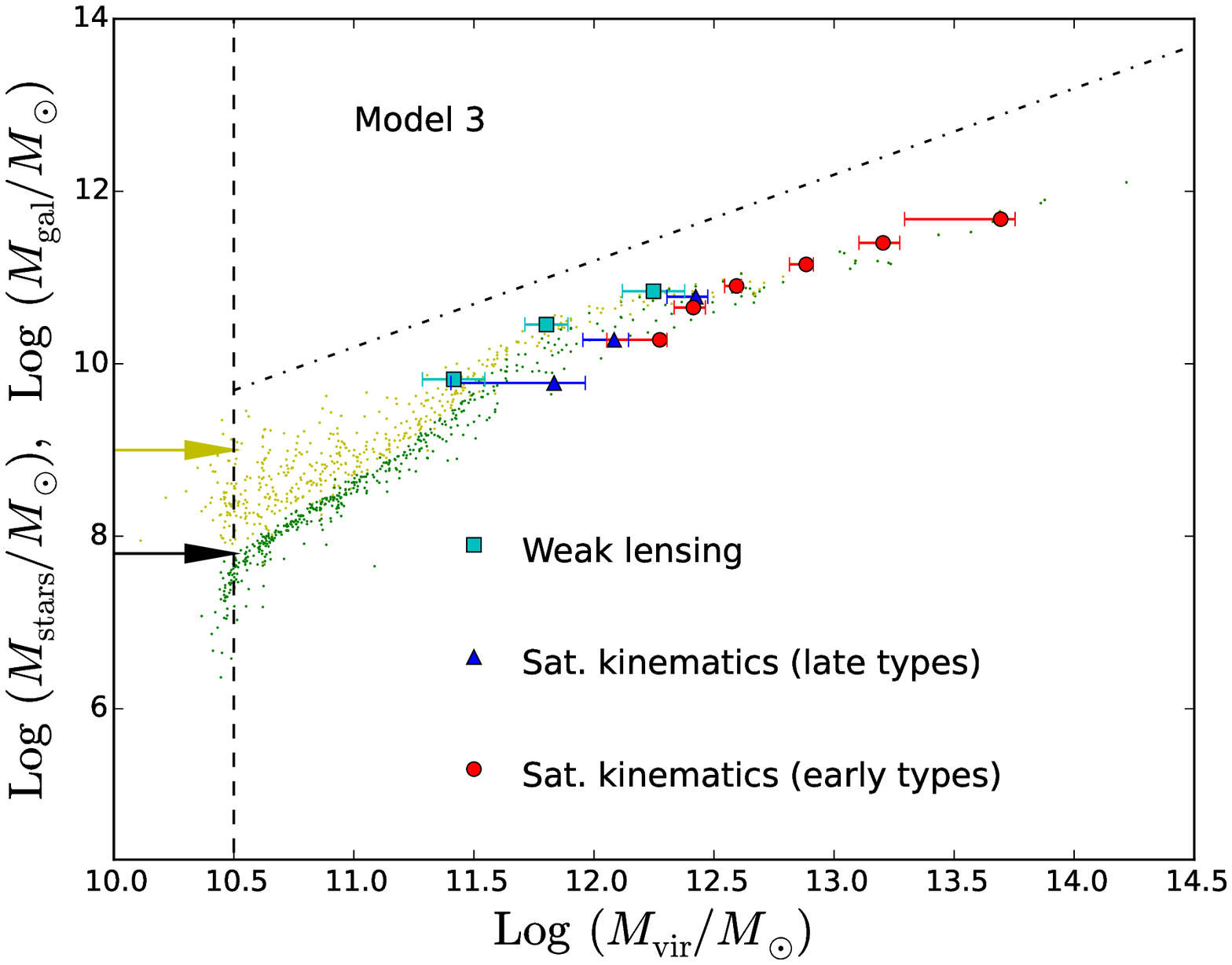}
\end{array}$
\end{center}
\caption{The $M_{\rm vir}$ - $M_{\rm stars}$ relation predicted by the default model (black point cloud, left) and model~3 (green point cloud, right) compared with observational determinations of halo masses from 
weak lensing (\citealp{reyes_etal12}; cyan squares) and the kinematics of satellite galaxies \citep{wojtak_mamon13}.
The results from the latter are shown separately for late-type galaxies (blue triangles) and early-type galaxies (red circles). The yellow point cloud 
shows the relation for the baryonic rather than the stellar mass.
The black and yellow arrows show our resolution in $M_{\rm stars}$ and $M_{\rm gal}$, respectively.
The dotted-dashed line corresponds to  $M_{\rm stars}=({\Omega_b/\Omega_M})M_{\rm vir}$. 
The vertical dashed line shows our halo-mass resolution.
}
\label{HOD}
\end{figure*}

\begin{figure*} M
\begin{center}$
\begin{array}{cc}
\includegraphics[width=0.5\hsize]{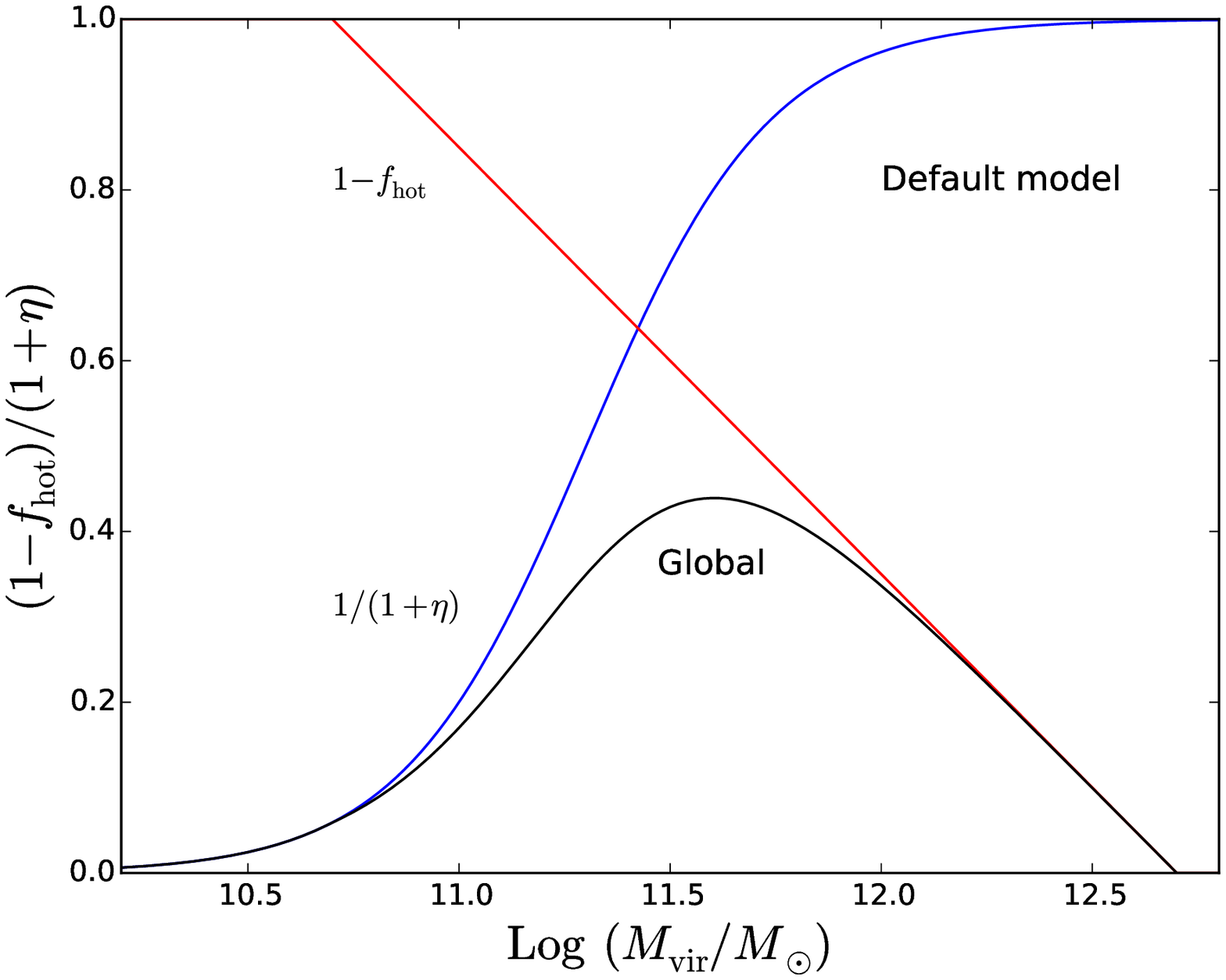}&
\includegraphics[width=0.5\hsize]{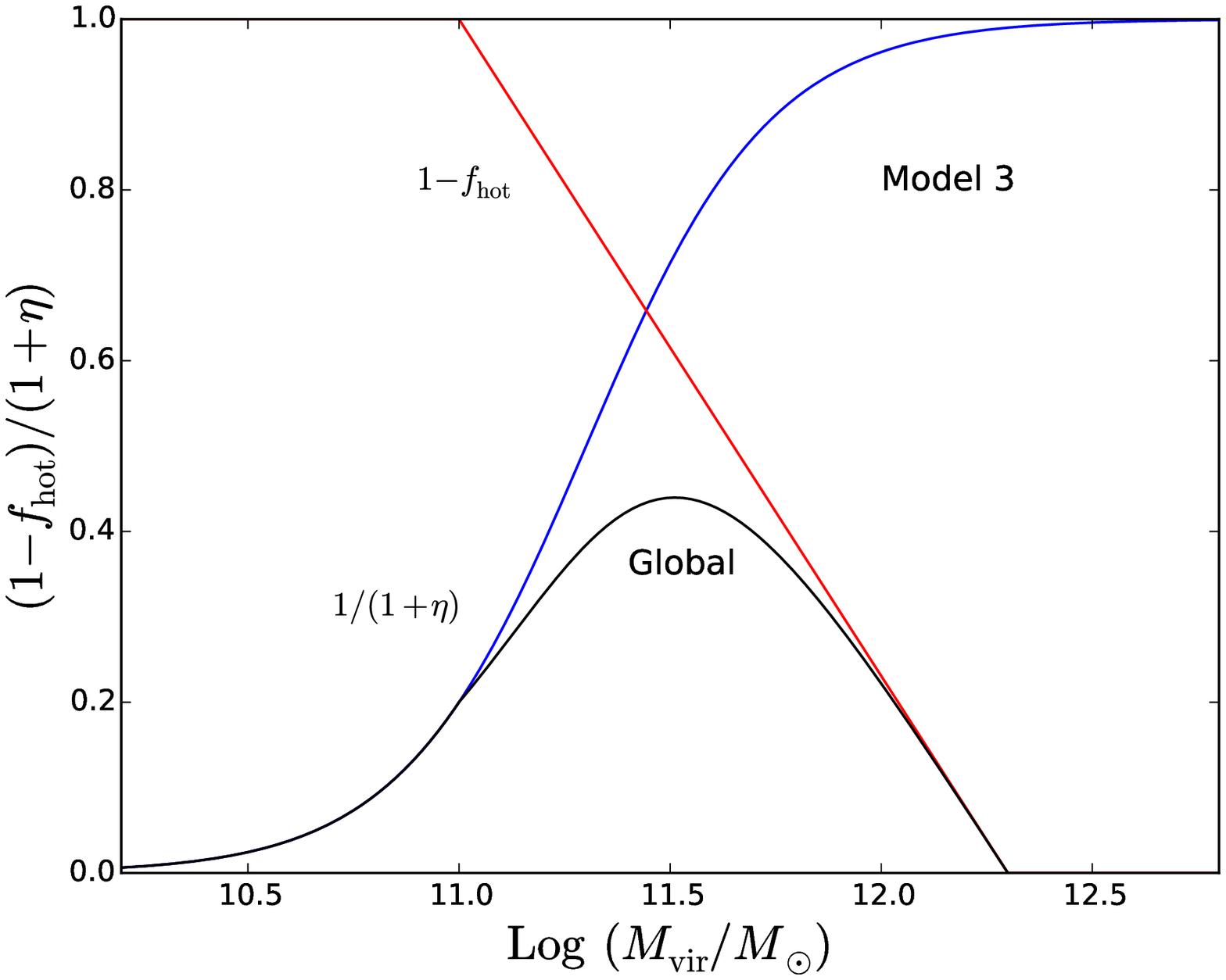}
\end{array}$
\end{center}
\caption{The baryon fraction available for star formation (black curve) is the product of the fraction that can accrete onto the galaxy
($1-f_{\rm hot}$; red curve) times
the one that is not ejected ($1/(1+\eta)$; blue curve).}
\label{gas_for_stars}
\end{figure*}
   
\begin{figure*}
\begin{center}$
\begin{array}{rr}
\includegraphics[width=0.5\hsize]{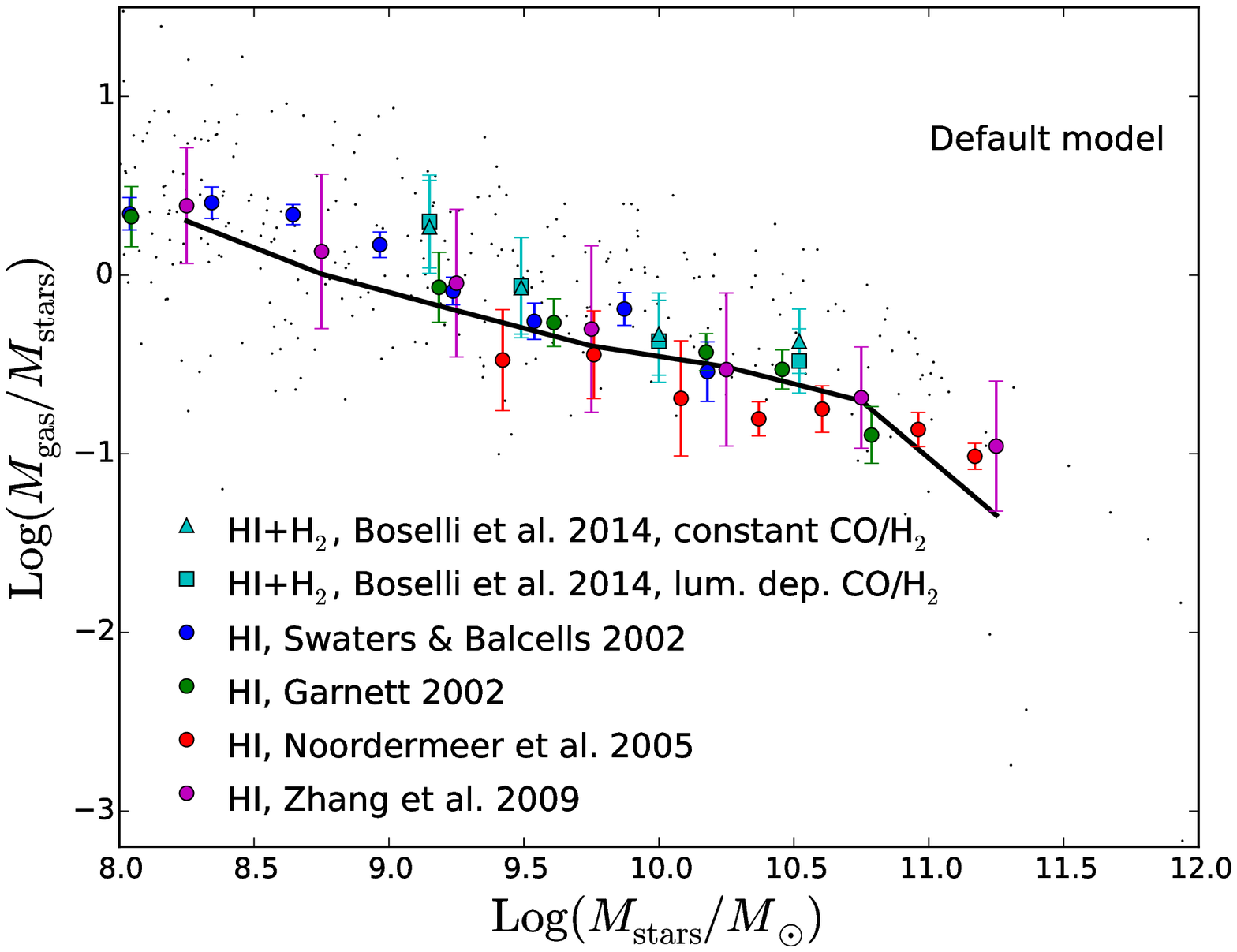}&
\includegraphics[width=0.5\hsize]{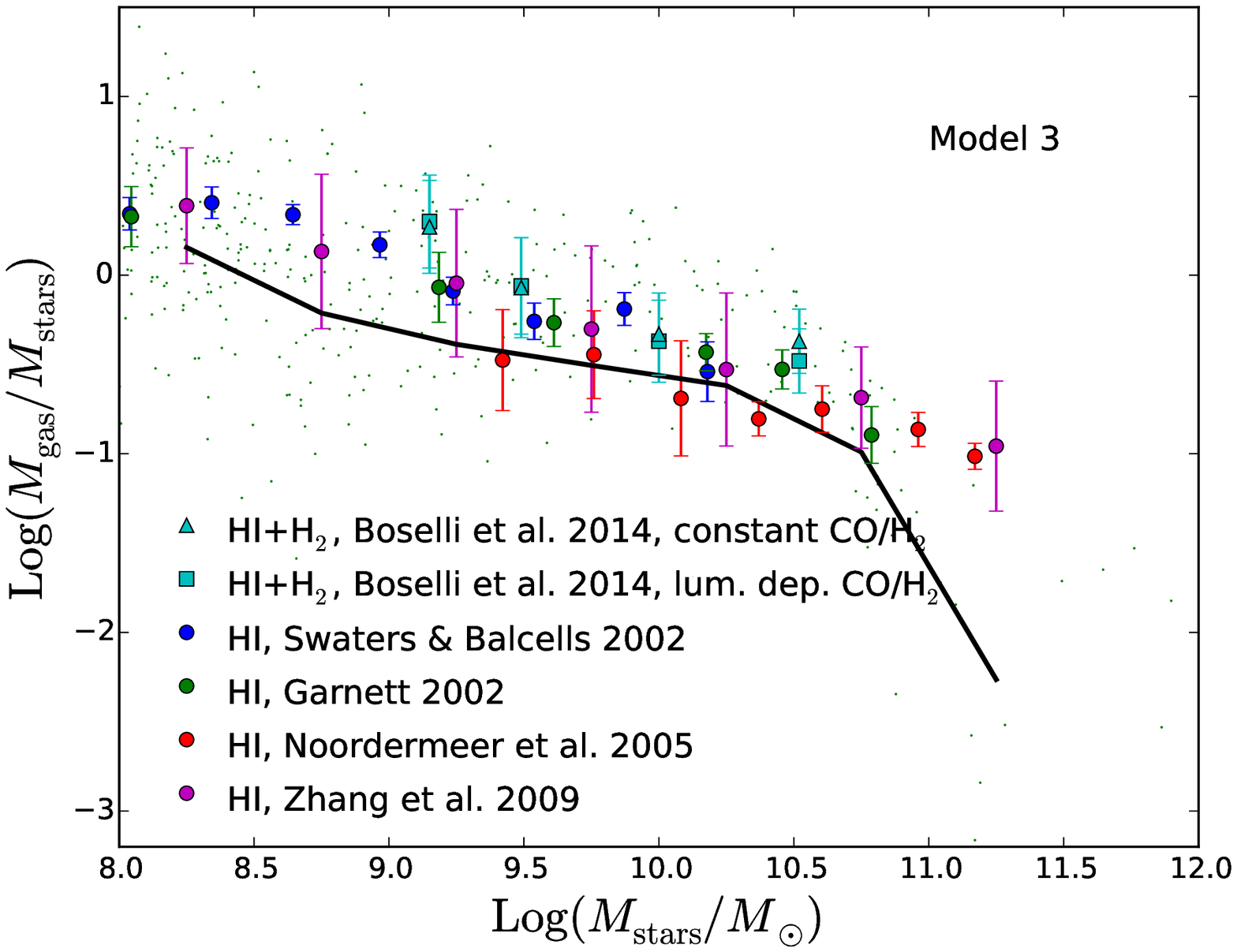}
\end{array}$ 
\end{center}
\caption{The local $M_{\rm gas}/M_{\rm star}$ - stellar mass relation in the default model (black point cloud, left), in model~3 (green point cloud, right) and in the observations (data points with error bars, identical in both panels).
The cyan triangles and squares are {\sc Hi}$+${\sc H}$_2$ data from \citet{boselli_etal14}.
The triangles are for a constant CO-to-{\sc H}$_2$ conversion factor. The squares are for a conversion factor that depends on luminosity.
The other symbols are observational determinations of  $M_{\rm gas}/M_{\rm star}$  based on {\sc Hi} data only
(blue: \citealp{swaters_balcells02}; green: \citealp{garnett02};  red: \citealp{noordermeer_etal05}; magenta: \citealp{zhang_etal09}).
The curves (black to the left, green to the right)
show the mean value of ${\rm Log}(M_{\rm gas}/M_{\rm stars})$ in bins of ${\rm Log}\,M_{\rm stars}$ for galaxies
that contain enough gas to be detected in {\sc Hi}.
The detection limit is about 
$M_{\rm gas}=10^{7.5}\,M_\odot$ at $M_{\rm stars}=10^8\,M_\odot$ and 
 $M_{\rm gas}=10^{7.5}\,M_\odot$ at $M_{\rm stars}=10^{11},M_\odot$ (A. Boselli, private communication). 
The {\sc Hi} detection limit for stellar masses intermediate between those above has been determined by logarithmic
interpolation.}
Only one model galaxy out of five has been shown not to overcrowd the
plot.

\label{gas_fraction}
\end{figure*}
\begin{figure*}
\begin{center}$
\begin{array}{rr}
\includegraphics[width=0.49\hsize]{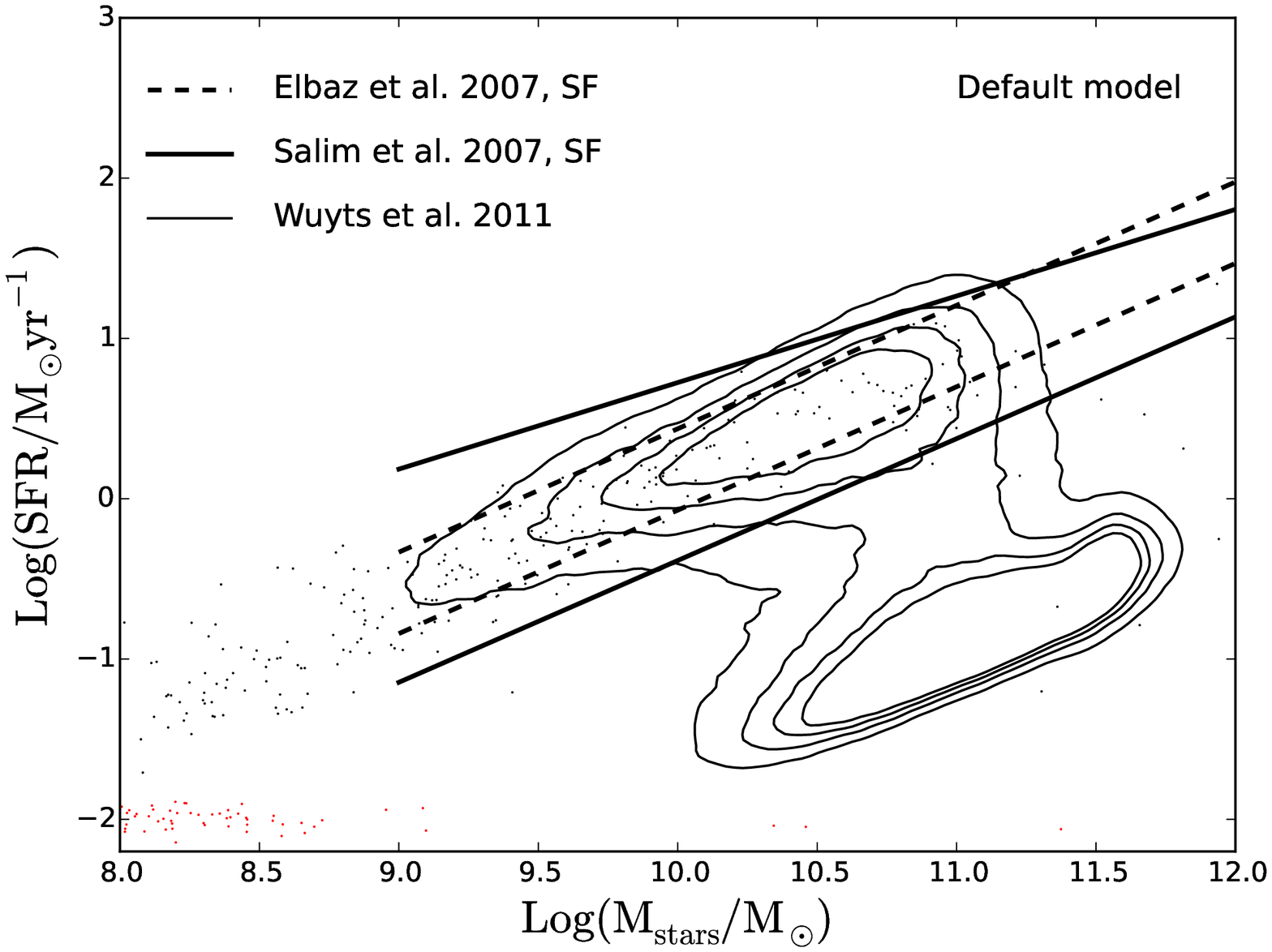}&
\includegraphics[width=0.49\hsize]{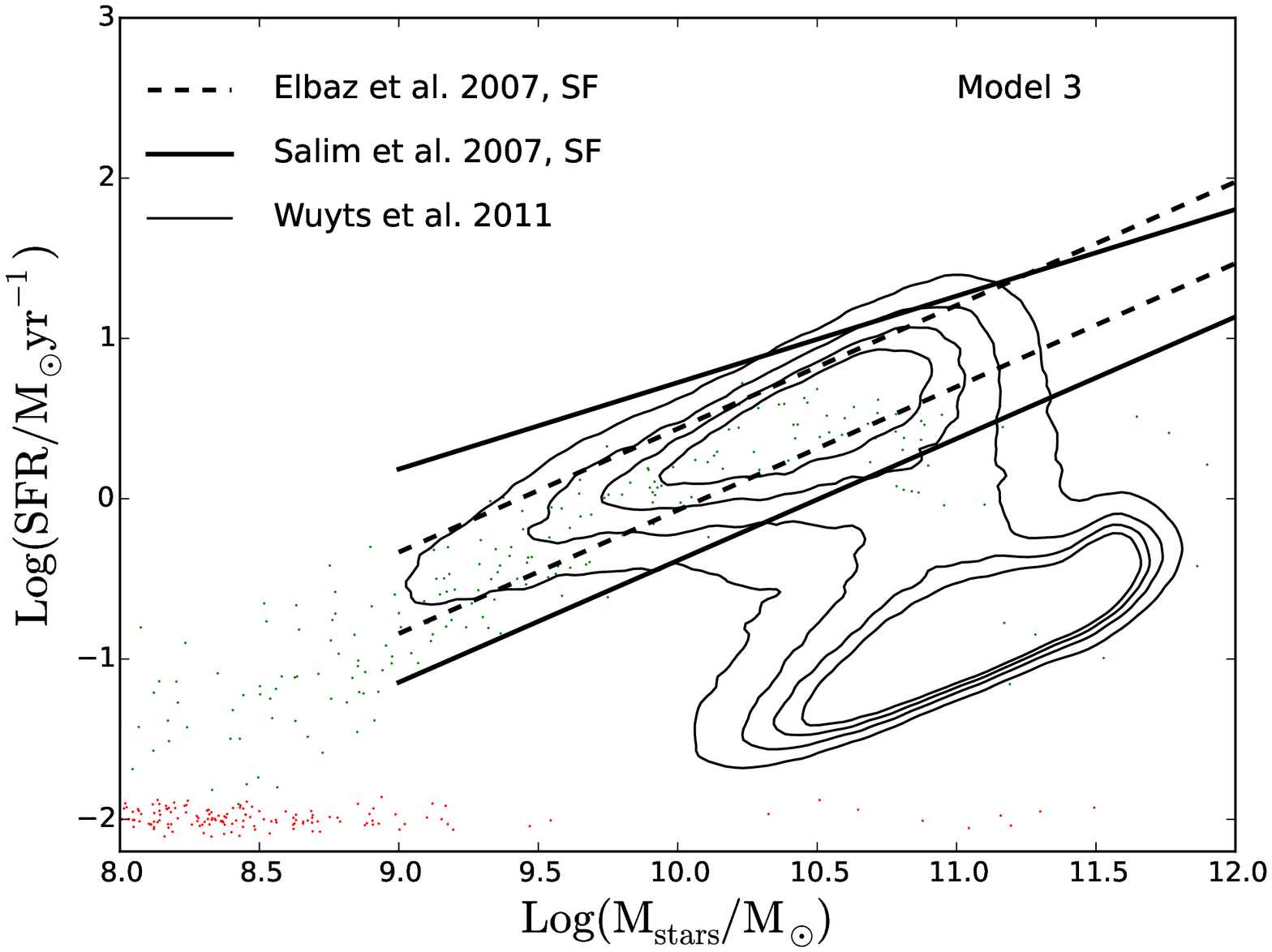} 
\end{array}$
\end{center}
\caption{The local star formation rate (SFR) - stellar mass relation in the default model (black point cloud, left) and model~3 (green point cloud, right) compared to observations.
The observations are the same in both panels.
The thick dashed and solid lines delimit the main sequence of star-forming (SF) galaxies as determined observationally by \citet{elbaz_etal07} and \citet{salim_etal07}, respectively.
The contours show the distribution of SDSS galaxies in the SFR - $M_{\rm stars}$ plane according to \citet{wuyts_etal11}.
Only one model galaxy out of five has been shown not to overcrowd the
plot.
The red dots are galaxies with zero SFR. 
We have assigned them an arbitrary SFR to be able to show them on the diagram.}
\label{SFR_mass}
\end{figure*}

\section{Comparison with observations}

In this section, we compare the models in Table~1 with observations. However, let us start with a foreword on our plotting conventions.
In most figures (SMFs, SFR function, cosmic SFR density, early-type fraction, disc sizes, stellar and baryonic TFR, Faber-Jackson relation), model predictions are shown by curves and observations by points with error bars.
In these figures, black solid curves correspond to the default model, red curves to model~1, black dashed curves to model~2 and green curves to model~3.
Some correlations ($M_{\rm stars}$ vs. $M_{\rm vir}$, gas-to-stellar mass ratio vs. $M_{\rm stars}$, SFR vs. $M_{\rm stars}$) are shown as scatter plots.
These figures are shown for the default model and model~3 only, the former with a black point cloud, the latter with a green one.

In this article, we compare our predictions to derived data (stellar masses, SFRs) rather than to primary data (magnitudes, colours)
because GalICS 2.0 has not been interfaced with stellar population synthesis models yet.
All data have been corrected for a Hubble constant of  $H_0=67.8{\rm\,km\,s}^{-1}{\rm Mpc}^{-1}$ and a \citet{chabrier03} inital mass function.

\subsection{Mass functions}

Fig.~\ref{SMFs} compare the galaxy SMFs predicted by the models in Table~1 (curves) to observations at different redshifts in the range $0<z<2.5$ (data points with error bars).
We begin our analysis from the local Universe ($z\simeq 0.1$), where 
there are also data for the baryonic mass function (Fig.~\ref{BMF}; the baryonic mass is the total mass of stars and cold neutral gas).
In Fig.~\ref{BMF}, we also show once and for all which line corresponds to each model.

Without adjusting any parameter besides $v_{\rm SN}$, $\alpha_v$ and $\alpha_z$, which are otherwise completely undetermined, the default model is
in good agreement with the local SMFs by \citet{baldry_etal12}, \citet{bernardi_etal13}, \citet{moustakas_etal13} and \citet{yang_etal09} in the mass range $10^9\,M_\odot<M_{\rm stars}<10^{11.8}\,M_\odot$.

The SMFs observed by different authors are overall fairly similar. The most noteworthy difference is that the SMF of
\citet{bernardi_etal13} contains a larger number of massive galaxies.
The reason is that \citet{bernardi_etal13} did not use the photometry from the SDSS pipeline.
They fitted the surface brightness profiles of galaxies with the combination of 
a \citet{sersic63} and an exponential profile, from which they computed magnitudes by extrapolating it to infinity.

At $M_{\rm stars}>10^{11.8}\,M_\odot$, we overpredict galaxy number densities by a factor of two
even with respect to  \citet{bernardi_etal13}. This could be an effect of cosmic variance
(the halo mass function shows an excess of objects by $\sim 40\%$ at $M_{\rm vir}\sim 3\times 10^{13}\,M_\odot$; Fig.\ref{HMF}), exacerbated by
overmerging because we have assumed that subhalo mergers result in
immediate galaxy mergers, though
tests based on a beta version with delayed merging show that this effect cannot be large.

Model~1 ($M_{\rm shock}=10^{11.3}\,M_\odot$ and $M_{\rm shutdown}=10^{12.4}\,M_\odot$; red curves) corresponds to a more abrupt shutdown of cold accretion,
reflected in a sharp change of slope of the SMF just below $M_{\rm stars}=10^{11}\,M_\odot$ (red curve at $z\simeq 0.1$). This model is in better agreement with the shape of the SMF measured by \citet{yang_etal09} and \citet{bernardi_etal13} 
at $10^9\,M_\odot<M_{\rm stars}<10^{11.6}\,M_\odot$.

Model~2 is identical to the default model except that we have limited the efficiency of SN feedback to $\epsilon_{\rm max}=0.12$. Capping the efficiency of SN feedback raises the slope of the galaxy SMF at low masses.
This improves the agreement with $z\simeq 0.1$ data  (compare the black dashes with the data at $M_{\rm stars}\lsim 10^9\,M_\odot$) but makes things worse at higher $z$.

The SMF of \citet{baldry_etal08,baldry_etal12} has a different shape from that of \citet{bernardi_etal13}, even though they are within each other's error bars everywhere except at the highest masses.
While the SMF of Bernardi et al. is consistent with a double power-law, that of Baldry et al. is steeper at $M_{\rm stars}<10^{9.5}\,M_\odot$, almost flat at
$10^{9.5}\,M_\odot<M_{\rm stars}<10^{10.5}$ and then drops more rapidly at higher masses.
Model~3 corresponds to a combination of parameters that was chosen to reproduce this behaviour.
We make the dependence of feedback on $v_{\rm vir}$ stronger (we pass from $\alpha_v=-4$ to $\alpha_v=-6.2$) so that the SMF becomes almost flat at intermediate masses but then we cap the efficiency of SN feedback
at $\epsilon_{\rm max}=0.12$ to increase the low-mass slope, like in model~2.
We also raise $M_{\rm shock}$ and lower $M_{\rm shutdown}$ (a bit like in model~1) to make the change of slope around $M_{\rm stars}\sim 10^{11}\,M_\odot$ more pronounced.

The turnovers seen at low masses in model 3 are caused by the limited resolution of the N-body simulation and provide a measure of the real stellar 
mass up to which resolution effects can propagate. This mass (of nearly $10^9\,M_\odot$) is an order of magnitude larger than the formal resolution limit (defined, in Section~3.2, as the $M_{\rm stars}$ corresponding to the minimum halo mass resolved by the N-body simulation).

At $0.5<z<0.8$, the agreement is still good. The default model and model~1 fit better the SMFs by \citet{ilbert_etal13} and \citet{muzzin_etal13}. 
Models~2 and~3 reproduce better the steep low-mass slope found by \citet{tomczak_etal14}.
The data of Ilbert et al. are closer to those of Muzzin et al. than to those of Tomczak et al.
However, Ilbert et al. and Tomczak et al. find the same behaviour at $0.5<z<0.8$ 
that Baldry et al. find in the local Universe:
their SMFs flatten around $M_{\rm stars}\sim 10^{10}\,M_\odot$ and steepen again at lower masses.
The only difference between the SMFs of Ilbert et al. and Tomczak et al. is that, 
at $M_{\rm stars}<10^{11}\,M_\odot$, the one of Tomczak et al. is shifted to higher masses by $0.3\,$dex on average.
In contrast, the SMF of Muzzin et al. displays a single slope at $M_{\rm stars}<10^{11}\,M_\odot$, like that of Bernardi et al. in the local Universe.

Once a major challenge for SAMs,
reproducing the number density of massive galaxies at high $z$ is no longer a problem
when we convolve our theoretical predictions with the observational errors to account for the Eddington bias.
\citet{ilbert_etal13} quote an error of $0.04(1+z)\,$dex on stellar masses, which they model with a Lorentzian distribution.
If we apply this assumption to our results, we find a small tail of galaxies the masses of which are overestimated by orders of magnitude.
We therefore make the conservative assumption that the errors are Gaussian.
While an error of $0.04(1+z)\,$dex may not apply to the data of other authors, who have not always stated how their errors vary with redshift,
we assume that the errors in the other datasets are of comparable magnitude.

The main discreapancy with the observations
is below the knee of the galaxy SMF. At $z>1$, models begin to overestimate the number density of galaxies with respect to all data sets.
The discrepancy is more severe when we limit the efficiency of SN feedback to $\epsilon_{\rm max}=0.12$ (models~2 and~3) and is a general problem of all SAMs (\citealp{fontanot_etal09};
\citealp{guo_etal11}; \citealp{henriques_etal12}; Asquith et al., in preparation).
{\sc Lgalaxies} \citep{henriques_etal13} is the only model that is marginally consistent with the observations because it combines high ejection rates with a reaccretion timescale that is inversely proportional to halo mass.
In GalICS 2.0, there is no reaccretion because there is no cooling, but simply reintroducing cooling, without a gradual return of gas to the halo, would not solve the problem because the reaccretion timescale required to make this picture work
($6\,$Gyr time for a halo with $M_{\rm vir}\sim 3\times 10^{10}\,M_\odot$) is much longer than the radiative cooling timescale.
Alternative explanations are overefficient star formation in dwarf galaxies in SAMs (but see Section~3.4) or that the observations are missing faint galaxies at high $z$.

Comparing GalICS 2.0 to the baryonic mass function is useful because we can see the extent to which our mass functions are affected by our star formation law.
Unfortunately, these data are only available for the local Universe. \citet{papastergis_etal12} determined the baryonic mass function from a sample for which both optical (SDSS) and {\sc Hi} (ALFALFA) data were available.
The baryonic mass was assumed to be  $M_{\rm gal}=M_{\rm stars}+1.4M_{\rm HI}$, where $M_{\rm stars}$ is the stellar mass derived from optical data, $M_{\rm HI}$ is the {\sc Hi} mass from radio data and the factor of $1.4$ accounts for the
presence of helium. This is a lower limit for $M_{\rm gal}$ because gas could be present and not be detected.
One can find an upper limit by giving to galaxies not detected in {\sc Hi} the maximum {\sc Hi} mass consistent with their non-detection.
The gray shaded area in Fig.~\ref{BMF} shows the region between these limits.

Massive galaxies have low gas fractions (Section~3.3). Therefore, the baryonic mass function is essentially identical to the SMF at high masses.
\citet{papastergis_etal12}'s SMF is intermediate between \citet{baldry_etal08}'s and \citet{yang_etal09}'s but closer to the former than to the latter.
Model~3 fits the baryonic mass function of Papastergis et al. at $10^{11}\,M_\odot<M_{\rm gal}<3\times 10^{11}\,M_\odot$ better than the other three models because it is calibrated on Baldry et al's data.
None of the models fits the baryonic mass function at $M_{\rm gal}>3\times 10^{11}\,M_\odot$ but neither do they fit the SMF of Baldry et al. at $M_{\rm stars}>3\times 10^{11}\,M_\odot$. 

The agreement of the default model, model~1 and model~2 with the baryonic mass function of \citet{papastergis_etal12}
is good down to $M_{\rm gal}\sim 10^{10}\,M_\odot$. At $10^9\,M_\odot<M_{\rm gal}<10^{10}\,M_\odot$, baryonic masses are underestimatee by 
$0.2\,$dex on average. $M_{\rm gal}\sim 10^9\,M_\odot$ is our resolution in baryonic mass, which produces the
turnover at low masses seen in all models. Hence, it would make no sense to extend the comparison with the data to lower masses.
In model~3, the baryonic mass function is reproduced correctly around $M_{\rm gal}\sim 10^9\,M_\odot$ but is underpredicted
around $M_{\rm gal}\sim 10^{10}\,M_\odot$.
Overall, the comparison with the baryonic mass function of \citet{papastergis_etal12} seems to suggest that, in GalICS 2.0,
 low-mass galaxies do not contain enough gas for their
stellar masses. See, however, Section~3.3 for a direct comparison with gas fractions.

\subsection{Halo masses}

Fig.~\ref{HOD} compares the $M_{\rm stars}$ - $M_{\rm vir}$ relation predicted by the default model (left, black point cloud) and model~3 (right, green point cloud)
with data from weak lensing \citep{reyes_etal12} and satellite kinematics \citep{wojtak_mamon13}. 
The excellent agreement (particularly with weak-lensing data)
provides additional and independent evidence that local haloes harbour galaxies with sensible stellar masses, at least for systems with $M_{\rm stars}\gsim 10^{10}\,M_\odot$.
On the other hand, Fig.~\ref{HOD} shows that these data are not very powerful to discriminate between models.

The dotted-dashed diagonal line in Fig.~\ref{HOD} corresponds to $M_{\rm stars}=(\Omega_b/\Omega_M)M_{\rm vir}$. It has been plotted to emphasize that the 
$M_{\rm stars}$ - $M_{\rm vir}$ relation is steeper than linear at $M_{\rm vir}<10^{12}\,M_\odot$ and shallower at higher masses.
The logarithmic slope of the relation varies from $\sim 2$ at low mass to $\sim 1/2$ at high mass.
This change is the reason why we observe a knee in the galaxy stellar mass function (Fig.~\ref{SMFs}).
In our model, it results from the combined effects of SN feedback (at low masses) and shock heating (at high masses), both of which limit the baryon mass that can be converted into stars.
 
This point is illustrated in Fig.~\ref{gas_for_stars}: $1/(1+\eta)$ is the retained (not ejected) gas fraction (blue curve); 
$1-f_{\rm hot}$ is the gas fraction that is able to accrete onto galaxies (red curve). Both are shown as function of $M_{\rm vir}$.
The gas that makes stars is the one that is able to accrete and to avoid ejection. Its mass fraction, $(1-f_{\rm hot})/(1+\eta)$, 
is shown by the black curve. The black curve peaks for $M_{\rm vir}\simeq 10^{11.6}\,M_\odot$ in the default model and
$M_{\rm vir}\simeq 10^{11.5}\,M_\odot$ in model~3.
Fig.~\ref{gas_for_stars} shows that the formation of galaxies is
efficient only in a narrow range of halo masses, between $10^{11}$ and
a few times $10^{12}\,M_\odot$,
in agreement with previous studies by \citet{bouche_etal10}, \citet{guo_etal10},
\citet{cattaneo_etal11}, \citet{behroozi_etal13} and \citet{birrer_etal14}.
Even in this mass range, it is very difficult for haloes to convert more than a third of the baryons into stars (Fig.~\ref{HOD}).

The yellow points in Fig.\ref{HOD} show the $M_{\rm gal}$ - $M_{\rm vir}$ relation for the baryonic rather than the stellar mass. 
The vertical dashed lines correspond to the resolution of the N-body simulation, while the black and the yellow arrows mark
the formal resolution masses for the stellar mass and the baryonic mass, respectively. We define them as the
masses $M_{\rm stars}$ and $M_{\rm gal}$ that corresponds to our halo-mass resolution ($M_{\rm vir}=10^{10.5}\,M_\odot$) on a galaxy mass - halo mass diagram (Fig.~\ref{HOD}).
Both the default model and model~3 have formal resolution $M_{\rm stars}\lsim 10^8\,M_\odot$.
Hence, the SMFs in Fig.~\ref{SMFs} should be well resolved over the entire plotting range, while the baryonic mass function in Fig.~\ref{BMF} is expected to be only at $M_{\rm gal}>10^9\,M_\odot$.
However, resolution effects can trickle above the formal resolution mass (e.g., \citealp{cattaneo_etal11}).
The turnovers in SMFs between $10^8\,M_\odot$ and $10^9\,M_\odot$ are a clear sign of that.
We therefore only trust results above $10^9\,M_\odot$ for both stellar and baryonic masses.

\begin{figure}
\begin{center}
\includegraphics[width=0.95\hsize]{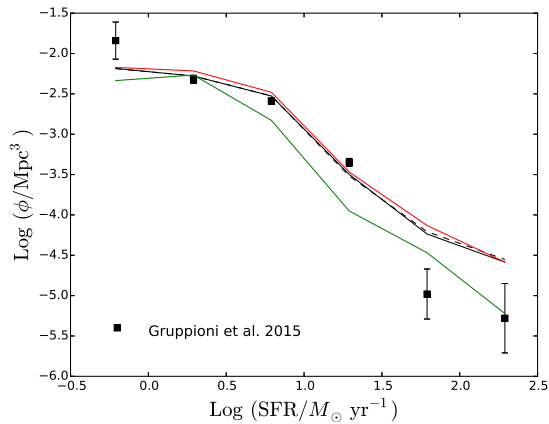} 
\end{center}
\caption{The SFR function in the local Universe. The curves show the prediction of GalICS 2.0 at $z\simeq 0.1$ 
(black solid: default model; red: model~1; black dashed: model~2; green: model~3). The data points with error bars are the 
observations by \citet{gruppioni_etal15} at $0<z<0.3$.}
\label{SFRf}
\end{figure}

\subsection{Gas fractions}

The default model's predictions for the gas-to-stellar mass ratio as a function of stellar mass are in good agreement with the measurements 
of $(M_{\rm HI}+M_{{\rm H}_2})/M_{\rm star}$
in a volume-limited, $K$-band-selected sample of nearby late-type galaxies (\citealp{boselli_etal14}; Fig.~\ref{gas_fraction}, left).
Unsurprisingly, these measurements find slightly higher gas-to-stellar mass ratios 
than studies based on {\sc Hi} data only (\citealp{swaters_balcells02,garnett02,noordermeer_etal05,zhang_etal09}; also shown in Fig.~\ref{gas_fraction}),
but the differences are small.

The {\sc Hi} measurements of \citet{swaters_balcells02}, \citet{garnett02} and \citet{noordermeer_etal05} are all 
for spiral or irregular galaxies, but  their results are not systematically different from those of
\citet{zhang_etal09}, who did not operate any morphological selection
(Zhang et al. did not measure the {\sc Hi} mass of each individual galaxy in their SDSS sample but inferred it from its colour and luminosity using an empirical relation  calibrated on $800$ galaxies with optical photometry from the SDSS and
{\sc Hi} masses from the HyperLeda catalogue of \citealp{paturel_etal03}).

While the point clouds that show the results of GalICS 2.0 in Fig.~\ref{gas_fraction} are composed of galaxies on which no
selection has been performed, galaxies with so little gas that would not be detected in {\sc Hi} have not been included in the calculation of the mean gas fraction (shown by the black curve for the default model and the green curve for model~3;
averages are logarithmic).
This selection, based on assuming a minimum detectable gas  mass of
$M_{\rm gas}=10^{7.5}\,M_\odot$ for a galaxy with $M_{\rm stars}=10^8\,M_\odot$ and 
 $M_{\rm gas}=10^{7.5}\,M_\odot$ for a galaxy $M_{\rm stars}=10^{11},M_\odot$ (A. Boselli, private communication),
is effectively equivalent to a morphological selection.

The direct measurements in Fig.~\ref{gas_fraction} should be compared to indirect constraints on gas-to-stellar mass ratios from the baryonic mass function \citep{papastergis_etal12} and the TFR relation in dwarf galaxies \citet{papastergis_etal16}.
Fig.~\ref{BMF} suggests that, at $M_{\rm gal}\sim 3\times 10^9\,M_\odot$, the default model underestimates $M_{\rm gal}$ by $\sim 0.2\,$dex on average,
possibly because the gas content is underpredicted, since the SMF by \citet{papastergis_etal12} is intermediate between those of
\citet{baldry_etal08,baldry_etal12} and \citet{yang_etal09}. 
TFR informs us on gas fractions because there are data sets for which we have both $M_{\rm stars}$ and $M_{\rm gal}$ as a function of rotation speed
but the interpretation of these data is not straightforward and will be discussed at length in Section~3.7.

\subsection{Star formation rates}

\begin{figure*}
\begin{center}
\includegraphics[width=0.95\hsize]{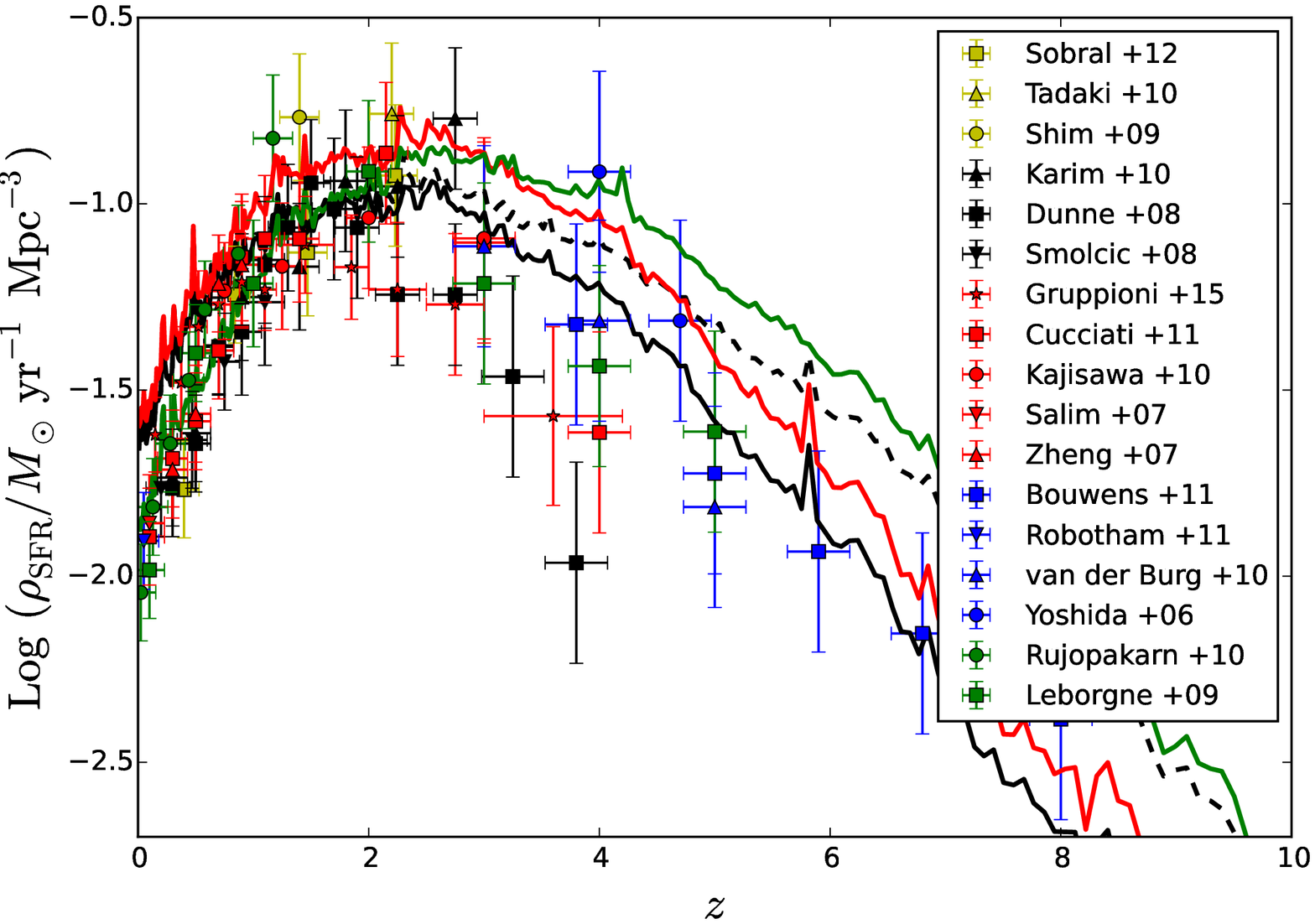} 
\end{center}
\caption{Evolution of the cosmic SFR density across the Hubble time in GalICS 2.0 (curves) and the observations (points with error bars from the compilation by \citealp{behroozi_etal13}). 
Yellow symbols show $H\alpha$ data (squares: \citealp{sobral_etal12} triangles: \citealp{tadaki_etal11}; circles: \citealp{shim_etal09}),
black symbols $1.4\,$GHz data (triangles: \citealp{karim_etal11}; squares: \citealp{dunne_etal09}; down-pointed triangles: \citealp{smolcic_etal09}),
red symbols combined UV/IR data (stars: \citealp{gruppioni_etal15}; squares: \citealp{cucciati_etal12}; circles: \citealp{kajisawa_etal10}; down-pointed triangles: \citealp{salim_etal07}; triangles: \citealp{zheng_etal07}),
blue symbols UV data only (squares: \citealp{bouwens_etal12}; down-pointed triangles: \citealp{robotham_driver11}; triangles: \citealp{vanderburg_etal10}; circles: \citealp{yoshida_etal06}), 
and green symbols FIR/IR data only (circles: \citealp{rujopakarn_etal10}; squares: \citealp{leborgne_etal09}).
The line styles for the models are the same as in Fig.~\ref{BMF}.}
\label{Madau}
\end{figure*}

Both the default model and model~3 are broadly consistent with the slope and the normalization of the $M_{\rm star}$ - SFR relation in the local Universe (\citealp{elbaz_etal07,salim_etal07,wuyts_etal11}; Fig.~\ref{SFR_mass}).
On a closer inspection, however, both have their shortcomings.
The slope of the main sequence of star-forming is correctly reproduced by the standard model but appears titlted in model~3.
In contrast, model~3 reproduces correctly the characteristic mass $M_{\rm stars}\sim 10^{10.7}-10^{10.8}\,M_\odot$ at which galaxies migrate from the star-forming population to the passive one according to \citet{wuyts_etal11}.
In the default model, this mass is overestimated because galaxies in the mass range 
$M_{\rm stars}\sim 10^{10.7}-10^{10.8}\,M_\odot$ have still got plenty of gas (Fig.~\ref{gas_fraction}).

In both cases, one remarks a small but non-negligible tail of star-forming galaxies at $M_{\rm stars}>3\times 10^{11}\,M_\odot$.
This tail is evidence that shutting down cold accretion may not be enough and that an additional quenching mechanism (e.g., quasar feedback) is probably needed (but see \citealp{bildfell_etal08}, who argued for reactivated star formation in cD galaxies).

Fig.~\ref{SFRf} compares the local ($z\simeq 0.1$) SFR functions in our four models with the data by \citet{gruppioni_etal15} at $0<z<0.3$.
The most interesting differences are between the default model and model~3.
Overall, model~3 predicts lower SFRs than the default model, which fits the SFR function better at all but the highest SFRs.

Having examined star formation in the local Universe, we move our attention to the evolution of the cosmic SFR density across the Hubble time (Fig.~\ref{Madau}).
Measured SFR densities were taken from the compilation by \citet{behroozi_etal13}, which includes 
$H\alpha$ \citep{sobral_etal12,tadaki_etal11, shim_etal09} 
radio ($1.4\,$GHz; \citealp{karim_etal11,dunne_etal09,smolcic_etal09}),
combined ultraviolet/infrared \citep{gruppioni_etal15,cucciati_etal12,kajisawa_etal10,salim_etal07,zheng_etal07},
ultraviolet data only \citep{bouwens_etal12,robotham_driver11,vanderburg_etal10,yoshida_etal06}, 
and far infrared data only \citep{rujopakarn_etal10,leborgne_etal09}.
The default model is in good agreement with observation at $z>2$ (especially with UV data) because it has strong feedback that suppresses star formation but forms too many stars at low $z$ (Figs.~\ref{SFR_mass} and~\ref{SFRf}).
In model~3, feedback is capped to $\epsilon_{\rm max}=0.12$. 
Therefore, even though the efficiency scales as $\epsilon_{\rm SN}\propto (1+z)^3$,
its increase at high $z$ cannot be as strong as in the default model and the cosmic SFR density is overpredicted.
However, the lower value of $M_{\rm shutdown}$ curbs star formation in massive galaxies much more effectively and this improves the fit at low $z$.
In fact, the agreement of model~3 with the observations at $0<z<2$ is unprecedented.

\begin{figure}
\begin{center}
\includegraphics[width=0.95\hsize]{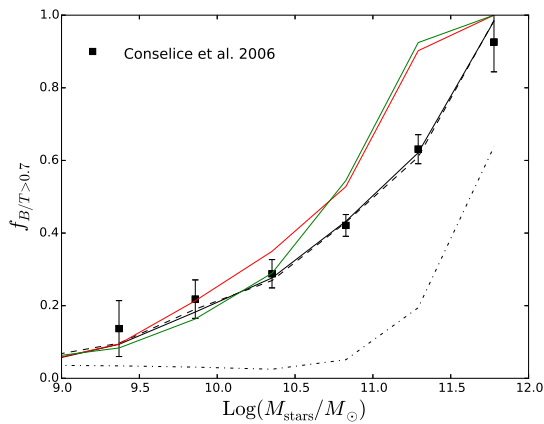} 
\end{center}
\caption{The fraction of galaxies with bulge-to-total mass ratio $B/T>0.7$ as a function of stellar mass in GalICS 2.0 (curves) compared to the fraction of elliptical galaxies in the observations of \citet{conselice06}.
As in Figs.~\ref{SMFs}, \ref{BMF} and~\ref{SFRf}, the black solid curve refers to the default model, while the red curve, the black dashed curve and the green curve refer to models~1, 2 and~3, respectively. Here we also show a fourth model, identical to the default one, except that we turn off disc instabilities (black dotted-dashed curve).}
\label{BT}
\end{figure}

\subsection{Morphologies}
 
The data points with error bars in Fig.~\ref{BT} show the fraction of elliptical galaxies. 
They are based on galaxies visually classified by \citet{conselice06}, who separated them into E, S and Irr types.
In order to compare the results of GalICS 2.0 to these observational data, it is necessary to determine to which bulge-to-total mass ratios $B/T$ this classification corresponds.
\citet{weinzirl_etal09} have shown that S-type galaxies have $B/T$ in the range $0-0.5$ (two thirds have $B/T<0.2$).
The range extends to $0-0.7$ if we broaden our definition of S-type galaxies to include S0s \citep{laurikainen_etal10}.
We therefore follow \citet{wilman_etal13} and \citet{fontanot_etal15} in using $B/T=0.7$ as the watershed bulge-to-total mass ratio that separates S- and E-type galaxies.

GalICS 2.0 computes $B/T$ ratios assuming that the bulge mass is the total stellar mass of the classical bulge and the pseudobulge. The total mass is the sum
of the stellar masses of the disc, the classical bulge and the pseudobulge.
The fraction of galaxies with $B/T>0.7$ increases with $M_{\rm stars}$, first more gently at $M_{\rm stars}<10^{11}\,M_\odot$, then more rapidly at $M_{\rm stars}>10^{11}\,M_\odot$,
where mergers become the main mechanism of galaxy growth (e.g., \citealp{cattaneo_etal11}) and bulge formation.
If mergers are the only mechanism to form bulges, a significant population of galaxies with $B/T>0.7$ appears only for $M_{\rm stars}>1-3\times 10^{11}\,M_\odot$
(the dotted-dashed curve in Fig.~\ref{BT} corresponds to the default model without disc instabilities; also see \citealp{lacey_etal16}).

When disc instabilities are activated alongside morphological transformation in major mergers,
the default model (black solid curve) and model~2 (black dashed curve) reproduce a good agreement to the data points 
in Fig.~\ref{BT}
for $\epsilon_{\rm m}=4$ and $\epsilon_{\rm inst}=0.9$. 
In contrast, model~1 (red curve) and model~3 (green curve) overestimate the fraction of massive ellipticals even when the critical mass ratio for major mergers is lowered to $\epsilon_{\rm m}=3$.
These differences come from the gas mass that accretes onto galaxies because the merger rate is set by the DM and is the same in all models
(in models with a larger value of $M_{\rm shutdown}$, even galaxies as large as the Milky Way have a chance to regrow a disc after a merger).

Fig.~\ref{BT} proves that our morphologies are reasonable.
However, as the association of visually classified elliptical galaxies to a critical bulge-to-total mass ratio
of $B/T=0.7$ is somewhat arbitrary,
a quantitative comparison with an observational sample with measured $B/T$ ratios is highly desirable.
Nevertheless, the only purpose of morphologies in this article is to select
spiral galaxies when comparing to observations for disc sizes (Section~3.6) and the TFR (Section~3.7).
As we have verified that neither disc sizes nor the TFR were sensitive 
to the critical $B/T$ used to select spiral galaxies (choosing $B/T<0.3$ or $B/T<0.7$ does not change the TFR significantly), 
we have decided to defer the comparison with quantitative morphologies to a future publication.

\begin{figure}
\begin{center}
\includegraphics[width=0.95\hsize]{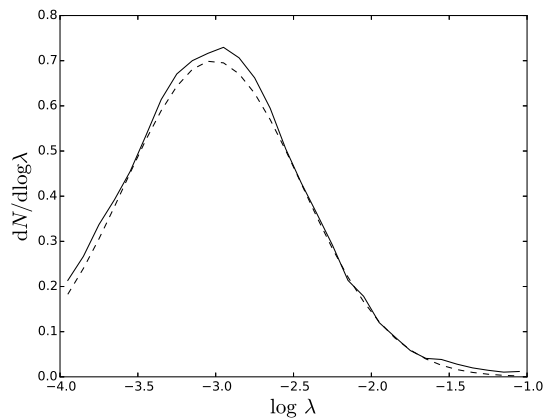} 
\end{center}
\caption{The distribution for the spin parameter $\lambda$ in our N-body simulation (solid curve) compared to a log-normal distribution with $\bar{\lambda}=0.049$ and $\sigma_{{\rm ln\,}\lambda}=0.57$ (dashed curve).}
\label{SpinDist}
\end{figure}

\begin{figure*}
\begin{center}$
\begin{array}{cc} 
\includegraphics[width=0.5\hsize]{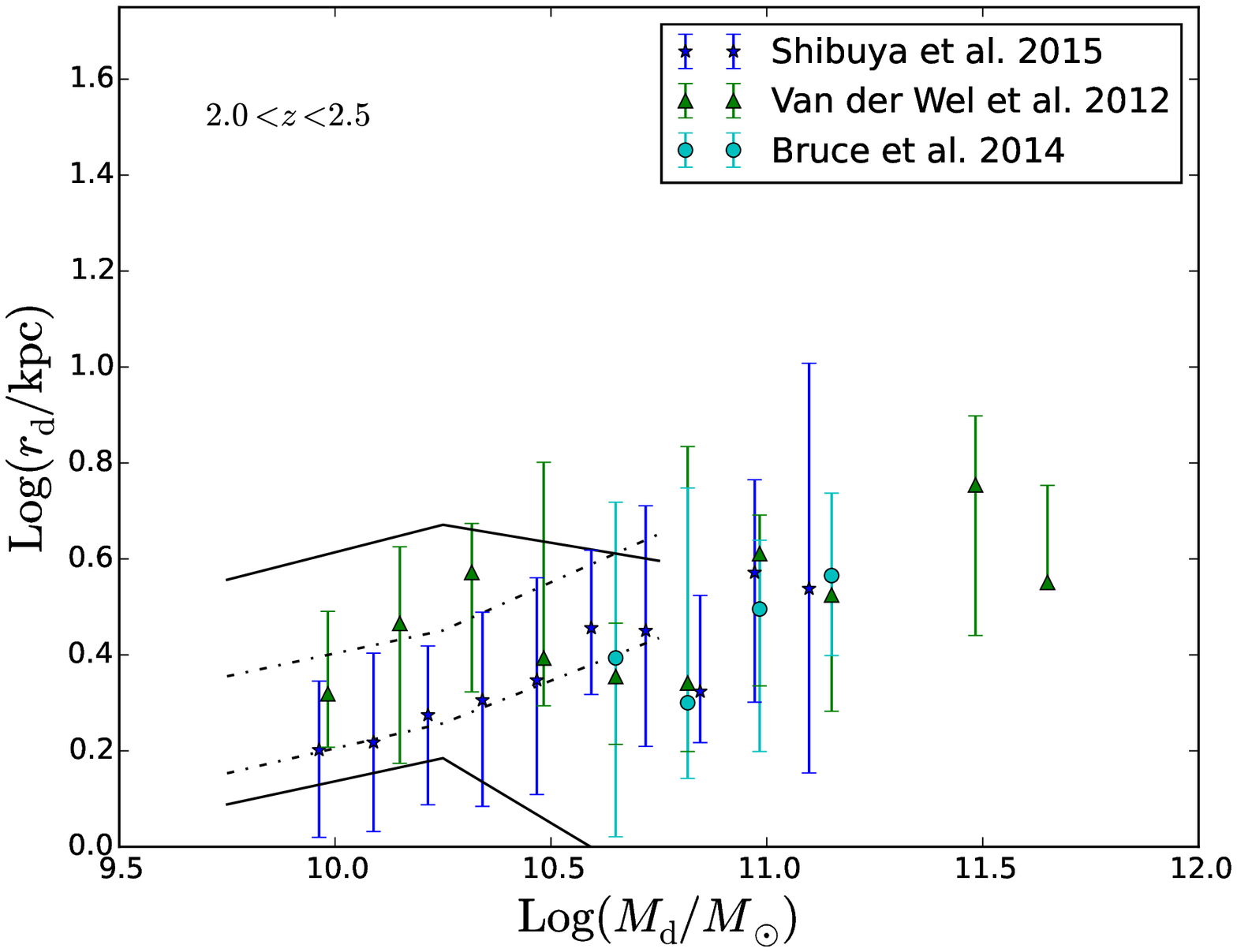} &
\includegraphics[width=0.5\hsize]{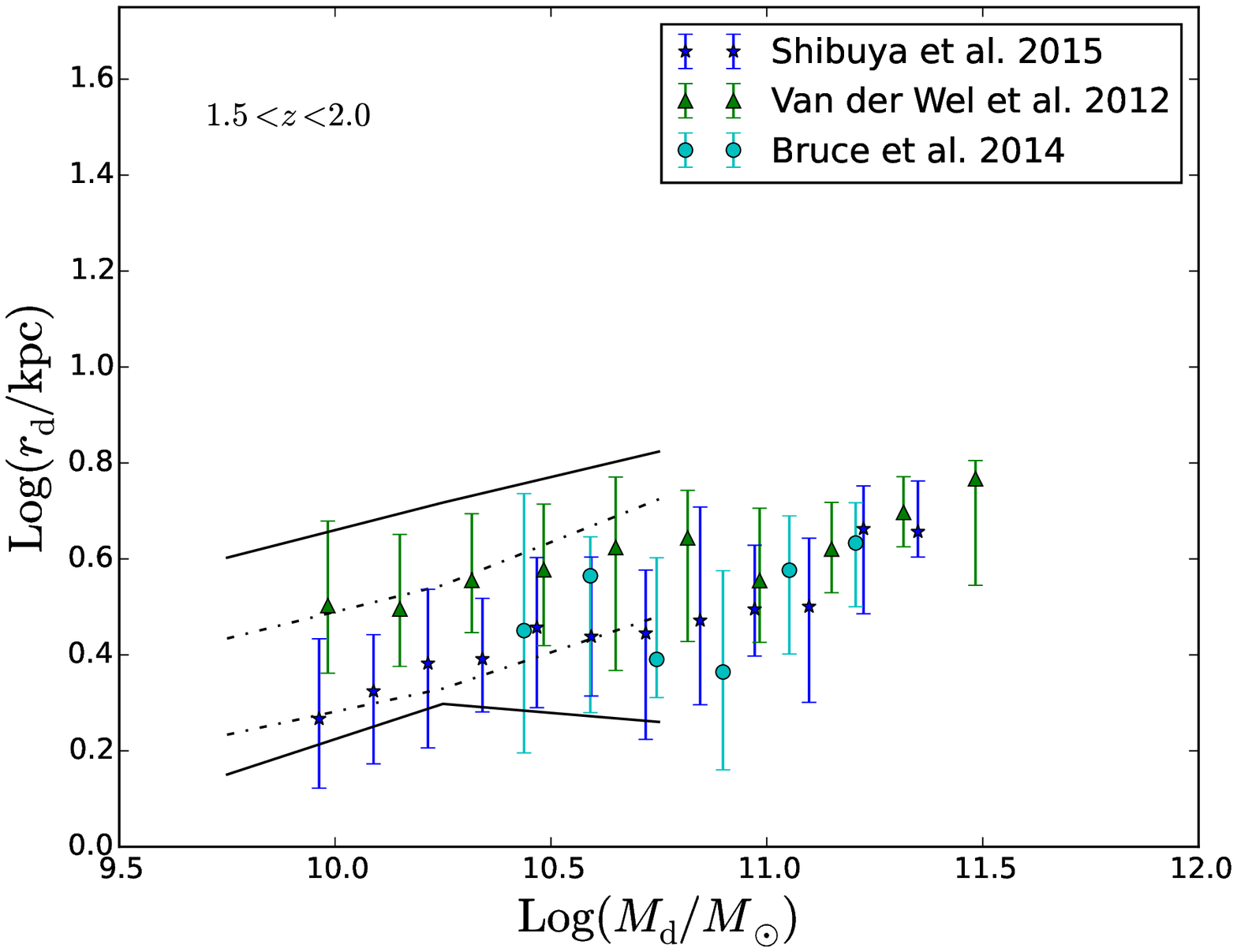} \\
\includegraphics[width=0.5\hsize]{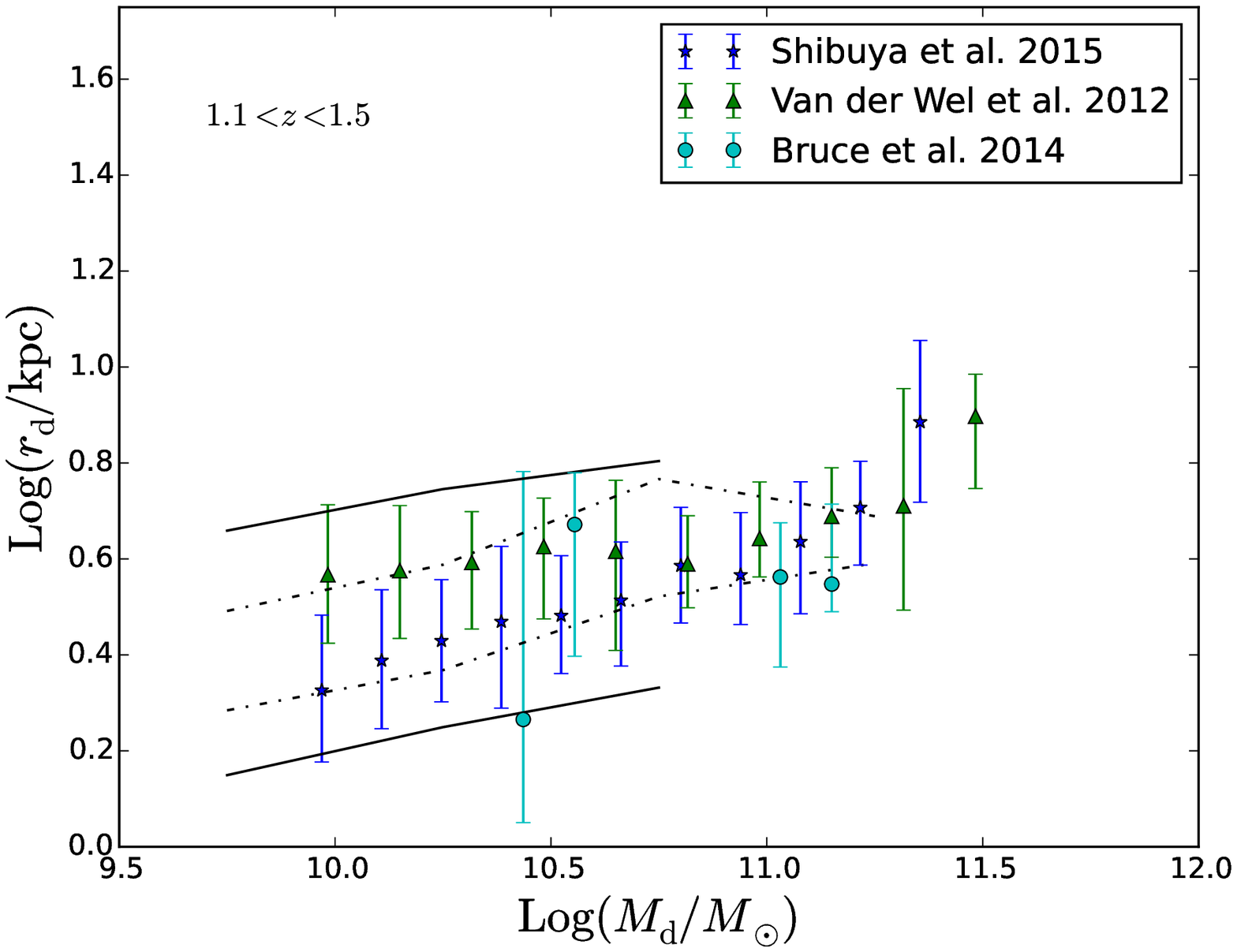}&
\includegraphics[width=0.5\hsize]{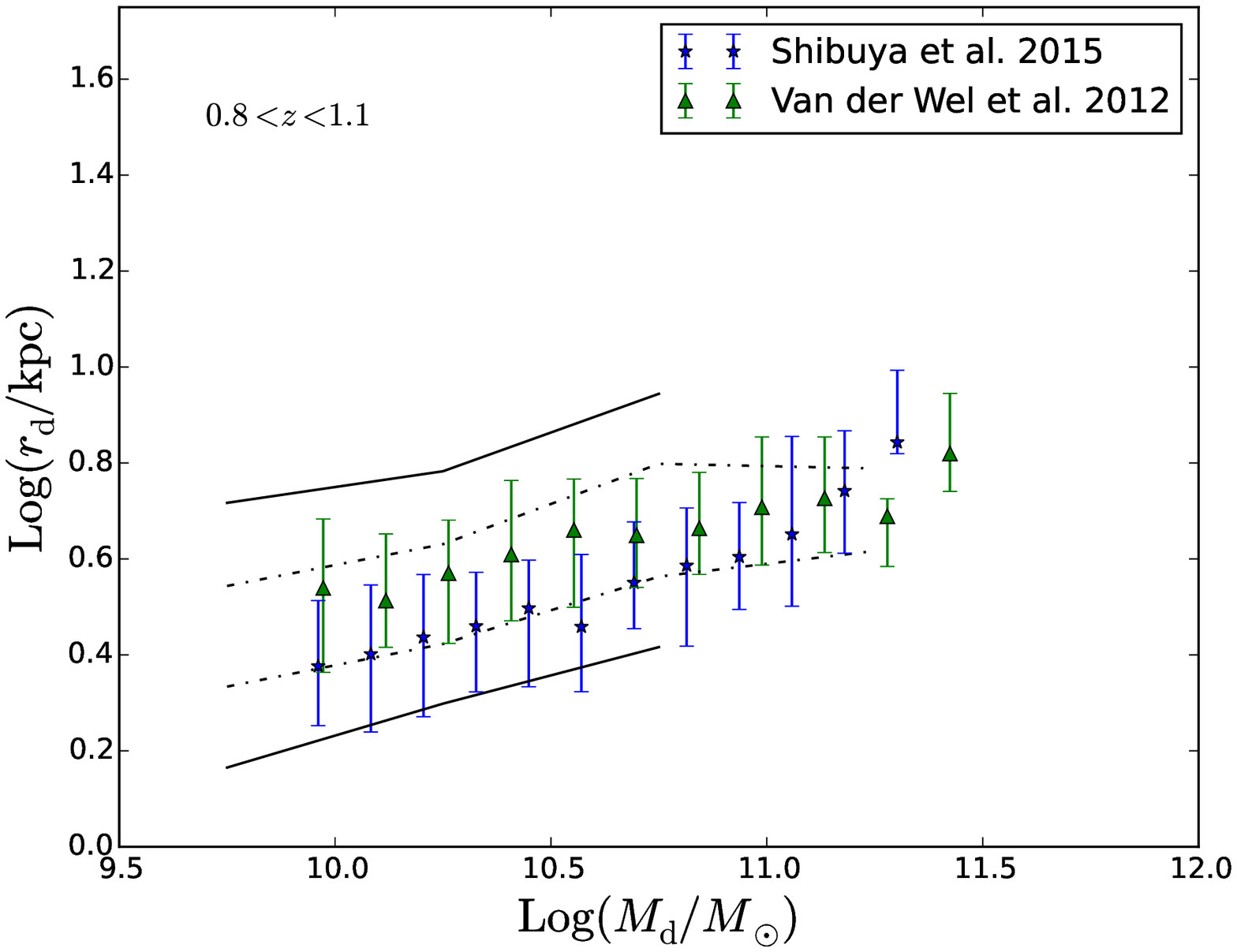} \\
\includegraphics[width=0.5\hsize]{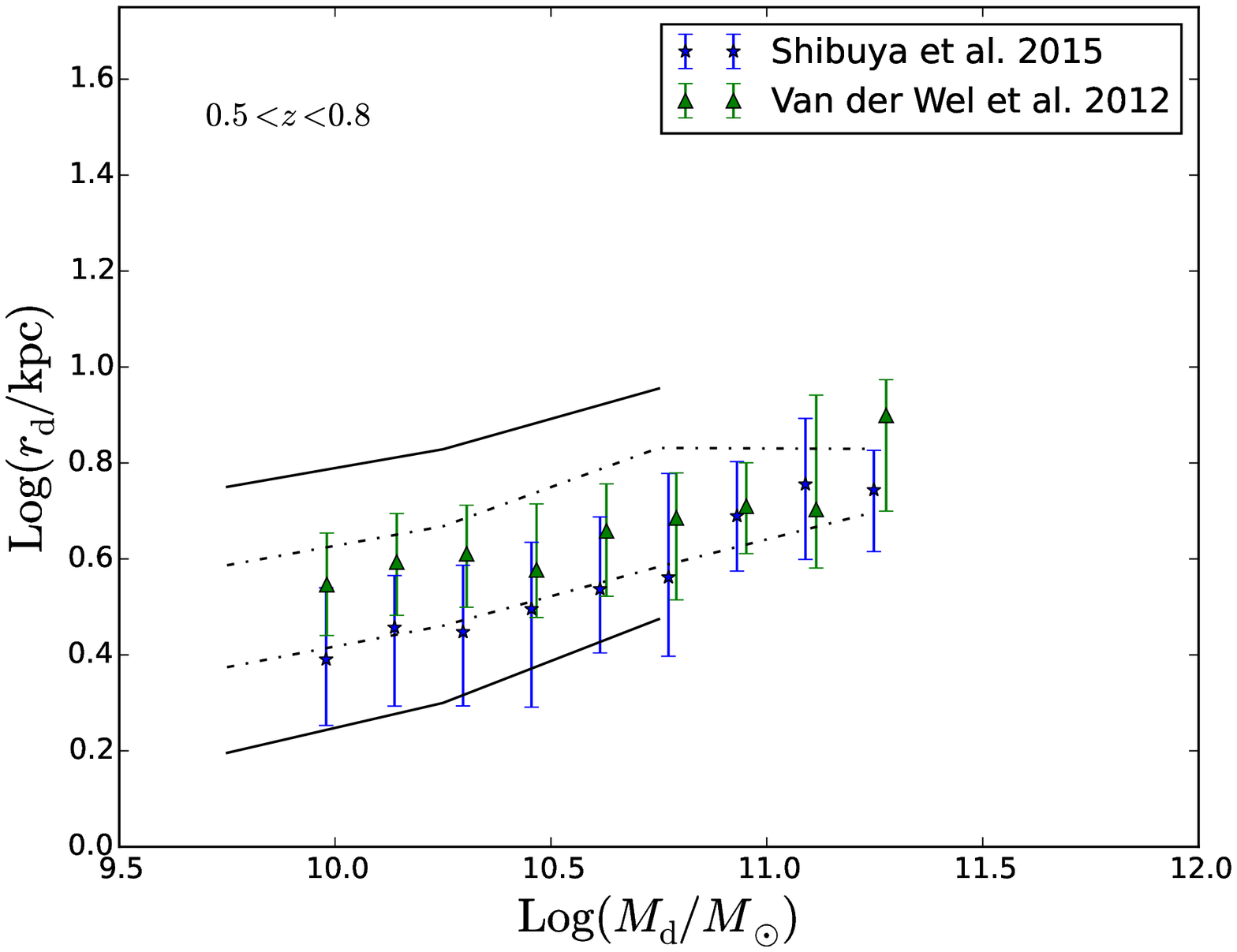} &
\includegraphics[width=0.5\hsize]{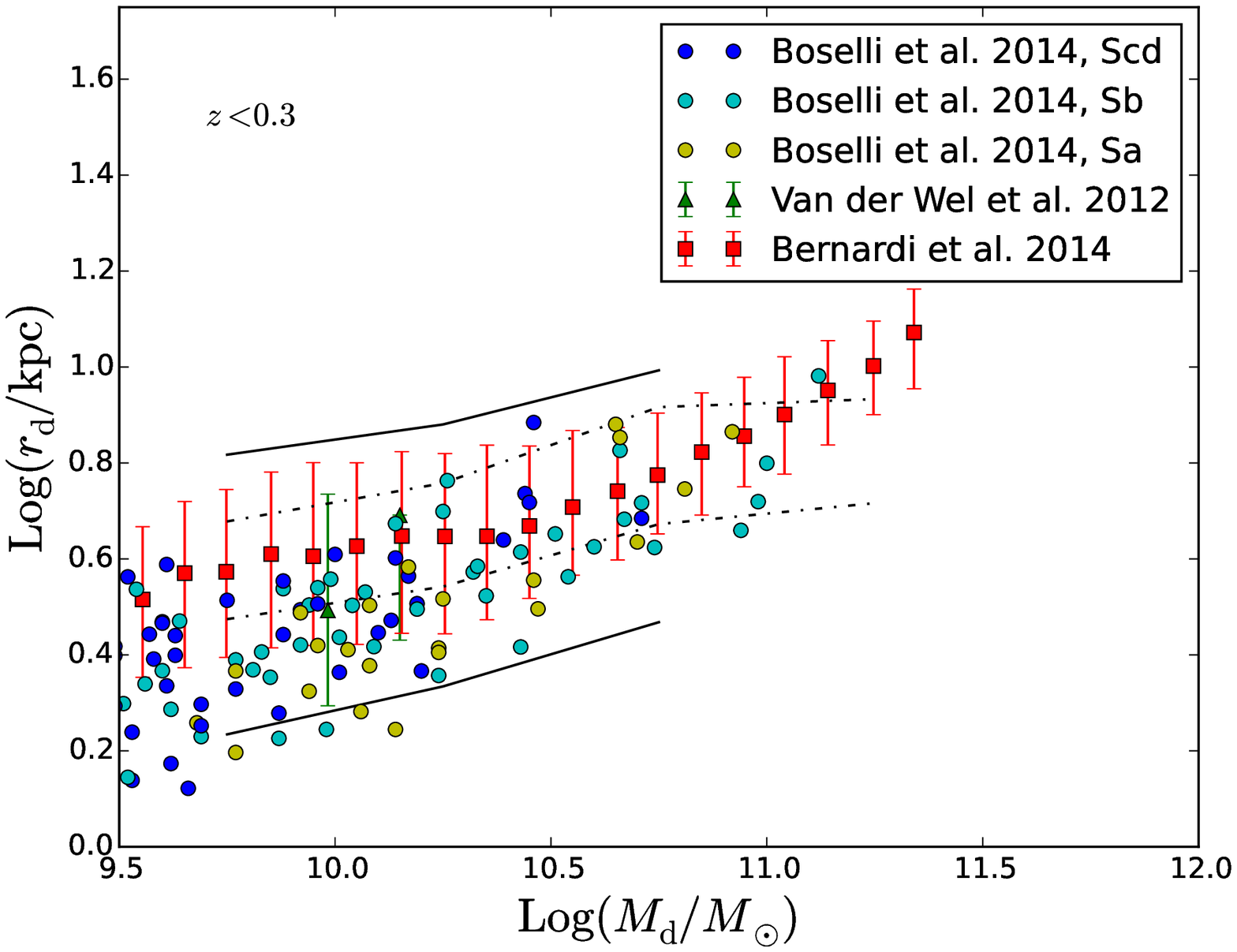} 
\end{array}$
\end{center}
\caption{The stellar mass - size (exponential scale-length) relation for discs (galaxies with $B/T<0.3$)
at different $z$ (curves: $\pm 1\sigma$ from the mean) in the default model
(solid curves) and a simplified version in which all discs have the same spin parameter $\lambda = 0.05$ (dotted-dashed curves).
The data points with error bars show the mean observational value of $r_d$ in a bin of stellar mass, while the circles without errors bars
(the local data by \citealp{boselli_etal14}) are measurements for individual galaxies and are colour-coded according to morphology
(blue: Sd or Sc; cyan: Sb; yellow: Sa).
The data of \citet{vanderwel_etal12} and \citet{shibuya_etal15} are for galaxies classified as late type, while \citet{bruce_etal14} 
and \citet{bernardi_etal14} performed a bulge/disc decomposition.
The solid curves stop at $10^{10.75}\,M_\odot$ because, in the default model,
there are not enough discs with $M_d>10^{10.75}\,M_\odot$
to compute meaningful averages.
}
\label{DiscSizes}
\end{figure*}

\subsection{Disc sizes}

In the last section, we checked that our morphologies are reasonable. Now, we focus on spiral galaxies and, in particular, on the disc mass-size relation.
The solid curves in Fig.~\ref{DiscSizes} show the predictions of the default model ($\pm 1$ standard deviation from the mean)
for galaxies with $B/T<0.3$ at $0<z<2.5$ (see Section~3.5 for a discussion of bulge-to-total mass ratios in GalICS 2.0).

The criterion $B/T<0.3$ has been chosen to match \citet{vanderwel_etal12}’s morphological selection by \citet{sersic63} index 
(M. Huertas-Company, private communication). Concerning the other data sets used for this comparison,
\citet{shibuya_etal15} selected late-type galaxies based on star formation, while
\citet{bruce_etal14} at high $z$ and \citet{bernardi_etal14} in the local Universe performed a bulge/disc decomposition.
Their results are therefore more directly comparable to ours. Also notice that both van der Wel et al. and Bruce et al. based their investigations on CANDELS data.
The local data from \citet{boselli_etal14} are individual spiral galaxies from the {\it Herschel} Reference Survey.

The data points sit comfortably in the range predicted by the default model, with the only possible exception of the most massive discs in the local Universe
(at high $M_d$, theoretical predictions are affected by poor statistics because of the decline of the galaxy SMF combined to the increase of $B/T$ with 
$M_{\rm stars}$).
However, as \citet{dejong_lacey00} had already found in an earlier SAM that computed disc radii from the halo spin distribution measured in N-body simulations,
the scatter is much larger in GalICS 2.0 than in the observations. 
In fact, it is so large that it covers any difference between models. Hence, in Fig.~\ref{DiscSizes}, only the default model has been shown.

Most of the scatter comes from the halo spin parameter.
To prove it, we have rerun the default model using $\lambda = 0.05$ for all haloes rather than the values 
measured in the N-body simulation. The results are shown by the dotted-dashed curves in Fig.~\ref{DiscSizes}.
Some scatter is still present because galaxies differ in halo concentration, $M_{\rm gal}/M_{\rm vir}$ (Fig.~\ref{HOD}) and $B/T$
(if any of these quantities increases, the rotation speed will increase, too; so, $r_d$ has to shrink if specific angular momentum is to be conserved).
However, the disc size-mass relation is much tighter when the scatter in $\lambda$ is removed.

This finding is puzzling because: a) there are many processes and sources of errors that could contribute to the observational scatter and that our model
does not include, and b) the spin distribution in our N-body simulation is in agreement with previous studies.
We fit our distribution for $\lambda$ with a log-normal distribution with $\bar{\lambda}=0.049$ and $\sigma_{{\rm ln\,}\lambda}=0.57$ (Fig.~\ref{SpinDist}), in agreement with \citealp{munoz_etal11}, who find $\bar{\lambda}=0.044$ and $\sigma_{{\rm ln\,}\lambda}=0.57$, and
\citet{burkert_etal16},  who find $\bar{\lambda}=0.052$ and $\sigma_{{\rm ln\,}\lambda}=0.46$.

The most likely explanation is that specific angular momentum is not conserved during infall.
\citet{sharma_steinmetz05} found that, in adiabatic cosmological simulations, the specific angular momentum distribution is narrower for gas than DM, owing to the presence of counter-rotating material in the latter but not in the former.
\citet{kimm_etal11} confirmed this finding for the specific angular momentum distribution of the gas and the stars in a Milky-Way-type galaxy, which they simulated both with and without feedback. 
At the end of their simulations ($z=3$), 
the galaxy and the halo had the same specific angular momentum on a global scale despite their different internal distributions,
but that was not true at all times and nobody has performed a
 systematic study on a representative volume to determine whether the scatter in the specific angular momentum distributions of discs and haloes is the same.
Fig.~\ref{DiscSizes} suggests that for some yet unexplained reason there is less scatter for discs than for haloes.

We remark that the model with $\lambda = 0.05$ for all haloes (corresponding to the dotted-dashed lines in Fig.~\ref{DiscSizes})
contains much fewer early-type galaxies than the default model (less than half).
Many of the galaxies that become unstable and develop a pseudobulge in the default model live in haloes with $\lambda\ll 0.05$.
They are systems in which the disc is very concentrated with respect to the DM.
Turning off disc instabilities causes the lower envelope of the disc mass-size (the lower solid line in the six panels of Fig.~\ref{DiscSizes}) 
to drop considerably because the spin distribution in Fig.~\ref{SpinDist} implies a tail of galaxies with very small radii.
When disc instabilities are activated, these galaxies cease to contribute to the disc mass-size relation because they develop massive
pseudobulges and therefore no longer satisfy the
$B/T<0.3$ selection criterion.
 
\begin{figure*}
\begin{center}$
\begin{array}{cc}
\includegraphics[width=0.5\hsize]{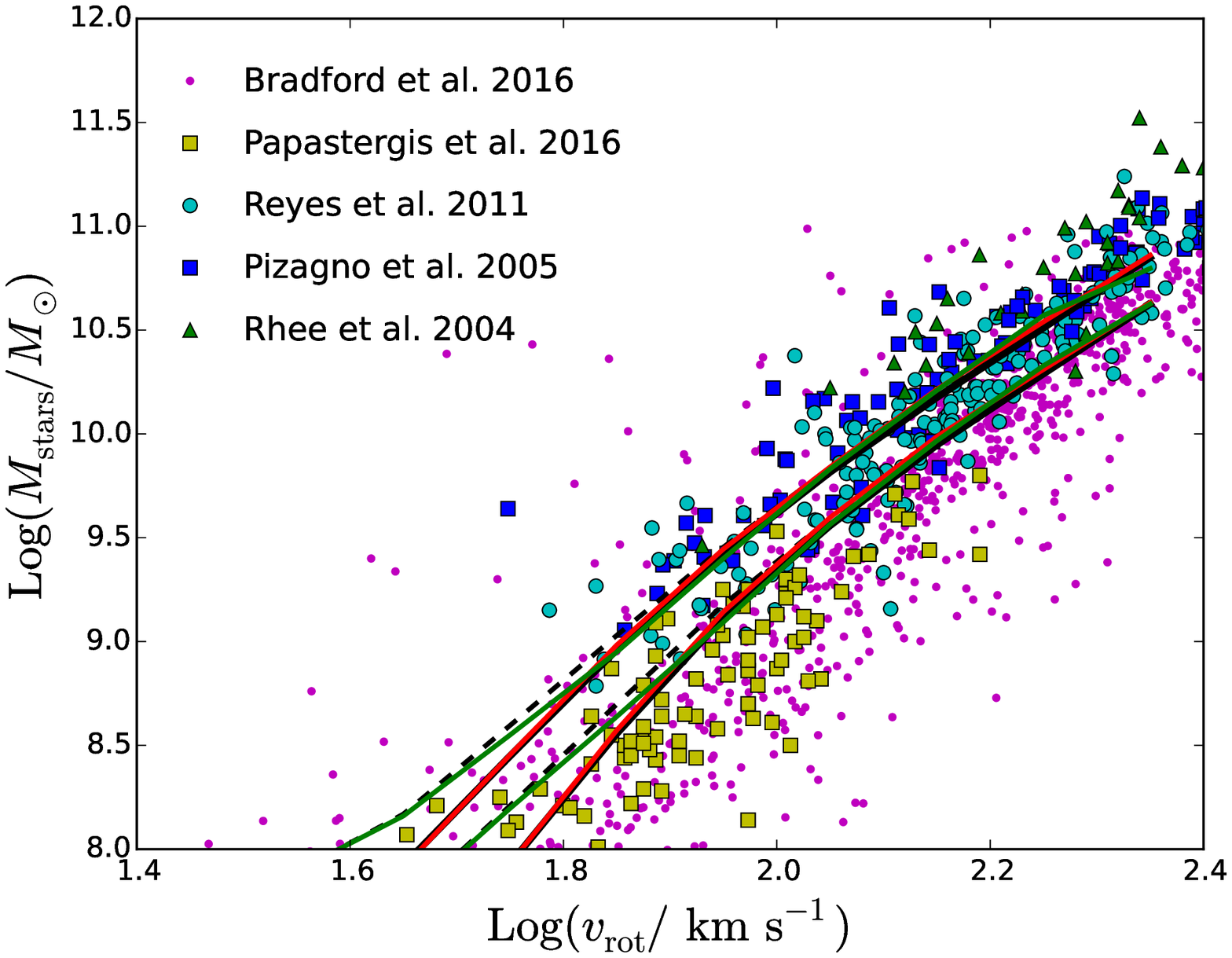} &
\includegraphics[width=0.5\hsize]{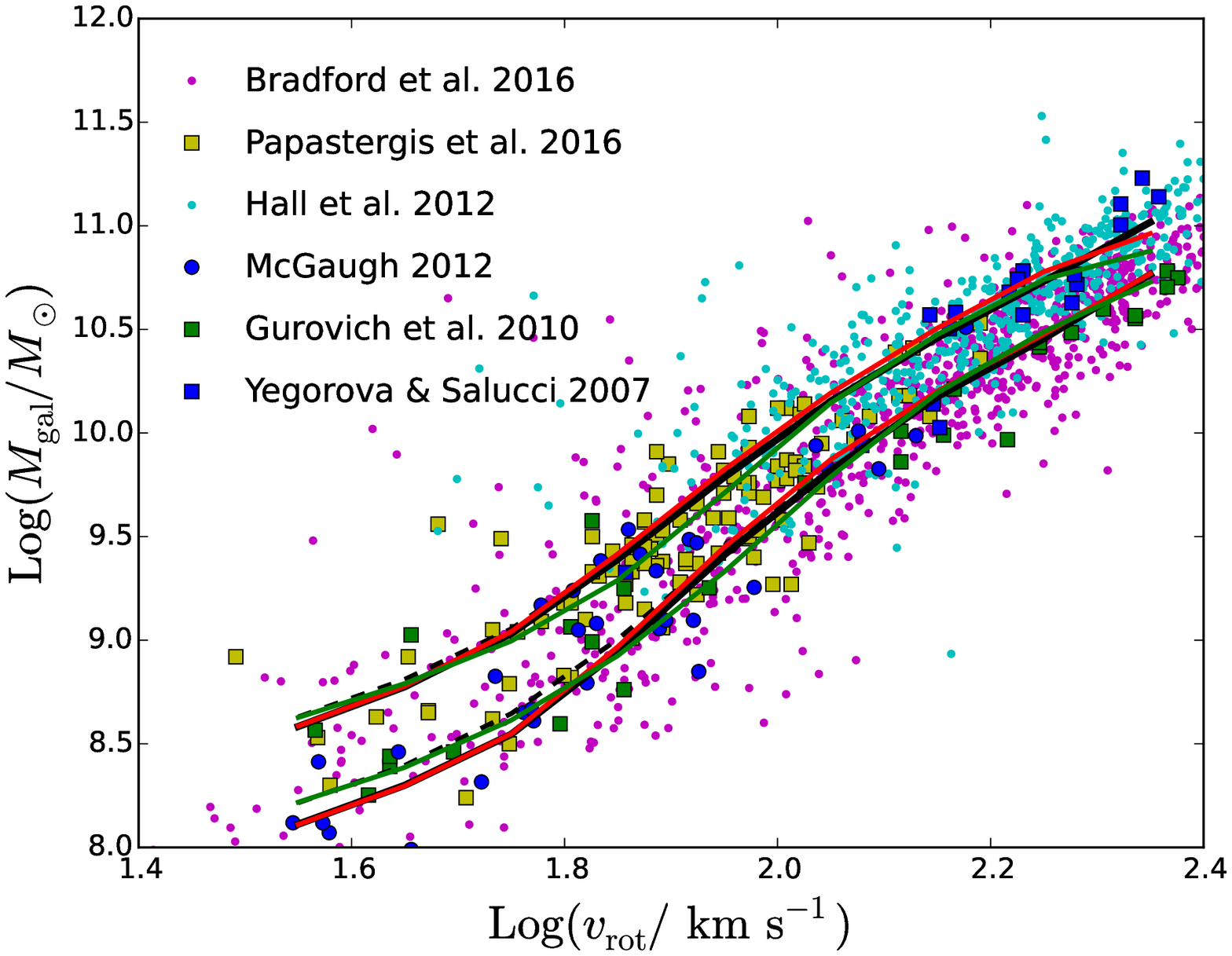} 
\end{array}$
\end{center}
\caption{The stellar (left) and baryonic (right) TFR for disc galaxies in the local Universe. The lines show the predictions of GalICS 2.0 (upper and lower quartiles).
They are colour-coded as in Fig.~\ref{BMF}: black solid, red, black dashed and green curves correspond to the default model and models~1, 2, 3, respectively.
We have selected disc galaxies using the criterion
$B/T<0.7$, where $B/T$ is computed considering both the classical and pseudobulge mass, as in Fig.~\ref{BT}.
The data points correspond to observed galaxies. Those for the stellar
TFR (left) are from \citet{bradford_etal16}, \citet{papastergis_etal16}, \citet{reyes_etal11}, \citet{pizagno_etal05} and \citet{rhee_etal04}.
In the case of the baryonic TFR (right), we have also shown data
from \citet{hall_etal12}, \citet{mcgaugh12}, \citet{gurovich_etal10} and \citet{yegorova_salucci07}.
The data points by Gurovich et al. have converted for their different estimate of $v_{\rm rot}$
(they inferred $v_{\rm rot}$ from the {\sc Hi} linewidth
at $20\%$ rather than $50\%$ of the peak height).
}
\label{TF}
\end{figure*}

\begin{figure*}
\begin{center}$
\begin{array}{cc}
\includegraphics[width=0.5\hsize]{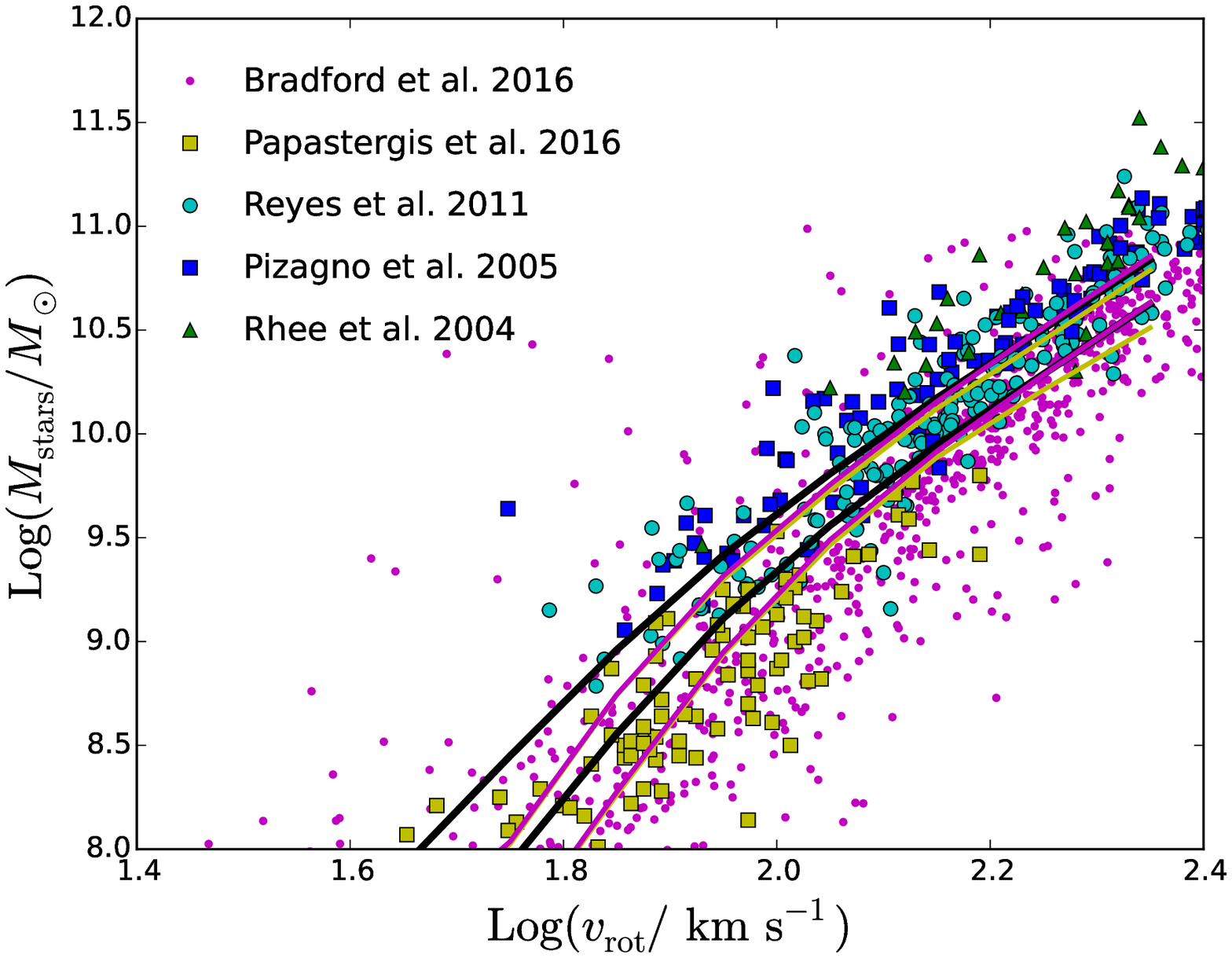} &
\includegraphics[width=0.5\hsize]{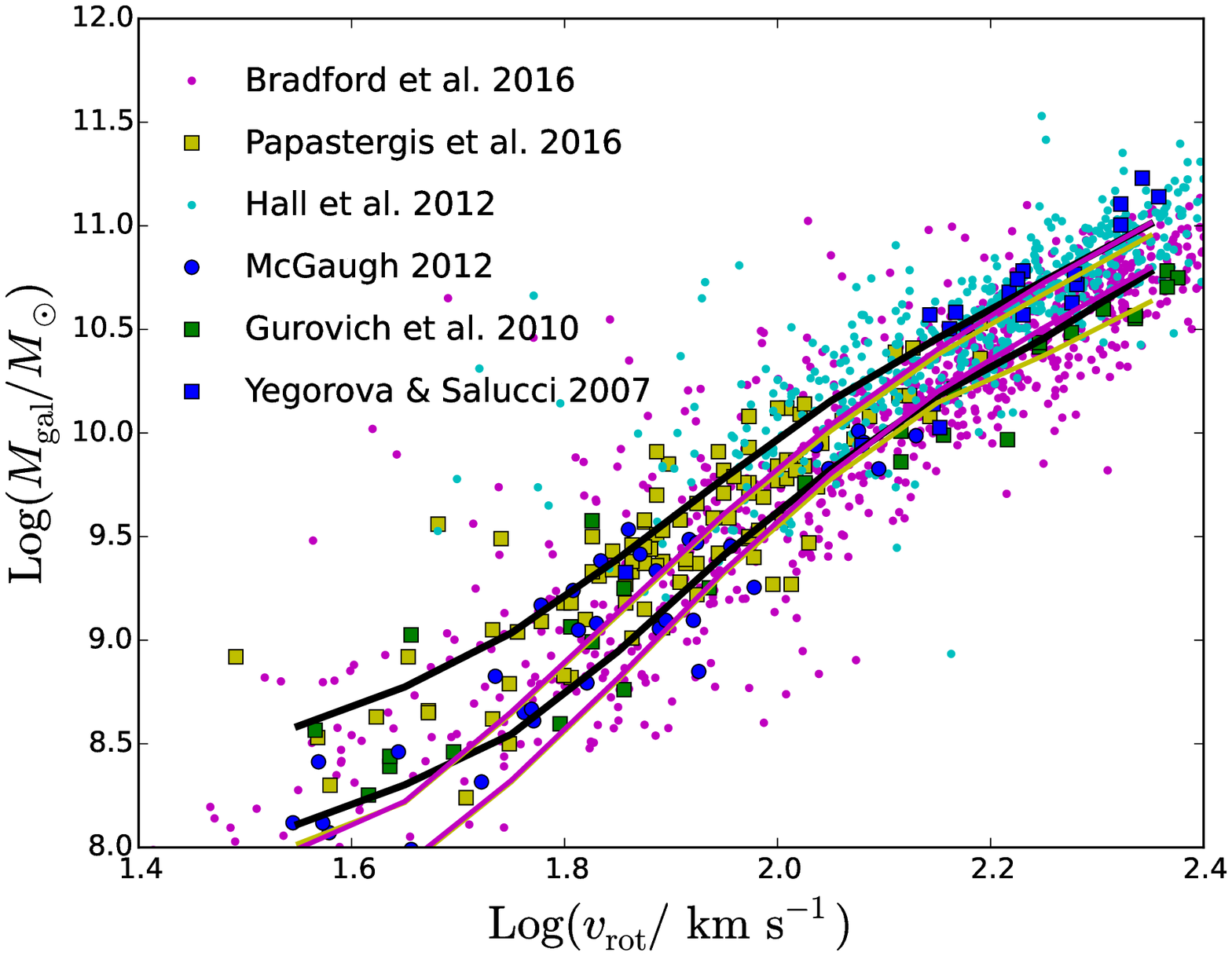} 
\end{array}$
\end{center}
\caption{Sensitivity of the TFR to errors on concentrations measurements and to adiabatic contraction.
The default model (the region between the black curves) and the data points are the same as in Fig.~\ref{TF}.
The magenta curves show how the stellar and the baryonic TFR vary when the halo concentrations that we measure in our N-body simulation are replaced by values obtained from our halo masses
by using the fitting formulae of \citet{dutton_maccio14}.
The gold curves show how adding adiabatic contraction (modeled with Eq.~\ref{adiabatic_contraction}) modifies the magenta curves.
}
\label{TFbis}
\end{figure*}
\begin{figure}
\begin{center}
\includegraphics[width=1.\hsize]{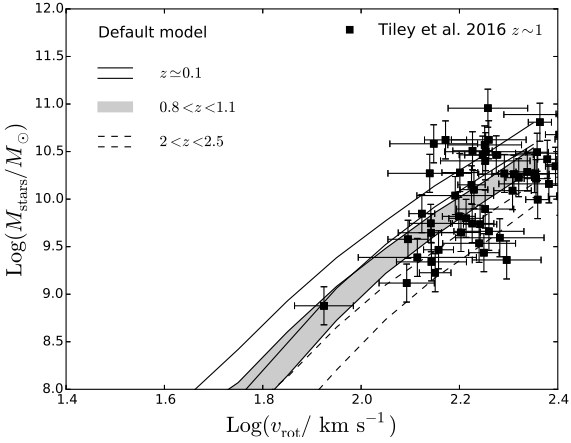} 
\end{center}
\caption{Predicted evolution of the stellar TFR from $z\sim 0$ to $z\sim 2$ (regions between lines/gray shaded area for $z\sim 1$) and comparison
to $z\sim 1$ data (\citealp{tiley_etal16}; points with error bars).}
\label{TF1}
\end{figure}

\begin{figure}
\begin{center} 
\includegraphics[width=0.95\hsize]{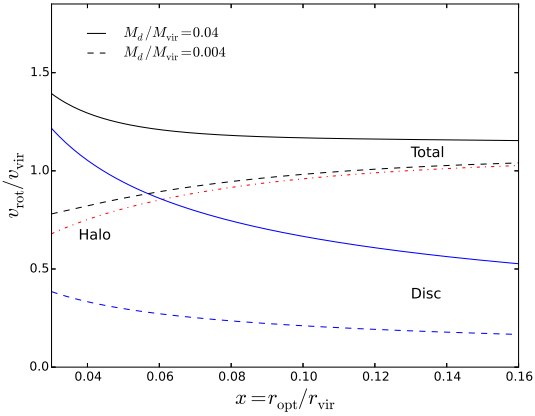}%& 
\end{center}
\caption{Dependence of $v_{\rm rot}/v_{\rm vir}$ on $r_{\rm opt}/r_{\rm vir}$ and $M_d/M_{\rm vir}$. The total value (black curves) 
is split in the contributions of the halo (red dotted-dashed curve, computed for a concentration parameter of $c=8$) and the disc (blue curves).
The disc contribution depends on $M_d/M_{\rm vir}$ and has been shown for two values: $M_d/M_{\rm vir}=0.04$ (blue solid curve) and  $M_d/M_{\rm vir}=0.004$ (dashed blue curve).
The black solid curve and the black dashed curve are obtained by summing the red dotted-dashed curve in quadrature 
with the blue solid curve and the blue dashed curve, respectively.
Notice that this figure is similar but not identical to the rotation curve ($v_{\rm rot}/v_{\rm vir}$ as a function of $r/r_{\rm vir}$) because the factor $1.11$ in Eq.~(\ref{vrot2}) 
is specific to the optical radius $r_{\rm opt}=3.2r_d$.}
\label{RC}
\end{figure}

\begin{figure}
\begin{center}
\includegraphics[width=0.95\hsize]{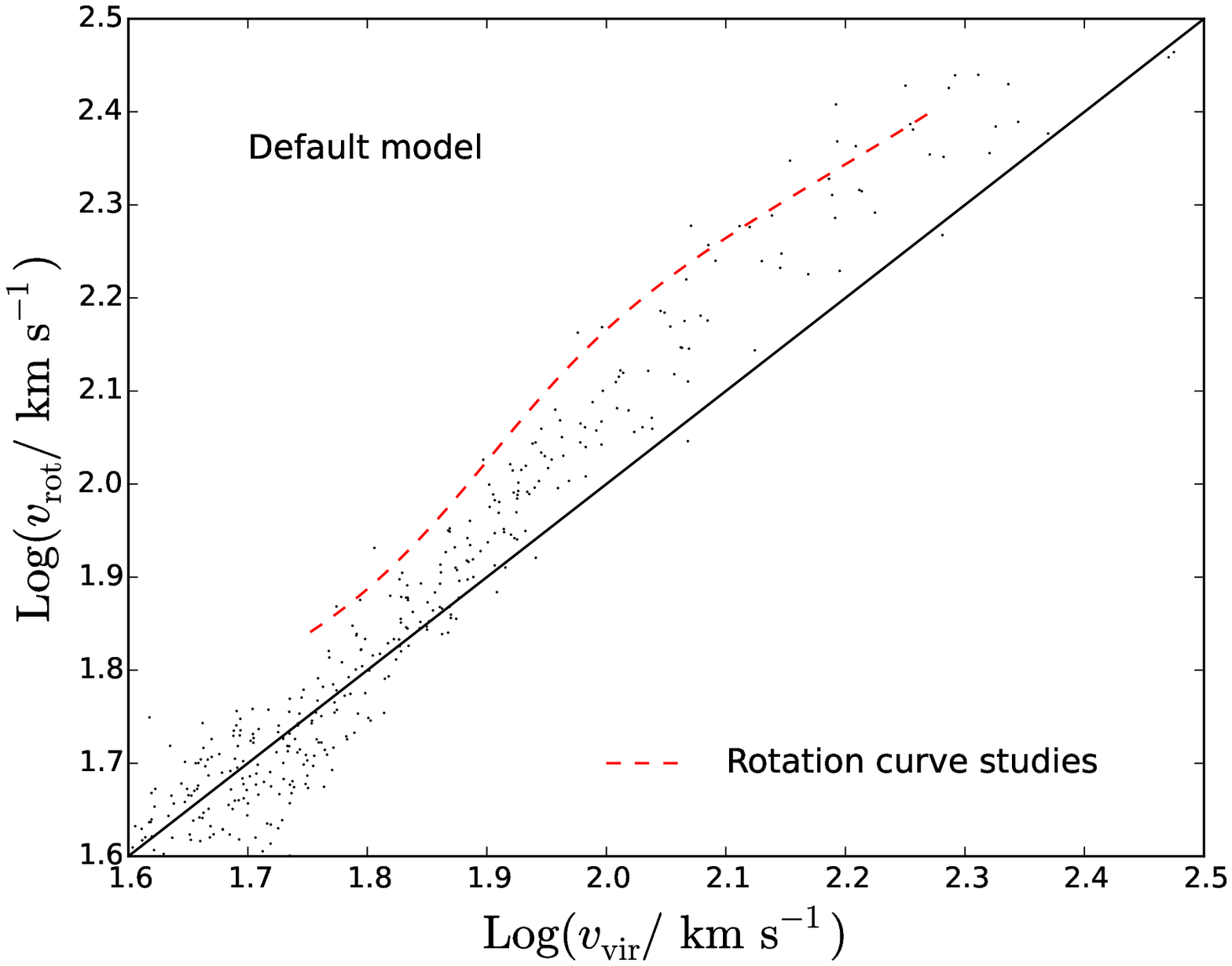}
\end{center}
\caption{The local disc rotation speed - virial velocity relation for model galaxies (point cloud).
Disc rotation speeds are given at the optical radius ($3.2r_{\rm d}$).  The black diagonal line is $v_{\rm disc}=v_{\rm vir}$.
The red dashed curve is the relation that \citet{cattaneo_etal14} infer from rotation curve studies.
The plot excludes galaxies with $B/T>0.3$ (the bulge-to-disc ratio $B/T$ is computed considering both the bulge and the pseudobulge mass).}
\label{vdisc_vvir}
\end{figure}

\subsection{The Tully-Fisher relation} 

In this section, we compare the TFR with observations at $z\sim 0$ (Fig.~\ref{TF}) and $z\sim 1$ (Fig.~\ref{TF1}),
and we explain the physical reason why its shape is reproduced correctly by GalICS 2.0, while it was not by previous models in 
which the rotation speed was proportional to the virial velocity (e.g., \citealp{kauffmann_etal93}; \citealp{cole_etal94}; \citealp{somerville_primack99}),
although the Durham model had already relaxed this assumption in \citet{cole_etal00}.

The TFR links the stellar mass $M_{\rm stars}$ or the baryon mass $M_{\rm gal}$ (i.e., the total mass of stars and cold neutral gas)
of a spiral galaxy to its rotation speed $v_{\rm rot}$. In this article,
$v_{\rm rot}$ is the circular velocity $v_c$ at the optical radius $r_{\rm opt}=3.2r_{\rm d}$,
which contains $83\%$ of the mass of an exponential disc.
We choose this definition for consistency with previous work \citep{cattaneo_etal14} and 
because measuring $v_{\rm rot}$ at the outer edge of the disc makes our results less sensitive to the real form of the DM density profile,
which is likely to differ at the centre from the NFW model assumed in GalICS 2.0 
(\citealp{moore94}; \citealp{flores_primack96}; \citealp{persic_etal96}; but also see \citealp{swaters_etal03}).

Observations of both the stellar (Fig.~\ref{TF}, left) and the baryonic (Fig.~\ref{TF}, right) TFR
show that each individual data set is consistent with a single power-law within its intrinsic scatter,
although different data sets differ in both slope and normalization.

In GalICS 2.0, $v_{\rm vir}\propto M_{\rm vir}^{1/3}$.
Hence, the $M_{\rm stars}$ - $v_{\rm vir}$ relation is entirely determined by the $M_{\rm stars}$ - $M_{\rm vir}$ relation,
which changes slope at $M_{\rm stars}\sim 2\times\sim 10^{10}\,M_\odot$ (Fig.~\ref{HOD}).
There is no such feature in the TFR.
Hence, understanding the TFR comes down to understanding how $v_{\rm rot}$ depends on $v_{\rm vir}$.

The models presented in Fig.~\ref{TF} (curves) include the presence of bulges but let us focus on pure discs to make
the interpretation simpler. In this case:
\begin{equation}
v_{\rm rot}^2 = v_c^2(r_{\rm opt})=
{{\rm G}M_{\rm dm}(r_{\rm opt})\over r_{\rm opt}}+1.11{{\rm G}M_d\over r_{\rm opt}},
\label{vrot1}
\end{equation}
where the first addend is the halo contribution
and the second addend is the square of the rotation speed at $r_{\rm opt}=3.2r_d$
for a self-gravitating exponential disc \citep{freeman70}.
By using Eq.~(\ref{M_nfw}) for the mass $M_{\rm dm}(r_{\rm opt})$ of the DM within $r_{\rm opt}$
and by defining $x\equiv r_{\rm opt}/r_{\rm vir}$, Eq.~(\ref{vrot1}) can be rewritten as:
\begin{equation}
{v_{\rm rot}^2\over v_{\rm vir}^2} = {{\log(1+cx)\over x} -{c\over 1+cx}\over \log(1+c)-{c\over 1+c}}+{1.1\over x}{M_d\over M_{\rm vir}}.
\label{vrot2}
\end{equation}
For $\lambda = 0.05$ and $c=8$ (the mean values of the spin and concentration parameters), Eq.~(\ref{r_d}) gives values of
$x$ that range from $x=0.083$ for a dwarf galaxy ($M_d/M_{\rm vir}=0.004$) to $x=0.063$ for a Milky Way ($M_d/M_{\rm vir}=0.04$).
For comparison, the standard model $r_d=\lambda r_{\rm vir}/2$ gives $x=0.08$.
The range of typical values extends to $0.03\lsim x\lsim 0.16$ when the scatter in $\lambda$ is considered.
In this range, the first term on the right hand side of Eq.~(\ref{vrot2}) is a function that grows slowly from
$0.74$ at $x=0.03$ to $1.1$ at $x=0.16$ (Fig.~\ref{RC}; red curve). 
The second term ranges from being a small correction for dwarf galaxies 
to being comparable to the first term for galaxies with masses comparable to the Milky Way (blue dashed and solid curves in Fig.~\ref{RC}, respectively).
The dependence on $M_d/M_{\rm vir}$ is the reason why GalICS 2.0 reproduces the $v_{\rm rot}$ - $v_{\rm vir}$ relation from
studies of rotation curves (Fig.~\ref{vdisc_vvir}; the default model and model~3 are indistinguishable in this respect)
and therefore the TFR (Fig.~\ref{TF}). 

The dependence of $v_{\rm rot}/v_{\rm vir}$ on $M_{\rm gal}/M_{\rm vir}$ is also the reason why, in GalICS 2.0, the TFR is very tight despite the large scatter
in disc radii. The second term on the right hand side of Eq.~(\ref{vrot2}) decreases with $x$ and, summed to first one,
conspires to a produce a rotation curve that is nearly flat at the optical radius (dashed and solid black curves in
Fig.~\ref{TF}).
Hence, $v_{\rm rot}/v_{\rm vir}$ is not very sensitive to $x$.

Having established that the dependence of $v_{\rm rot}/v_{\rm vir}$ on $M_{\rm gal}/M_{\rm vir}$ is the physical reason why GalICS 2.0 is able to reproduce the shape of the TFR,
we can now look in closer details at the predictions of our different models.
Model~1 is very similar to the default model and model~2 is very similar to model~3.
Hence, we shall focus our discussion of Fig.~\ref{TF} on the default model (black solid curves) and model~3 (green solid curves).

We start by noting that the TFR in the two models is very similar down $M_{\rm stars}\sim 10^9\,M_\odot$.
As our results below this mass may be affected by halo-mass resolution (discussion in Sections~3.1 and~3.2),
the differences between the default model and model~3 should not be overinterpreted.
Nevertheless, these differences make sense when one considers the SMFs predicted by these models.
In the default model,  the SMF has a constant shallow slope in the mass range
$10^8\,M_\odot<M_{\rm stars}<10^{10}\,M_\odot$ and the stellar TFR has a small concavity below $M_{\rm stars}\sim 10^9\,M_\odot$.
In model~3, the SMF has a steeper faint-end slope in better agreement with the observations
 ($M_{\rm stars}/M_{\rm vir}$ decreases less rapidly at low masses) and the concavity is absent.

The baryonic TFR becomes slightly concave below $M_{\rm gal}\sim 10^{10}\,M_\odot$ but the concavity turns to a convexity below $M_{\rm gal}\sim 10^9\,M_\odot$
because, below this mass, the gas surface density 
is lower than the threshold $\Sigma_{\rm th}$ for star formation (Section~2.4.1 and Table~1). Hence, there is neither star formation nor ejection of gas.
It is possible that these concavities and convexities would not be noticeable if measurement errors on masses and speeds were included in our analysis.

The greatest uncertainty in our predictions for the TFR derives from our incapacity to measure concentrations accurately for haloes with $M_{\rm vir}<10^{12}\,M_\odot$ (Section~2.1.1).
To understand the impact that errors on concentration measurements could have on our results, we have rerun the default model replacing our measurements with
concentrations computed with the fitting formulae of Dutton \& Macci{\`o} (2014; Fig.~\ref{TFbis}, magenta curves).
The difference with respect to the TFR obtained using our concentrations (black curves) is negligible at $M_{\rm stars}>10^{9.5}\,M_\odot$.
Below this mass, the higher concentrations from the fitting formulae of \citet{dutton_maccio14} cause galaxies with $M_{\rm stars}>10^{8.5}\,M_\odot$ to turn with rototation speeds $\sim 15\%$ higher.
Hence, the concavity of the stellar TFR in default model becomes more pronounced.
 
All the models considered until now assume no adiabatic contraction (our standard assumption throughout this article).
However, the role of adiabatic contraction on the zero point of the TFR has been a matter of discussion in previous SAM studies \citep{cole_etal00,guo_etal11}.
We have therefore considered another variant of the default model, in which not only do we use concentrations computed with the fitting formulae of \citet{dutton_maccio14} but also
we include adiabatic contraction modeled as in Blumenthal et al. (1986; Section~2.3.3). 
The TFR for this model is shown by the region within the gold curves in Fig.~\ref{TFbis}.
Adiabatic contraction produces measurable effects only in galaxies with $M_{\rm gal}>10^{10}\,M_\odot$.
At lower masses, $M_{\rm gal}/M_{\rm vir}$ is too low for the baryons to produce any noticeable effect on the DM,
even when the higher concentration of the former with respect to the latter is accounted for.
Above $M_{\rm gal}>10^{10}\,M_\odot$, adiabatic contraction increases the DM mass within the galaxy and thus the disc rotation speed.
The increase is stronger at higher masses. Hence, adiabatic contraction tilts the slope of the TFR (it causes it to be less steep at high masses).

Assessing the agreement with observations  is not straightforward because the answer depends on the data used for the comparison.
The measurements by \citet{papastergis_etal16} at low masses provide the main observational hint for a concave stellar TFR
when combined with those by \citet{bradford_etal16} and \citet{reyes_etal11} at higher masses,
but the same data are perfectly consistent with a single power-law if taken in conjunction with those by \citet{rhee_etal04}, instead.
This stresses the danger of combining different data sets without fully understanding their systematics.

The TFRs by \citet{bradford_etal16}, \citet{papastergis_etal16}, \citet{mcgaugh12} and \citet{gurovich_etal10} are based on {\sc Hi} linewidths
(McGaugh uses the rotation speed in the outermost region, where the rotation curve is approximately flat) and are all more or less consistent with one another.
In contrast, Hall et al. (2012; {\sc Hi} linewidths), Yegorova \& Salucci (2007; resolved rotation curves from {\sc H}$\alpha$ and {\sc Hi} data),
Pizagno et al. (2005; {\sc H}$\alpha$ linewidths) and Rhee et al. (2005; {\sc H}$\alpha$ linewidths) find a normalization that is higher by 
$\sim 0.3\,$dex in mass
(the systematic shift in $M_{\rm stars}$ between the SMFs of \citealp{baldry_etal08} and \citealp{bernardi_etal14}).
\citet{yegorova_salucci07} are the only ones among the aforementioned authors to measure $M_{\rm gal}$ dynamically by performing a halo/disc decomposition (the other authors measured $M_{\rm stars}$
by assuming a mass-to-light ratio or by using a stellar population synthesis model).
By using resolved rotation curves, they could measure $v_{\rm rot}$ at exactly $r_{\rm opt}=3.2r_d$.
The {\sc H}$\alpha$ measurements by \citet{reyes_etal11} are intermediate but closer to the results of other  {\sc H}$\alpha$ studies.
Therefore, for a same $M_{\rm stars}$ or $M_{\rm gal}$, the rotation speeds from {\sc Hi} studies tend to be systematically
higher than those from {\sc H}$\alpha$ studies.

This finding has a simple explanation.
The {\sc H}$\alpha$ emission probes star formation, which is mainly concentrated in the inner regions of galaxies ($r<r_{\rm opt}$),
while the {\sc Hi} emission extends beyond the limit $r_{\rm opt}$ of the stellar disc.
The rotation curves of discs with $M_{\rm stars}\lsim 10^{10.5}\,M_\odot$ are usually rising \citep{persic_etal96}.
Hence, the gas seen in {\sc Hi} has higher rotation speeds than the gas seen in {\sc H}$\alpha$.
This also explains why
the stellar TFR in GalICS 2.0 is intermediate between the {\sc Hi} determination by \citet{bradford_etal16} and the {\sc H}$\alpha$ determination by \citet{pizagno_etal05}, 
though all our models are consistent with the {\sc H}$\alpha$ measurements by \citet{reyes_etal11}.

An important question in relation to the dwarf-galaxy data 
by \citet{papastergis_etal16} is why GalICS 2.0 appears to be in good agreement with them for the baryonic 
TFR but not for the stellar one.
A straightforward interpretation of this finding is that the baryonic masses are computed correctly but star formation has not been efficient enough.
This interpretation would suggest that gas-to-stellar mass ratios are overestimated, not underestimated, as the baryonic mass function seemed to
suggest
(direct measuremts of $M_{\rm gas}/M_{\rm stars}$ suggested that gas fractions are reproduced correctly, 
particularly in the default model).
Another explanation is that the discrepancy is a selection effect. \citet{papastergis_etal16} selected their galaxies to be heavily gas-dominated.
Thus, it is normal that, for a fixed baryon mass, their galaxies contain less stars than average.
The difficulty with this explanation is that we see a similar behaviour in the data by \citet{bradford_etal16}, 
which should be less biased in this sense.
A third explanation is that dwarf galaxies have rising rotation curves. Therefore, {\sc Hi} data overestimate the rotation speed at $r_{\rm opt}$
(see above). If one followed this explanation, one should apply the same consideration 
to the baryonic TFR and conclude that the gas fractions of dwarf galaxies are indeed
underestimated (a possibility that we had previously argued based on the baryonic mass function by \citealp{papastergis_etal12}).

A final point to consider when comparing models to observations is the implicit assumption that discs are infinitely thin and cold,
so that the rotation speed $v_{\rm rot}$ is equal to the circular velocity $v_c$ required to support a particle on a circular orbit against gravity.
However, $v_{\rm rot}$ can be lower than $v_c$ if random motions contribute to the support against gravity.
Therefore, a model like the one with adiabatic contraction,
which appears to get the wrong slope for the TFR (gold curves in Fig.~\ref{TFbis}), 
might not be incompatible with the data if there were a lesser degree of rotational support in the discs
of massive spiral/S0 galaxies.

Fig.~\ref{TF1} shows the evolution of the stellar TFR in the default model from $z\sim 0$ (region between the solid curves)
to $z\sim 1$ (gray shaded area) and $z\sim 2-2.5$ (region between the dashed curves).
The evolution in the other models is similar. Only the detailed shape of the relation at low masses differs.
This evolution is driven not only by a decrease in $M_{\rm stars}/M_{\rm vir}$ when moving to higher redshifts
\citep{behroozi_etal13,birrer_etal14,coupon_etal15,mccracken_etal15,skibba_etal15}, which, in GalICS 2.0, is
not strong enough to  explain the evolution of the GSMF with $z$ (Fig.~\ref{SMFs}),
but also by the increase of $v_{\rm vir}$ with $z$ for a fixed $M_{\rm vir}$.

The stellar TFR at $z\sim 1$ can be compared to {\sc H}$\alpha$ data 
from the KMOSS Redshift One Spectroscopic Survey (KROSS; \citealp{tiley_etal16}).
To quantify the evolution of the TFR from $z\sim 0$ to $z\sim 1$, Tiley et al. fitted
the local data by \citet{pizagno_etal05}, \citet{rhee_etal04} and \citet{reyes_etal11} with a slope $3.68$,
which they then used to fit their $z\sim 1$ data and measure the change in normalization.
Any conclusion from this comparison should be tempered by the scatter in the observations. However, it is encouraging that
when we fit our default model at $z\sim 1$ with a slope $3.68$ over the $v_{\rm rot}$ range covered by the KROSS data
we find exactly the same normalization as \citet{tiley_etal16}, possibly because their procedure to measure $v_{\rm rot}$ is very similar to ours.
They use the radius that contains $80\%$ of the light. We use the radius that contains $83\%$ of the mass.
 
\subsection{The Faber-Jackson relation}

The Faber-Jackson relation (hereafter, FJR; \citealp{faber_jackson76})
 links the luminosity (in our case, the stellar mass) of an elliptical galaxy to its stellar velocity dispersion.
The FJR is to elliptical galaxies what the TFR is to spirals,
although the scatter of the FJR relation is larger than that of the TFR.
Our analysis and discussion of elliptical galaxies will not be as in depth as for spiral galaxies because elliptical galaxies are not the focus of this article
and because potential errors in the modelling of elliptical galaxies have no repercussions on the properties of spirals (the converse is not true).
We nevertheless include this paragraph on the FJR for completeness and as a sanity check
to demonstrate that our modelling of elliptical galaxies is reasonable.

\begin{figure}
\begin{center}
\includegraphics[width=0.95\hsize]{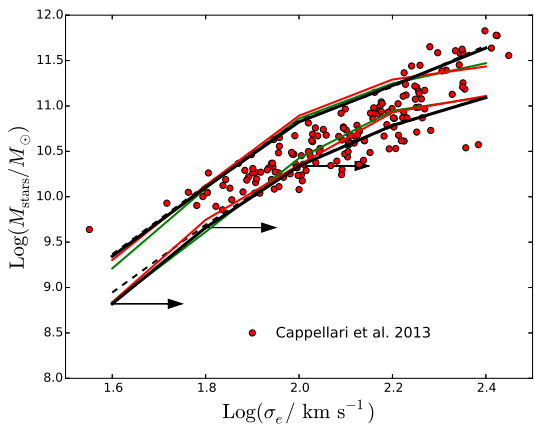} 
\end{center}
\caption{The FJR for galaxies with $B/T>0/7$ in GalICS 2.0 and in the data points of Cappellari et al. (2013, red circles).
$M_{\rm stars}$ is the total stellar mass and $\sigma_e$ is the stellar velocity dispersion within an aperture $R_e$.
The predictions of the default model, model~1, model~2 and model~3 correspond to the region of the diagram between the pair of black solid curves,
red curves, black dashed curves and green curves, respectively. For each model, the curves correspond to upper and lower quartiles for $M_{\rm stars}$ in bins of $\sigma_e$.
The arrows correspond to a displacement of $0.15\,$dex. They are based on the default model and they show the maximum increase in $\sigma_e$
that can be reasonably attributed to dissipation in gas-rich mergers.}
\label{FJR}
\end{figure}

Fig.~\ref{FJR} compares the FJR predicted by the four models in Table~1 with the data points of \citet{cappellari_etal13}.
The structure coefficient $C$ (Eq.~\ref{sigma_e})
used to pass from the mass and the effective $R_e$ of an elliptical galaxy to the velocity dispersion $\sigma_e$ within an aperture $R_e$
has been calibrated on the data of \citet{cappellari_etal06,cappellari_etal13}. They find $C=2.5$.
The structure coefficient predicted by the Hernquist model assuming isotropic velocity dispersion
returns values of $\sigma_e$ that are systematically lower than those found using $C=2.5$
by a factor of $0.87$ (\citealp{courteau_etal14}, chapter~5, section~B).
Therefore, what our comparison with the FJR really probes is the dependence of $R_e$ on $M_{\rm stars}$.

GalICS 2.0 computes 
bulge radii from energy conservation in major mergers (Eq.~\ref{en_cons2}).
This assumption gives results in reasonable agreement with the FJR (Fig.~\ref{FJR}).
The agreement is better for the default model, which allows the accretion of gas onto the central galaxies of haloes with masses up to $M_{\rm shutdown}=10^{12.7}\,M_\odot$,
than it is for models~1 and~3, where gas accretion shuts down at a lower halo mass.

However, Fig.~\ref{FJR} shows that our SAM tends to overestimate the lower envelope of the FJR for
 $M_{\rm stars}<10^{11}\,M_\odot$.
This tendency has a simple physical explanation. Gas can radiate its internal energy, invalidating the assumption of energy conservation.
Hydrodynamic simulations by \citet{covington_etal08} show that,
in gas-rich mergers, dissipation can cause the radii of bulges to contract by up to a factor of two.
 At constant mass,
shrinking radii translate into higher velocity dispersions.

Dissipation will be more significant for ellipticals with $M_{\rm stars}\lsim 10^{11}\,M_\odot$, 
the cuspy profiles of which are consistent with a dissipational origin in gas-rich (`wet') mergers. In contrast,
ellipticals with $M_{\rm stars}\gg 10^{11}\,M_\odot$ have cores that are consistent with their formation through dissipationless (`dry') mergers
\citep{kormendy_etal09,bernardi_etal11}.
\citet{cattaneo_etal11} confirmed this picture by showing that the transition from a dissipative regime
to a dissipationless one at $M_{\rm stars}\sim 10^{11}\,M_\odot$
is a direct consequence of the shutdown of gas accretion above a critical halo mass.
The arrows in Fig.~\ref{FJR} show the maximum increase in $\sigma_e$ that dissipation could plausibly cause in bulges with  $M_{\rm stars}\lsim 10^{11}\,M_\odot$.
They correspond to a contraction in radius by a factor of two (the maximum value allowed by \citealp{covington_etal08}).
Fig.~\ref{FJR} shows that dissipation could easily explain any systematic difference between the default model and the data points.

Incidentally, we note that the data points of \citet{cappellari_etal13} overlap with those for the stellar TFR relation 
if we plot $M_{\rm stars}$ on the $y$ axis and $\sqrt{3}\sigma_e$ on the $x$ axis.
The overlap of the FJR with the TFR after correcting $\sigma_e$ by a factor of $\sqrt{2}$ or $\sqrt{3}$ was first remarked by
\citet{kassin_etal07}.

\section{Conclusion}

GalICS 2.0 is a new semianalytic code that we run on merger trees from an N-body simulation in a {\it Planck} cosmology.
The simulation is used to follow the evolution of DM haloes in mass, position, angular momentum and concentration.
In the version used for this article, GalICS 2.0 assumes a one-to-one correspondence between galaxies and haloes/subhaloes.
A beta version that includes the possibility of a delay between halo and galaxy mergers shows no difference with respect to our conclusions.

The masses of luminous galaxies within haloes are determined by: 
i) the rate at which gas flows to the centre and accretes onto galaxies, 
ii) the rate at which the accreted gas is converted into stars, and
iii) the rate at which gas is blown out of galaxies.
 
In GalICS 2.0, galaxies grow through accretion of cold flows \citep{keres_etal05,dekel_birnboim06,dekel_etal09}.
Shock-heated gas never cools and ejected gas is never reaccreted.
Hence, the shock-heated fraction determines the accretion rate
(reionization heating is included but neglible on the scales resolved by our N-body simulation).
In the Horizon/Mare Nostrum cosmological hydrodynamic simulation \citep{ocvirk_etal08}, shock heating begins above $M_{\rm vir}=10^{10.7}\,M_\odot$ and
is complete when $M_{\rm vir}=10^{12.7}\,M_\odot$. This is the default assumption 
to compute the accretion rate onto galaxies in GalICS 2.0.
A sharper transition from cold accretion to shock heating (model~1) improves the fit to the SMFs by 
\citet{yang_etal09} and
\citet{bernardi_etal13} but not to other data sets (Fig.~\ref{SMFs}, $z\simeq 0.1$). 
Assuming no cooling and no reaccretion is admittedly extreme but we deliberately intended to explore how far it can take us.
The effects of reintroducing these processes will be explored in a future publication, where we shall also discuss which observations can constrain their importance.

The assumption ${\rm SFR}=\epsilon_{\rm sf}M_{\rm gas}/t_{\rm dyn}$ is standard in SAMs
and has been calibrated on observational data (\citealp{boselli_etal14}; Fig.~\ref{tSFR})
In contrast, outflow rates are considerably uncertain.
We have computed them assuming that a fraction $\epsilon_{\rm SN}$ of the power output from SNe is used to heat gas to $T_{\rm vir}$ and/or 
to blow it out of haloes. This fraction
depends on the depth of gravitational potential well (measured by $v_{\rm vir}$) with an exponent tuned to fit the local SMF.

In agreement with previous studies (\citealp{cole_etal94,somerville_etal08,benson_bower10,guo_etal11,benson12}; but see \citealp{henriques_etal13}),
we find that $\eta_{\rm SN}$ must depend strongly on $v_{\rm vir}$
to reproduce the shallow slope of the local SMF at $10^9\,M_\odot<M_{\rm stars}<10^{11}\,M_\odot$.

Below $M_{\rm stars}\sim 10^9\,M_\odot$, the slope of the observed SMF steepens again \citep{baldry_etal08,baldry_etal12}.
This is unlikely to a reionization signature because 33 out 38 dwarf galaxies in the Local Group formed the bulk of their stars after cosmic reionization \citep{weisz_etal14}.
In contrast, GalICS 2.0 shows that this feature arises naturally when we require that $\epsilon_{\rm SN}$ cannot exceed a maximum value.
The efficiency of SN feedback increases as $\epsilon_{\rm SN}\propto v_{\rm vir}^{-4}$ to keep the slope of the SMF shallow until it reaches a saturation value 
$\epsilon_{\rm max}$, after which it the slope starts rising again.
For our default choice $\epsilon_{\rm max}=1$, the saturation scale is lower than the N-body resolution.
However, lowering the maximum feedback efficiency to $\epsilon_{\rm max}=0.12$ 
\footnote{Our efficiency parameter is likely to underestimate the real feedback efficiency required by our model (Section~2.4.2) but the difference
is only a factor $\sim 1.5$ if the ejected gas is heated to the virial temperature and mixed with the hot atmosphere.} (model~2)
changes the slope at $M_{\rm stars}\lsim 10^{9.5}\,M_\odot$, which becomes steeper (Fig.~\ref{SMFs}).
This is beneficial for the agreement with the local SMF but worsens the problem at high $z$, where even the default model overestimates the number density of low-mass galaxies.
This is a general problem of SAMs (\citealp{guo_etal11}; \citealp{henriques_etal12}; Asquith et al., in preparation; but see \citealp{henriques_etal13}).
In contrast, GalICS 2.0 has no difficulty with the number density of massive galaxies that are already in place at $z\sim 2-2.5$,
which used to be a major challenge for SAMs until a few years ago.

If we look at the SFR density integrated over all galaxy masses, then, despite overpredicting the number densities of low-mass galaxies,
the default model is consistent with the observations within the errors at $2<z<8$
(Fig.~\ref{Madau}). In contrast, model~2 is clearly above the data points because the low-mass galaxy excess is too large this time.
Overall, our analysis favours strong feedback at high $z$ (the default model is the best in relation to both the SMF and the cosmic SFR density)
and a lower efficiency at low $z$ to explain why the slope of the SMF rises again below $M_{\rm stars}\sim 10^9\,M_\odot$.

The main problem of the default model with respect to the cosmic SFR density is
that it does not seem to be capable to  reproduce its decline by a factor of $\sim 20$ from $z\sim 2$ to $z\sim 0$.
This is another general problem of SAMs (see \citealp{knebe_etal15} for an extensive comparison of all major codes).
In GalICS 2.0, the local SFR density excess (a factor of $\sim 2$ with respect to the observations)
is due to an excess in the number density of galaxies with ${\rm SFR}>30\,M_\odot{\rm\,yr}^{-1}$ (Fig.~\ref{SFRf}).
This excess is also visible on the SFR - $M_{\rm stars}$ diagramme (Fig.~\ref{SFR_mass}). 
We can curb it by lowering the shutdown mass, above which shock heating is complete,
as in model~3, a parameter combination with $M_{\rm shutdown}=10^{12.3}\,M_\odot$ and $\epsilon_{\rm max}=0.12$ designed to fit the local SMF by \citet{baldry_etal08,baldry_etal12}. This choice is in better agreement with
the number density of galaxies with ${\rm SFR}>30\,M_\odot{\rm\,yr}^{-1}$,
though it underpredicts the number density of galaxies with  ${\rm SFR}=10-20\,M_\odot{\rm\,yr}^{-1}$, 
which the default model reproduces it correctly (Fig.~\ref{SFRf}). Model~3 also
reproduces correct bimodality scale at which the transition from the main sequence of star-forming galaxies to the quiescent population occurs 
(Fig.~\ref{SFR_mass}) and the evolution of the cosmic SFR density in the redshift range $0<z<2$. In fact, we are not aware of another SAM that reproduces
so well the decline of the cosmic SFR density between $z\sim 2$ and $z\sim 0$.

The assumption that the masses and growth histories of DM haloes determine the properties of galaxies is at the heart and foundation of SAMs.
Halo quantities determine the accretion rate onto galaxies, the dynamical time, which sets the star formation timescale, and the feedback efficiency.
Hence, direct probes of the galaxy - halo connection are particularly significant.
GalICS 2.0 is in good agreement with the halo mass estimates from weak lensing for late-type galaxies \citep{reyes_etal12} 
and satellite kinematics for early-type galaxies (\citealp{wojtak_mamon13}; Fig.~\ref{HOD}).
We also find an unprecedented fit to the baryonic and stellar TFR at $z\sim 0$ (Fig.~\ref{TF}), as well as the correct evolution when passing from $z\sim 0$ to $z\sim 1$
(Fig.~\ref{TF1}; recent data by \citealp{tiley_etal16}).

Reproducing the SMF and the TFR simultaneously has been a major challenge for SAMs since their inception \citep{kauffmann_etal93,heyl_etal95}.
This difficulty has a simple explanation.
The characteristic knee of the galaxy SMF around $M_{\rm stars}\sim 10^{10.5}\,M_\odot$ arises because the slope of the $M_{\rm stars}$ - $M_{\rm vir}$ relation changes from
approximately $M_{\rm stars}\propto M_{\rm vir}^2$ to approximately  $M_{\rm stars}\propto M_{\rm vir}^{1/2}$ around $M_{\rm vir}\sim 7\times 10^{11}\,M_\odot$ (Fig.~\ref{HOD}).
Therefore, assuming $v_{\rm rot}\propto v_{\rm vir}\propto M_{\rm vir}^{1/3}$ will necessarily predict a bend in the TFR if a model is in agreement with the SMF.

GalICS 2.0 is not affected by this problem because we do not assume $v_{\rm rot}\propto v_{\rm vir}$. We compute the disc rotation speed from the disc self-gravity,
the gravity of the central bulge and the gravitational potential of the DM halo (we fit NFW profiles to the haloes identifed in the N-body simulation).
Interestingly, \citet{cole_etal00} followed an approach similar to ours (they computed the rotation speeds of spiral galaxies at the disc half-mass radius
taking the gravity of the disc, the bulge and the halo properly into account) and they did find a much better agreement with the shape of the TFR than previous studies.
In those days, however, the offset of the zero point received more attention. For a same $I$-band magnitude, the disc rotation speeds predicted by \citet{cole_etal00} are systematically higher than those of the Sd-Sc galaxies of \citet{dejong_lacey00} by $\sim 30\%$ on average.

\citet{cole_etal00}'s model for adiabatic contraction (based on \citealp{blumenthal_etal86}) is partially responsible for this offset
because it overestimates the phenomenon \citep{gnedin_etal04,abadi_etal10} and thus the DM mass with the disc.
Besides the limitations of the model used to describe adiabatic contraction, it possible that the phenomenon itself may be largely compensated by
adiabatic expansion during massive outflows, the extent of which is not easily quantified.

The models presented in this article neglect adiabatic contraction but we have made the experiment to run our default model with adiabatic contraction modeled as in \citet{blumenthal_etal86}. 
The only noticeable differences were in the TFR (Fig.~\ref{TF}), which shifted to higher rotation speeds but not by a large amount.
The differences that adiabatic contraction makes are larger at high masses, where $M_{\rm gal}/M_{\rm vir}$ is larger and thus we expect a stronger effect of the baryons on the central density of the DM
in the model with adiabatic contraction.

In recent versions (e.g., \citealp{guo_etal11}), the Munich model, too, has relaxed the assumption $v_{\rm rot}\simeq v_{\rm vir}$
by assuming that $v_{\rm rot}$ is the maximum circular velocity of the DM halo.
However, even with this improvement, the agreement with the shape of the TFR is not as good as the one in Fig.~\ref{TF} because the model does not
consider the dependence of $v_{\rm rot}/v_{\rm vir}$ on $M_{\rm gal}/M_{\rm vir}$,
and this is the essential point to reproduce the galaxy SMF and the TFR
simultaneously.

Fig.~\ref{RC} shows that $v_{\rm rot}\simeq v_{\rm vir}$ at $r_{\rm opt}=3.2\lambda r_{\rm vir}/2\sim 0.08 r_{\rm vir}$ 
for dwarf galaxies with $M_{\rm gal}/M_{\rm vir}\sim 10^{-3}-10^{-2}$ (black dashed curve), while  $v_{\rm rot}\simeq 1.5v_{\rm vir}$ at the same radius
for galaxies with masses comparable to the Milky Way, in which $M_{\rm gal}/M_{\rm vir}\sim 0.04$ (Fig.~\ref{HOD}).
The relation between $v_{\rm rot}$ and $v_{\rm vir}$ that the dependence of  $v_{\rm rot}/v_{\rm vir}$ on $M_{\rm gal}/M_{\rm vir}$ produces (Fig.~\ref{vdisc_vvir})
is in agreement with previous findings from rotation curve studies \citep{cattaneo_etal14} and 
compensates the relation between $M_{\rm stars}$ and $v_{\rm vir}$,
producing a nearly straight TFR.

Fig.~\ref{TF} shows that the compensation is not perfect because the stellar TFR deviates from a single power-law at $M_{\rm stars}<10^9\,M_\odot$.
In the default model and model~1, which correspond to a shallow low-mass end of the galaxy SMF (as in \citealp{bernardi_etal13}), 
the predicted TFR is slightly concave below $v_{\rm rot}\sim 80{\rm\,km\,s}^{-1}$.
In models~2 and~3, which correspond to the steep low-mass slope of \citet{baldry_etal08,baldry_etal12}, the predicted TFR is slightly convex below $v_{\rm rot}\sim 80{\rm\,km\,s}^{-1}$.
Three caveats must be taken into account when discussing these differences. First,
observational errors on $M_{\rm stars}$ and $v_{\rm rot}$ may blur these differences.
Second, the comparison with observations at $v_{\rm rot}<80{\rm\,km\,s}^{-1}$ should also have in made that the data for dwarf galaxies are all based on {\sc Hi} measurements.
The {\sc Hi} disc may extend well beyond the optical radius at which we measure $v_{\rm rot}$ in GalICS 2.0. 
This concern is more significant for dwarf galaxies because of their rising rotation curves.
Last but not least, the turnovers at low masses in model SMFs (Fig.~\ref{SMFs}) show that resolution effects can propagate up to $M_{\rm stars}\sim 10^9\,M_\odot$ (discussion in Sections~3.1 and~3.2).
Hence, any result at $M_{\rm stars}\lsim 19^9\,M_\odot$ (below the horizontal dashed-dotted line in Fig.~\ref{TF}) should be taken with caution to avoid the risk of overinterpretation.

Fig.~\ref{vdisc_vvir} differs from the relation between $v_{\rm rot}$ and $v_{\rm vir}$ that \citet{sales_etal16} measure in the {\sc eagle} and {\sc apostle} cosmological hydrodynamic simulations.
They find $v_{\rm rot}\simeq 1.15v_{\rm vir}$ over a broad range of virial velocities ($v_{\rm vir}=30-200{\rm\,km\,s}^{-1}$)
possibly because their $M_{\rm gal}/M_{\rm vir}$ ratios do not vary with $M_{\rm vir}$ as much as ours (a natural consequence of less extreme feedback).
The proportionality of $v_{\rm rot}$ with $v_{\rm vir}$ (Fig.~5 of \citealp{sales_etal16} is the reason why the {\sc eagle} SMF (\citealp{schaye_etal15}; simulation Ref-L100N1504 at $z=0.1$)
is steeper than the observations at low $M_{\rm stars}$ and shallower than the observations at high $M_{\rm stars}$ (Fig.~\ref{SMFs}).
Although the {\sc eagle} simulation was calibrated on the SMF and its agreement
with the data of \citet{baldry_etal12} is quite reasonable 
considering that a hydrodynamic simulation does not have the same freedom as a SAM,
this is the typical behaviour of models calibrated on the TFR.

The optical radius $r_{\rm opt}=3.2r_d$ at which we measure $v_{\rm rot}$ in GalICS 2.0 depends on scale-lengths computed from angular momentum conservation.
This assumption may be inaccurate on a galaxy-by-galaxy basis because hydrodynamic simulations have shown that angular momentum is not conserved for individual fluid elements \citep{sharma_steinmetz05,kimm_etal11}.
However, the assumption is consistent with the observed spin distribution of disc galaxies both at $z\sim 0$ \citep{tonini_etal06} and $z\sim 1-3$ \citep{burkert_etal16}, and
the predictions of GalICS 2.0 are in good agreement with disc sizes over the entire redshift range $0<z<2.5$, 
though we tend to see more scatter in GalICS 2.0 than in the observations (Fig.~\ref{DiscSizes}).
In any case, rotation curves around the optical radius are usually nearly flat, so the dependence of $v_{\rm rot}$ on $r_{\rm opt}$ is weak (Fig.~\ref{RC}) and any error on disc sizes
has limited impact on our predictions for the TFR.

Fig.~\ref{HOD} implies a formal resolution of $M_{\rm stars}\lsim 10^8\,M_\odot$ but we know from experience (e.g., \citealp{cattaneo_etal11})
that resolution effects could trickle up to scales an order of magnitude larger.
Therefore, it is possible that the SMF at $M_{\rm stars}< 10^9\,M_\odot$ could have contained more galaxies if we had used merger trees from an N-body simulation with $1024^3$ rather than $512^3$ particles.
As the difference would come from the number of low-mass haloes, we expect it to have limited impact on the TFR. 
We intend to make a resolution study and to present it in a future publication but we are confident that our results are robust at $M_{\rm stars}> 10^9\,M_\odot$.

The FJR is the equivalent of the TFR for elliptical galaxies. 
The focus of this article is on spiral galaxies but we added a section on the FJR for completeness.
In the same way that the radii of discs are computed by assuming that the infall of gas conserves angular momentum,
the radii of bulges are computed by assuming that mergers conserve the total mechanical energy of the merging galaxies.
Accounting for systematic departures of observed galaxies from the Hernquist model \citep{cappellari_etal06,courteau_etal14}, 
this assumption is
in reasonably good agreement with observations of the FJR \citep{cappellari_etal13}, although it tends to underestimate the velocity dispersions of lower-mass
ellipticals.
This tendency can be explained as an effect of dissipation, which plays an important role in major mergers at $M_{\rm stars}< 10^{11}\,M_\odot$ but not 
in major mergers at larger masses 
\citep{kormendy_etal09,bernardi_etal11,cattaneo_etal11}.

\section*{Acknowledgements}

We thank P.~S.~Behroozi, A.~Boselli, J.~D.~Bradford, J.~Freundlich, M.~Huertas-Company, O.~Ilbert, E.~Papastergis, P.~Salucci, and A.~L. Tiley 
for providing electronic data that have assisted us in the comparison with observations.
We also thank C.~Pichon for useful conversation.

\bibliographystyle{mn2e}

\bibliography{ref_av}
%%%%%%%%%%%%%%%%%%%%%%%%%%%%%%%%%%%%%%%%%%%%%%%%%

\label{lastpage}
\end{document}